\documentclass[12pt]{iopart}
\usepackage{iopams}
\usepackage{theorem}
\usepackage{algorithm}
\usepackage{graphicx}
\usepackage[usenames]{color}
\usepackage{multirow}
\usepackage{dsfont}

\newcommand{\eqref}[1]{\eref{#1}}

\newtheorem{Lem}{Lemma}[section]

\newenvironment{ifelse}{%
  \begin{list}{}{%
      \setlength{\topsep}{0pt}\setlength{\parskip}{0pt}
      \setlength{\partopsep}{0pt}\setlength{\itemsep}{0pt}}}%
  {\end{list}}

\makeatletter

  \gdef\listctr{list\romannumeral\the\@listdepth}\expandafter
  
\makeatother

\newenvironment{AlgorithmSteps}[1][1]{%
  \begin{list}{\csname label\listctr\endcsname}{%
      \usecounter{\listctr}
      
      \settowidth{\labelwidth}{\textsc{Step\ #1.}}%
      \setlength{\leftmargin}{\labelwidth}\addtolength{\leftmargin}{\labelsep}}}%
  {\end{list}}

\begin{document}

\title[Accelerated gradient methods for the X-ray imaging of solar flares]{Accelerated gradient methods for the X-ray imaging of solar flares}

\author{S Bonettini$^1$ and M Prato$^2$}
\address{$^1$ Dipartimento di Matematica e Informatica, Universit\`a di Ferrara, Via Saragat 1, 44122 Ferrara, Italy}
\address{$^2$ Dipartimento di Scienze Fisiche, Informatiche e Matematiche, Universit\`a di Modena e Reggio Emilia, Via Campi 213/b, 41125 Modena, Italy}
\eads{\mailto{silvia.bonettini@unife.it}, \mailto{marco.prato@unimore.it}}

\begin{abstract}
In this paper we present new optimization strategies for the reconstruction of X-ray images of solar flares by means of the data collected by the Reuven Ramaty High Energy Solar Spectroscopic Imager (RHESSI). The imaging concept of the satellite is based of rotating modulation collimator instruments, which allow the use of both Fourier imaging approaches and reconstruction techniques based on the straightforward inversion of the modulated count profiles. Although in the last decade a greater attention has been devoted to the former strategies due to their very limited computational cost, here we consider the latter model and investigate the effectiveness of different accelerated gradient methods for the solution of the corresponding constrained minimization problem. Moreover, regularization is introduced through either an early stopping of the iterative procedure, or a Tikhonov term added to the discrepancy function, by means of a discrepancy principle accounting for the Poisson nature of the noise affecting the data.
\end{abstract}

\submitto{\IP}

\section{Introduction}

For any imaging problem, the wavelength of the radiation emitted by the object under investigation is the crucial parameter which addresses the hardware design of the instrument devoted to collect that radiation. While a wide range of wavelengths can be handled by means of optical systems, for ultraviolet, X-ray and $\gamma$-ray emission different strategies are mandatory, involving collimators, masks and/or grids \cite{Hurford2010,Ramsey1994}.
An example of a similar imaging system is the one provided by the Reuven Ramaty High Energy Solar Spectroscopic Imager (RHESSI), launched from Cape Canaveral on February 5 2002 with the aim to recover temporally, spatially and spectrally resolved X-ray images of solar eruptions \cite{Krucker2008,Lin2002}. The RHESSI spacecraft is the last descendant of a generation of collimator-based satellites (HINOTORI, YOHKOH/HXT, HEIDI) originated in 1980 from the hard X-ray imaging spectrometer (HXIS), carried in the Solar Maximum Mission. It observes X-ray emission from the entire solar disc through a set of nine rotating modulation collimators (RMCs) \cite{Hurford2002}, each one being a pair of co-aligned identical grids which allow only a fraction of the incoming radiation to be collected by the corresponding detector. Since each collimator is made up of grids with different pitches, the signals provided by the nine detectors are characterized by different resolutions and signal-to-noise (SNR) ratios. \\
Thanks to the RMCs hardware, two strategies are available to reconstruct the desired image:
\begin{enumerate}
\item the straightforward inversion of the modulated count profiles provided by RHESSI, or
\item the construction (through a fitting procedure of the count profiles) of a set of amplitude and phase pairs of the detected radiation, also known in optical interferometry with the term {\em visibilities}, corresponding to spatial frequencies depending on the angular resolutions of each RMC pair (see the appendices of \cite{Hurford2002,Bonettini2013b}), followed by the application of a Fourier-based inversion technique.
\end{enumerate}
The very first algorithms designed for the RHESSI imaging followed the first strategy, and included CLEAN, Pixon and Maximum Entropy (see \cite{Hurford2002} and references therein). However, in the last ten years the visibility approach became more attractive, since the use of the Fast Fourier Transform led to a computational cost of the reconstruction algorithms of the order of few seconds. Examples of techniques developed according to this strategy are given by the visibility-based Maximum Entropy Method \cite{Bong2006}, the uv-smooth algorithm \cite{Massone2009} and the Space-D approach \cite{Bonettini2010a,Prato2013b}. The price to pay when using the visibilities is that the input of the reconstruction algorithms are not the raw data provided by the spacecraft, but the result of a fitting preprocessing step, which introduces systematic errors whose size increases as the count noise grows up.\\
In this paper we follow the idea proposed in \cite{Benvenuto2013}, in which the RHESSI imaging problem from count profiles has been approached by means of the expectation maximization (EM) algorithm. In particular, numerical simulations showed that EM is able to equal the accuracy in the reconstructions of Pixon with, in general, one fourth of the computational time. The aim of this paper is to further improve the computational cost without losing in accuracy through several accelerated versions of EM. In fact, since the iterations of EM converge to a minimizer of the Kullback-Leibler (KL) divergence, also known as the Csisz\'{a}r I-divergence \cite{Csiszar1991}, one can reformulate the RHESSI imaging problem as the non-negative minimization of the KL divergence and address its solution by means of a very powerful constrained optimization scheme. Here we adopt some recently proposed schemes as the scaled gradient projection (SGP) algorithm \cite{Bonettini2009}, the gradient projection method with extrapolation (GPE) \cite{Bertsekas2012} and the affine scaling interior point cyclic Barzilai-Borwein (AS\_CBB) approach \cite{Hager2009}.\\
A further advantage of these methods is that they can be easily extended to allow for the addition of regularization terms to the KL divergence. Due to the smooth and extended geometries which are typically observed in RHESSI images, here we do not consider edge-preserving or sparsity-inducing priors, but we analyze the effectiveness of a Tikhonov regularization term \cite{Tikhonov1977}. As for the choice of the stopping rule for the iterations (in the case of KL minimization) and the regularization parameter (in the case of KL + Tikhonov minimization), we adopt the discrepancy principle for Poisson data proposed by Bertero et al. \cite{Bertero2010}. In particular, we propose to include in the discrepancy principle a simple procedure which gives a better estimate of the expected value of the counts also for high noise levels. Our numerical experiments show that the accelerated methods are able to provide a substantial gain in the number of iterations needed to provide the regularized solution, and with lower reconstruction errors. Moreover, the presence of the Tikhonov regularization seems not to provide any substantial improvement in the reconstructions, thus the time-consuming step of calculating the optimal regularization parameter seems avoidable.\\
The plan of the paper is the following: in Section \ref{sec2} the RHESSI imaging problem from the modulated count profiles is formulated. In Section \ref{sec3} the main features of the accelerated gradient methods are briefly described, while in Section \ref{sec4} we analyze possible ways to introduce regularization and recover a meaningful image. Section \ref{sec5} is devoted to our numerical simulations and the discussion of the results we obtained. Finally, some comments and conclusions are provided in Section \ref{sec6}.

\section{The RHESSI imaging problem}\label{sec2}

\begin{figure}[ht]
\begin{center}
\begin{tabular}{cc}
\includegraphics[width=0.35\columnwidth]{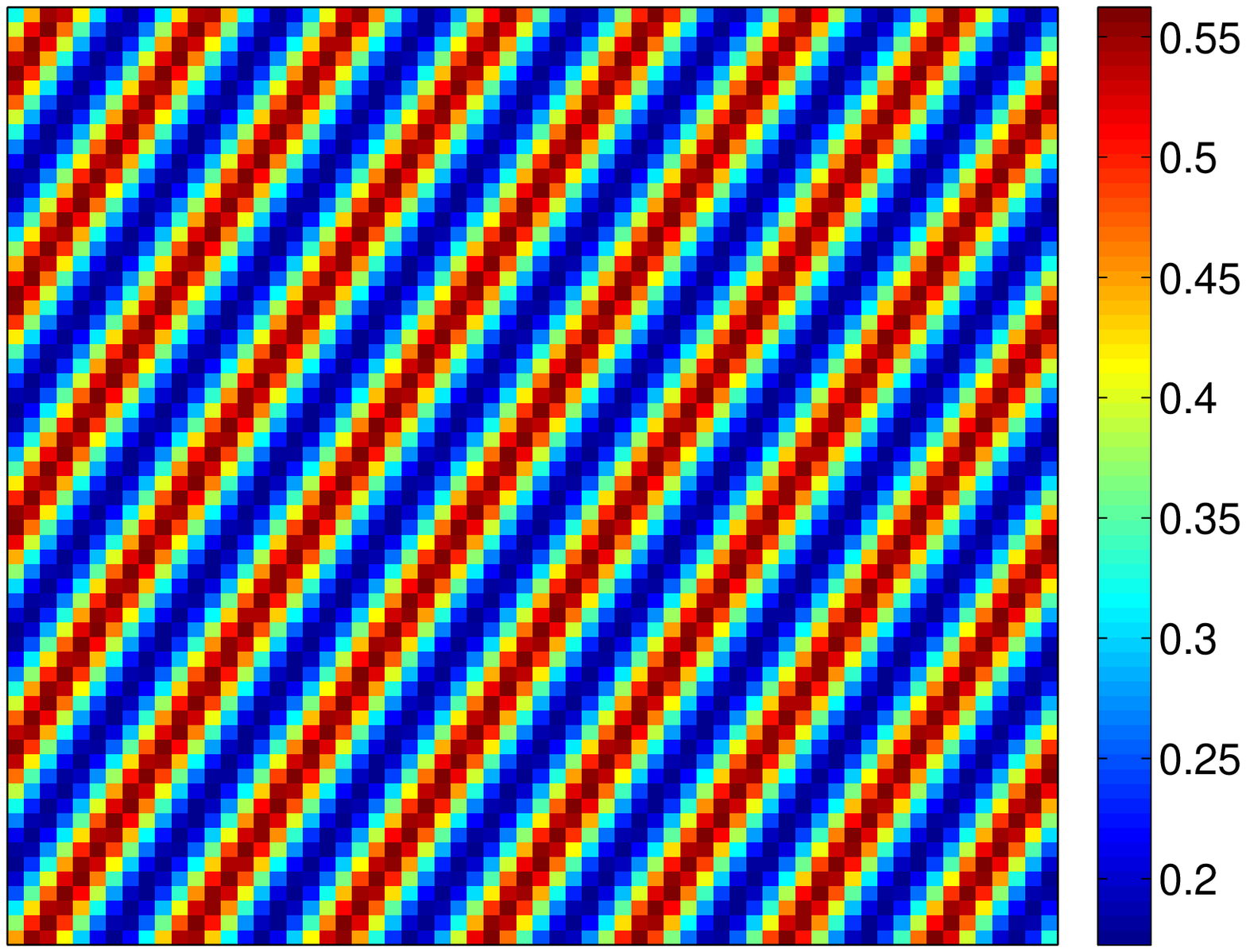} &
\includegraphics[width=0.35\columnwidth]{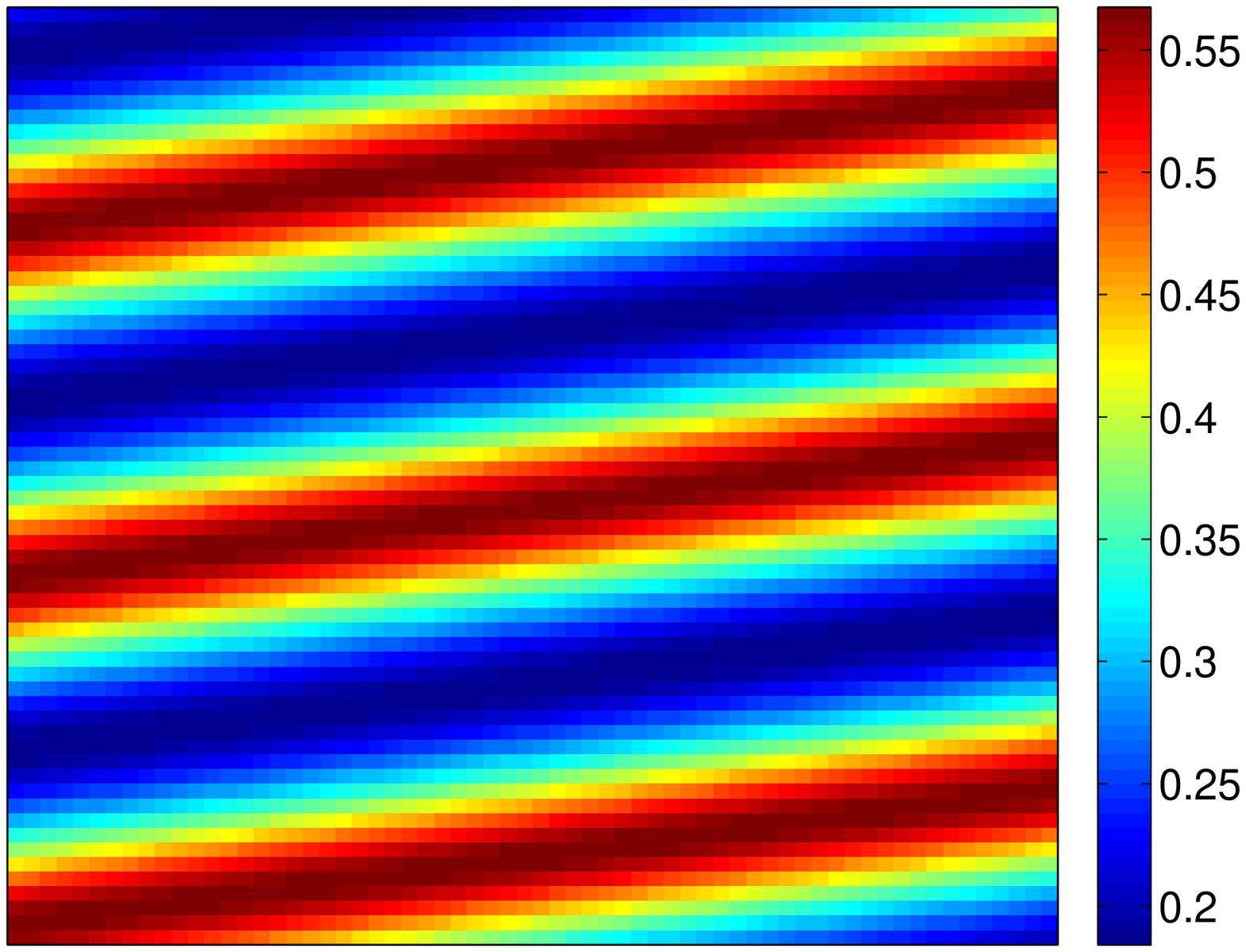}
\end{tabular}
\caption{Examples of transmission probability maps for subcollimators 2 (left) and 4 (right) at different time instants.}
\label{figmodpat}
\end{center}
\end{figure}

Let $\mathcal{D}$ be the domain of the unknown image, and let $f(x_s,y_s)$ be the intensity of the radiation as a function of the source position $(x_s,y_s) \in \mathcal{D}$. For the $j$-th subcollimator ($j=1,\ldots,9$), let us denote by $P^{(j)}(x_s,y_s;t)$ the fraction of the incoming radiation from position $(x_s,y_s) \in \mathcal{D}$ which will reach the detector at time $t$ (called {\em transmission probability}). Two examples of transmission probability maps for subcollimators 2 and 4 are shown in figure \ref{figmodpat}. It follows that the detected flux at time $t$ is proportional to the whole emission modulated by the corresponding transmission probability
\begin{equation}\label{cphiell}
C^{(j)}(t) = K^{(j)}\int_{\mathcal D} f(x_s,y_s)P^{(j)}(x_s,y_s;t)dx_sdy_s, \ \ \ j=1,...,9,
\end{equation}
where $K^{(j)}$ is some known constant parameter related to the system features (detector area, efficiency,...). For simplicity we assume that the field of view coincides with the domain $\mathcal D$, which is equivalent to assume that there is no radiation from points outside the map.\\

\noindent Equation \eqref{cphiell} is a Fredholm integral equation of the first kind whose kernel is the transmission probability $P^{(j)}(x_s,y_s;t)$, which, unfortunately, is not space invariant.\\
In a more realistic frame, each RHESSI collimator actually provides a set of discrete data $c^{(j)} = (c^{(j)}_1,...,c^{(j)}_{N_j})^T$ which can be modeled as the integral of \eqref{cphiell} in a certain number of time intervals
\begin{equation}
\label{C0k}
c^{(j)}_k = \int_{t_{k-1}}^{t_k}C^{(j)}(t)dt, \ \ k=1,...,N_j, \ \ j=1,...,9.
\end{equation}
The RHESSI imaging problem consists in finding an approximation of the flux distribution $f(x,y)$ given the data $c^{(j)}_k$, the transmission probabilities $P^{(j)}(x_s,y_s;t)$ and the parameters $K^{(j)}$. We remark that such inverse problem is ill-posed in the sense of Hadamard \cite{Hadamard1923}, due to the boundedness of the kernels $P^{(j)}$. In fact, if we denote with $|{\mathcal D}|$ the area of the domain ${\mathcal D}$ and with $\overline{P}$ the maximum of $P^{(j)}(x_s,y_s;t)$ over $(x_s,y_s) \in \mathcal{D}$, $t \in [t_{k-1},t_k]$ and $j=1,\ldots,9$, from the inequality
\begin{equation*}
\int_{t_{k-1}}^{t_k}\int_{\mathcal D}P^{(j)}(x_s,y_s;t)^2dx_sdy_sdt \leq \overline{P}^2 |{\mathcal D}| (t_k - t_{k-1})
\end{equation*}
it follows that each kernel $P^{(j)}(x_s,y_s;t)$ belongs to $L^2({\mathcal D},[t_{k-1},t_k])$, which means that the operator mapping $f(x,y)$ into $C^{(j)}(t)$ is Hilbert-Schmidt \cite{Reed1972}. Since the Hilbert-Schmidt condition is sufficient for an operator to be compact, we have the ill-posedness of the inverse problem \cite{Kirsch1996}. \\
In order to define a discrete counterpart of \eqref{C0k}, we assume $t_{k}-t_{k-1}= \Delta t$ and we define gridpoints $(x_h,y_\ell)$, $h,\ell=1,...,n$ on $\mathcal D$, with uniform spacing $\Delta x$, $\Delta y$ along the two dimensions.
Then, by applying the rectangular rule on \eqref{C0k} we obtain the following approximation
\begin{equation*}
c^{(j)}_k\approx K^{(j)} \Delta t\Delta x\Delta y \sum_{h,\ell=1}^nf(x_h,y_\ell)P^{(j)}(x_h,y_\ell;t_k).
\end{equation*}
If we define the vector $f=(f_1,...,f_{n^2})^T$ by a lexicographic reordering of $f_{h\ell}\equiv f(x_h,y_\ell)\Delta x\Delta y$, we can write the discrete formulation of the RHESSI imaging problem as follows
\begin{equation}\label{dp}
c = Pf
\end{equation}
where
\begin{equation*}
c = \left(\begin{array}{c} c^{(1)}\\ \vdots \\ c^{(9)}\end{array}\right) \ \ ; \ \
P = \left(\begin{array}{c} P^{(1)}\\ \vdots \\ P^{(9)}\end{array}\right)
\end{equation*}
and $P^{(j)}$ is a $N_j\times n^2$ matrix whose components are given by $K^{(j)}P^{(j)}(x_h,y_\ell;t_k)\Delta t$ in the suitable order. {{Problem \eqref{dp} is again ill-posed since in general its solution does not exists or is not unique, according to the thickness of the discretization of the observation time in time bins (i.e., the number $N_j$, which can be chosen by the user). This pathology is typically eluded by computing the generalized solution of \eqref{dp}, which for discrete problems guarantees the well-posedness of the procedure. However, it is well-known that the discretization of a continuous problem which lacks of continuity of the solution in the data leads to a highly ill-conditioned problem \cite{Bertero1998b}, which needs to be addressed by means of some regularization approach.}}\\
Indeed, we must also take into account that the data suffer from the loss of information due to measurement errors which occur in the acquisition process, yielding a perturbation of the data which is usually defined as \emph{noise}. The noise is a statistical feature in the acquisition process, and, since our data correspond to photon counts, it is reasonable to assume that it obeys to Poisson statistics. More precisely, the actual detected data $c_k^{(j)}$ should be considered as a realization of a Poisson random variable, whose expected value and variance are given by the noise-free data
\begin{equation*}
c^{(j)}_k \propto \mbox{Poisson }\left(\int_{t_{k-1}}^{t_k}C^{(j)}(t)dt\right), \ \ k=1,...,N_j, \ \ j=1,...,9.
\end{equation*}
Therefore, maximizing the likelihood (or, equivalently, minimizing the negative logarithm of the likelihood) of measuring the count profiles $c$ given a flux distribution $f$ yields to the minimization of the Kullback-Leibler (KL) divergence of the vector $Pf$ from the vector $c$, defined by
\begin{equation*}
D_{KL}(c,Pf) = \sum_{j=1}^9\sum_{k=1}^{N_j}\left\{c^{(j)}_k\ln\left(\frac{c^{(j)}_k}{(P^{(j)}f)_k}\right) + (P^{(j)}f)_k - c^{(j)}_k\right\}.
\end{equation*}
Since the entries of the flux distribution $f$ must be non-negative, the RHESSI imaging problem can be formulated as the following constrained minimization problem:
\begin{equation}\label{minpro}
\min_{f \geq 0} J(f) := D_{KL}(c,Pf).
\end{equation}
In a recent paper \cite{Benvenuto2013}, the authors propose to solve \eqref{minpro} with the well-known expectation-maximization (EM) algorithm, defined by
\begin{equation*}
f^{(k+1)} = \frac{f^{(k)}}{P^T\mathds{1}} \cdot P^T\left(\frac{c}{Pf^{(k)}}\right), \quad k = 1,2,\ldots,
\end{equation*}
where $\mathds{1}$ denotes a constant array of ones and $\cdot$ , $/$ are the componentwise product and quotient, respectively. In particular, they show that the method is able to provide reconstructions of the same accuracy level of the Pixon algorithm, which is generally considered as the most reliable technique in providing accurate image photometry \cite{Dennis2009}. The key result on the use of EM is that is considerably faster than Pixon, with an average gain of a factor of four (potentially growing up to a factor of twenty if EM is employed with the highly optimized Interactive Data Language (IDL) routine available in the Solar SoftWare (SSW) to compute the transmission probabilities -- see \cite{Hurford2002}). In this work we want to go one step further and address the solution of \eqref{minpro} by means of recently proposed accelerations of EM. In the following sections we recall the main features of these schemes and discuss the regularization issue, which is a crucial point to contrast the corruptive effect of the noise on the reconstructed image.

\section{Optimization methods}\label{sec3}

In this section we focus on some optimization methods which are able to address the minimization problem \eqref{minpro}. In the recent literature, a number of methods have been proposed for non-negatively constrained minimization and also for specific image restoration problems involving the Kullback--Leibler divergence. Among them, we mention the first order methods in \cite{Bonettini2009,Hager2009}; the second order Newton--like methods \cite{Landi2012,Bonettini2009b}; the especially tailored versions of the alternating direction method of multipliers (ADMM) proposed in \cite{Setzer10,Figueiredo2010}; the gradient projection with extrapolation methods \cite{Beck2009b,Nesterov2005}, which are proved to be optimal in the sense that the decrease of the objective function with respect to its minimum value is quadratic with the iteration number. To devise the appropriate approaches for the count-based image reconstruction problem \eqref{minpro}, its special features must be taken into account, in particular, one should remember that
\begin{itemize}
\item the matrix $P$ and the Hessian $\nabla^2 J(f)$ are very large, dense and have no special structures;
\item the Lipschitz constant of $\nabla J(f)$ is not available.
\end{itemize}
Indeed, unlike the deconvolution case, here the computation of matrix--vector products involving $P$ and $P^T$ is very expensive. Moreover, there is no effective way to solve a linear system related to the Hessian $\nabla^2 J(f)$ or also to the matrix $I+P^TP$: this makes the ADMM methods \cite{Setzer10,Figueiredo2010} and the Newton--like methods \cite{Landi2012,Bonettini2009b} very impractical. On the other side, the explicit knowledge of the Lipschitz constant of $\nabla J(f)$ is needed for the implementation of the Nesterov algorithm \cite{Nesterov2005}. \\
For the reasons above we restrict the choice of the optimization method for solving \eqref{minpro} to gradient-based methods: in particular, we compare the scaled gradient projection (SGP) method \cite{Bonettini2009}, the affine scaling interior point cyclic Barzilai-Borwein (AS\_CBB) method \cite{Hager2009} and the gradient projection method with extrapolation (GPE) \cite{Bertsekas2012,Beck2009b}, which can be implemented also when the Lipschitz constant of $\nabla J(f)$ is unknown.\\
It is worth stressing that the image reconstruction problem underlying \eqref{minpro} is ill-posed and an appropriate regularization is needed, as explained in detail in Section \ref{sec4}. Thus, the optimization methods should be evaluated also focusing on their behaviour in the regularization context.

\subsection{The SGP algorithm}\label{subsec3.1}

The SGP algorithm is a gradient projection method proposed by Bonettini et al. \cite{Bonettini2009} to solve a minimization problem of a differentiable function over a convex set. The scheme of the SGP method in the case of non-negative constraints is given in Algorithm \ref{SGPalg}, where $\mathcal{D}$ is the set of the $n^2 \times n^2$ diagonal positive definite matrices, whose diagonal elements have values between $L_1$ and $L_2$, for given thresholds $0 < L_1 < L_2$. We do not discuss here the choice of all the several parameters of SGP, which can be found, for example, in \cite{Prato2012}. We only describe the three main features of SGP, which are the scaling matrix multiplying the gradient, the steplength parameter and the projection on the feasible set:
\begin{itemize}
\item $D_k$ is a diagonal scaling matrix defined as
\begin{equation*}
D_k = \mbox{diag}\left(\min\left(L_2,\max\left(L_1,\frac{f^{(k)}}{P^T\mathds{1}}\right)\right)\right),
\end{equation*}
$(L_1,L_2)$ being given constants estimated from min/max values of one EM iteration. It has been shown that scaling the gradient might lead to a more faithful recovery of a regularized solution \cite{Bonettini2013,Cornelio2013};
\item the steplength $\alpha_k$ is selected by alternating the following generalized Barzilai-Borwein (BB) rules \cite{Bonettini2009,Barzilai1988}
\begin{eqnarray}
& & \alpha_k(D_k)^{{(BB1)}} = \frac{(s^{(k-1)})^T D_k^{-1} D_k^{-1}(s^{(k-1)})}{(s^{(k-1)})^T D_k^{-1} (z^{(k-1)})}, \label{BB1}\\
& & \alpha_k(D_k)^{{(BB2)}} = \frac{(s^{(k-1)})^T D_k (z^{(k-1)})}{(z^{(k-1)})^T D_k D_k (z^{(k-1)})},\nonumber
\end{eqnarray}
where $s^{(k-1)} = f^{(k)}- f^{(k-1)}$ and $z^{(k-1)} = \nabla J(f^{(k)}) - \nabla J(f^{(k-1)})$, and introducing suitable upper and lower bounds. The alternation of the BB rules allows a faster convergence with respect to gradient methods exploiting only one of the two rules (see e.g. \cite{Frassoldati2008});
\item $\mathcal{P}_+$ is the projection onto the non-negative orthant. Since the scaling matrix is diagonal, it consists in simply setting to zero the negative components.
\end{itemize}

\begin{algorithm}[ht]
\caption{Scaled gradient projection (SGP) method}
\label{SGPalg}
Choose the starting point $f^{(0)} \geq 0$ and set the parameters $\beta, \theta\in (0,1)$, $0< \alpha_{min} <\alpha_{max}$.\\[.2cm]
{\textsc{For}} $k=0,1,2,...$ \textsc{do the following steps:}
\begin{itemize}
\item[]
\begin{AlgorithmSteps}[4]
\item[1] Choose the parameter $\alpha_k \in [\alpha_{min},\alpha_{max}]$ and the scaling matrix $D_k\in \mathcal{D}$.
\item[2] Projection: \begin{equation*} y^{(k)} = \mathcal{P}_+(f^{(k)}-\alpha_kD_k\nabla J(f^{(k)})). \end{equation*}
\item[3] Descent direction: $d^{(k)} = y^{(k)} - f^{(k)}$.
\item[4] Set $\lambda_k = 1$.
\item[5] Backtracking loop:
\begin{ifelse}
\item let $J_{new} = J(f^{(k)} + \lambda_k d^{(k)})$;
\item \textsc{If} $J_{new}\leq J(f^{(k)}) + \beta\lambda_k\nabla J(f^{(k)})^T d^{(k)}$
      \textsc{Then} \\  \hspace*{.3cm} go to step 6;
\item \textsc{Else} \\  \hspace*{.3cm} set $\lambda_k = \theta \lambda_k$ and go to step 5.
\item \textsc{Endif}
\end{ifelse}
\item[6] Set $f^{(k+1)} = f^{(k)} + \lambda_k d^{(k)}$.
\end{AlgorithmSteps}
\end{itemize}
\textsc{End}
\end{algorithm}

\paragraph{Theoretical convergence properties of SGP.} The convergence properties of SGP are stated in \cite{Bonettini2009} for general, possibly nonconvex problems with convex constraint set; in particular, it is proved that any limit point of the sequence $f^{(k)}$ is a stationary point for the constrained problem. As a consequence, for convex problems such as \eqref{minpro}, any limit point of $f^{(k)}$ is a minimum point; if, in addition, the objective function is strictly convex with bounded level sets, then $f^{(k)}$ converges to the unique minimum point. A stronger result for the convex case can be found in \cite{Iusem03}, where the convergence of the whole sequence is proved for the nonscaled method ($D_k=I$) with the only assumption that a solution exists. As far as we know, a theoretical convergence rate estimate for the gradient projection method with variable scaling matrix and linesearch along the descent direction has not been well investigated in the literature.

\subsection{The AS\_CBB method}\label{subsec3.2}

The AS\_CBB method proposed by Hager et al. in \cite{Hager2009} shares with SGP the scaling idea and the exploitation of the BB rules. Indeed, AS\_CBB can be framed in the scheme of Algorithm \ref{SGPalg} in which the computation of the search direction in steps 2 and 3 is substituted by $d^{(k)} = - D_k\nabla J(f^{(k)})$, where:
\begin{itemize}
\item the diagonal entries of the diagonal scaling matrix $D_k$ are defined by
\begin{equation}\label{HMZscaling}
(D_k)_{ii} = \frac{f_i^{(k)}}{\alpha_kf_i^{(k)}+[\nabla_iJ(f^{(k)})]^+},
\end{equation}
where $[\cdot]^+$ denotes the positive part of a real number, i.e. $[x]^+=\max(x,0)$. In this case, the diagonal entries of the scaling matrix are not bounded away from zero;
\item the parameter $\alpha_k$ in \eqref{HMZscaling} is computed by a cyclic update of the first BB rule as $\alpha_k = \alpha_{q\cdot c}(I)^{(BB1)} $ in \eqref{BB1} when $k=q\cdot c+r$ for a given, fixed cycle length $c>0$ and $q=1,2,...$, $r=0,1,...,c-1$ (in the numerical experiments of Section \ref{sec5} we set $c=3$).
\end{itemize}
Actually, the scaling matrix \eqref{HMZscaling} implies that, when the algorithm is initialized by a point with positive entries, then also the point $f^{(k)} - D_k\nabla J(f^{(k)})$ is still strictly positive for any $k$.

\paragraph{Theoretical convergence properties of AS\_CBB.} The AS\_CBB method applies to box constrained problems and convergence can be stated if the objective function is strongly convex and has bounded level sets. Moreover, in \cite{Hager2009} the authors prove also the local R-linear convergence of the sequence $f^{(k)}$ to the unique minimum point.

\subsection{The GPE method}\label{subsec3.3}

While SGP and AS\_CBB exploit gradient scaling and the BB rules to improve the convergence speed, the GPE method uses a completely different acceleration strategy which dates back to Polyak \cite{Polyak1987} and consists in adding an extrapolation term to the gradient direction. Several recently proposed methods, for example the very popular Nesterov's method \cite{Nesterov2005} and FISTA algorithm \cite{Beck2009b}, include such extrapolation step.
Here we consider the general scheme described in \cite[\S 6.9.2]{Bertsekas2012} applied to \eqref{minpro}, with an automatic estimate of the Lipschitz constant of $\nabla J(f)$ (see Algorithm \ref{GPE}).
\begin{algorithm}[ht]
\caption{Gradient projection method with extrapolation (GPE).}
\label{GPE}
Choose the starting point $f^{(0)} \geq 0$ and the parameter $\theta\in (0,1)$; set $f^{(-1)} = 0 $.\\[.2cm]
{\textsc{For}} $k=0,1,2,...$ \textsc{do the following steps:}
\begin{itemize}
\item[]
\begin{AlgorithmSteps}[4]
\item[1] Extrapolation: \begin{equation*} y^{(k)} = f^{(k)}+\beta_k(f^{(k)}-f^{(k-1)}). \end{equation*}
\item[2] Choose an initial estimate of the Lipschitz constant $\alpha_k$.
\item[3] Backtracking loop:
\begin{ifelse}
\item Compute $f(\alpha_k) = \mathcal{P}_+(f^{(k)}-\alpha_k\nabla J(f^{(k)})) $.
\item \textsc{If}
          $$J(f(\alpha_k))\leq J(f^{(k)}) + \alpha_k\nabla J(y^{(k)})^T (f(\alpha_k)-y^{(k)})+\frac{1}{2\alpha_k}\|f(\alpha_k)-y^{(k)} \|^2$$
      \textsc{Then} \\ \hspace*{.3cm} go to step 4;
\item \textsc{Else} \\ \hspace*{.3cm} set $\alpha_k = \theta \alpha_k$ and go to step 3.
\item \textsc{Endif}
\end{ifelse}
\item[4] Set $f^{(k+1)} = f(\alpha_k)$.
\end{AlgorithmSteps}
\end{itemize}
\textsc{End}
\end{algorithm}
Possible choices for the extrapolation parameter $\beta_k$ are
\begin{equation}\label{betabert}
\beta_k = \left\{ \begin{array}{ll} 0&\mbox{ if } k=0\\ \frac{k-1}{k+2} &\mbox{ if } k>0\end{array}\right.
\end{equation}
(see \cite[p. 397]{Bertsekas2012}), or also
\begin{equation}\label{betafista}
\beta_k = \frac{t_k -1}{t_{k+1}}, \ \ \mbox{ where } t_{k+1} = \frac{1+\sqrt{1+4t_k^2}}{2}
\end{equation}
with $t_0=1$, which corresponds to the FISTA algorithm \cite{Beck2009b}.

\paragraph{Theoretical convergence properties of GPE.} The GPE method is specific for convex problems, and its more interesting property is the estimate
\begin{equation*}
J(f^{(k)}) - J^* = \mathcal{O}\left(\frac 1{k^2}\right),
\end{equation*}
where $J^*$ is the minimum of $J$, which can be proved for both the choices \eqref{betabert}--\eqref{betafista}. It is worth stressing that this estimate concerns the objective function value, instead of the sequence of the iterates.

\section{Handling with regularization}\label{sec4}

The presence of noise on the measured data makes any minimizer of \eqref{minpro} meaningless. Two main strategies are typically adopted to overcome this drawback: a) arrest the iterative procedure employed to solve \eqref{minpro} before convergence, or b) add a (suitably weighted) regularization term to the KL which preserves certain features of the unknown target distribution and solve the resulting minimum problem exactly. Both approaches then need a criterion for the choice of the so-called regularization parameter, which in the former case is the number of iterations to be performed while in the latter case is the constant multiplying the regularization term. In our approach we investigated the effectiveness of a recently proposed discrepancy principle for data affected by Poisson noise \cite{Bertero2010,Zanella2009,Zanella2013}, whose main result is the following Lemma.
\begin{Lem}\label{lemma}
Let $Y_\lambda$ be a Poisson random variable with expected value $\lambda$ and consider the following function of $Y_\lambda$:
\begin{equation*}
F(Y_\lambda) = 2\left\{Y_\lambda\ln\left(\frac{Y_\lambda}{\lambda}\right)+ \lambda - Y_\lambda \right\}.
\end{equation*}
Then, for large $\lambda$, the following asymptotic estimate of the expected value of $F(Y_\lambda)$ holds true:
\begin{equation*}
\mathbb{E}[F(Y_\lambda)] = 1 + O\left(\frac{1}{\lambda}\right).
\end{equation*}
\end{Lem}
Therefore, according to this discrepancy principle, an optimal regularized solution $f_\eta$ should be the one satisfying
\begin{equation}\label{onepluseps}
\frac{2}{\sum_{j=1}^9N_j}D_{KL}(c,Pf_\eta) = 1 + \varepsilon,
\end{equation}
where $\varepsilon$ depends on the noise level of the measured data. For medium/high level of statistics, the default value $\varepsilon = 1/m$, where $m = {\rm{mean}}(c)$, provided within the SGP code results to be a reliable choice. This is not true for count profiles characterized by a low level of statistics. In this case, the term $O\left(\frac{1}{\lambda}\right)$ in Lemma \ref{lemma} has to be accurately estimated.\\
We tried to estimate $\varepsilon$ by following the proof of Lemma \ref{lemma}, which is based on the Taylor formula of the function
\begin{equation*}
\ln\left(\frac{Y_\lambda}{\lambda}\right) = \ln\left(1 + \frac{Y_\lambda - \lambda}{\lambda}\right) =: \ln\left(1 + \xi\right).
\end{equation*}
Since
\begin{equation*}
\ln\left(1 + \xi\right) = \sum_{\ell=1}^\infty \frac{(-1)^{\ell+1}}{\ell}\xi^\ell,
\end{equation*}
after some computations, we get the following expression for $F(Y_\lambda)$:
\begin{equation*}
F(Y_\lambda) = \sum_{\ell=1}^\infty \frac{2(-1)^{\ell+1}}{\ell(\ell+1)}\cdot\frac{(Y_\lambda-\lambda)^{\ell+1}}{\lambda^\ell}.
\end{equation*}
By introducing the central moments
\begin{equation*}
\mu_\ell = \mathbb{E}[(Y_\lambda-\lambda)^\ell]
\end{equation*}
and by recalling that, for a Poisson random variable, $\mu_2 = \mu_3 = \lambda$ and $\mu_{\ell+1}$, $\ell > 2$, can be derived from the recursive rule \cite{Johnson1993}
\begin{equation*}
\mu_{\ell+1} = \ell \lambda \mu_{\ell-1} + \lambda \mu_\ell',
\end{equation*}
the expected value of $F(Y_\lambda)$ becomes
\begin{equation}\label{epsproof}
\mathbb{E}[F(Y_\lambda)] = \sum_{\ell=1}^\infty \frac{2(-1)^{\ell+1}}{\ell(\ell+1)}\cdot\frac{\mu_{\ell+1}}{\lambda^\ell} = 1 + \sum_{\ell=2}^\infty \frac{2(-1)^{\ell+1}}{\ell(\ell+1)}\cdot\frac{\mu_{\ell+1}}{\lambda^\ell}.
\end{equation}
Therefore, the quantity $\varepsilon$ in \eqref{onepluseps} can be estimated from the last term in \eqref{epsproof}. We implemented a routine in Matlab\textsuperscript{\textregistered} (The MathWorks, Natick, MA) which approximates $\varepsilon$ by truncating the series at a given number of terms $\overline{\ell}$. In our tests we decided to arrest the series at $\overline{\ell} = 10$ since we noticed some numerical instability for a higher numbers of terms.\\
{{As concerns the way of introducing regularization, we first investigate the use of the proposed gradient methods as iterative regularization methods for solving \eqref{minpro} (in this case the role of the regularization parameter is played by the iteration number $k$); then we use the SGP method to solve the minimization problem
\begin{equation}\label{minproreg}
\min_{f \geq 0} J_R(f) := D_{KL}(c,Pf) + \eta\|\mathcal{L}f\|_2^2,
\end{equation}
where $\mathcal{L}$ is the identity matrix (zero order Tikhonov regularization) or the discrete gradient operator (first order Tikhonov regularization), for different values of $\eta > 0$.}} In the former case, the value $1+\varepsilon$ estimated from \eqref{epsproof} is used as stopping criterion for alle the gradient algorithms, which are arrested at the first iteration $k$ satisfying the condition
\begin{equation}\label{discr1}
\frac{2}{\sum_{j=1}^9N_j}D_{KL}(c,Pf^{(k)}) \leq 1 + \varepsilon.
\end{equation}
When the Tikhonov term is added in the objective function, we compute the solution $f_\eta$ of \eqref{minproreg} with SGP for different values of $\eta$ and choose the regularized solution $f_{\eta^*}$ satisfying
\begin{equation}\label{discr2}
\frac{2}{\sum_{j=1}^9N_j}D_{KL}(c,Pf_{\eta^*}) \approx 1 + \varepsilon.
\end{equation}
For the computation of the Tikhonov solutions, the stopping criterion
\begin{equation*}
|J_R(f^{(k+1)}) - J_R(f^{(k)})| \leq 10^{-9}|J_R(f^{(k+1)})|
\end{equation*}
has been used.\\
We point out that the modifications needed by SGP to be applied to \eqref{minproreg} are minimal, since we only have to change $J$ with $J_R$ in Algorithm \ref{SGPalg} and in the computation of the steplength $\alpha_k$ and to adopt the scaling matrix
\begin{equation}\label{scalreg}
D_k = \mbox{diag}\left(\min\left(L_2,\max\left(L_1,\frac{f^{(k)}}{P^T\mathds{1}+\ell\eta f^{(k)}}\right)\right)\right)
\end{equation}
suggested by the scaled gradient method proposed by Lant\'{e}ri et al \cite{Lanteri2001,Lanteri2002}. The constant $\ell$ in \eqref{scalreg} is equal to 1 for zero order regularization while $\ell=4$ has to be set if a first order Tikhonov term is adopted.

\section{Numerical experiments}\label{sec5}

In this section we test the effectiveness of the considered optimization schemes in providing good reconstructions of realistic images with a notable reduction in the number of iterations required with respect to EM. As a further comparison, we also consider the images recovered by a visibility-based strategy. Among the possible choices, we chose the uv-smooth algorithm \cite{Massone2009}, which is the approach providing the most reliable results among the algorithms implemented in SSW which use the visibilities as input data. For sake of completeness, we summarize the main features of uv-smooth in the following section.

\subsection{The uv-smooth algorithm}

The uv-smooth method consists of a two-step process: a) interpolation to generate a smooth continuum of visibilities within the disk in the Fourier plane spanned by the available data; b) out-of-band extrapolation through a FFT-based iterative method with the imposition of image positivity. More in details, the measured visibilities are interpolated in the Fourier plane through a thin-plate spline algorithm, by means of the IDL routine {\em grid\_tps.pro} \cite{Barrodale1993}. The resulting surface is then resampled on a uniform grid. The advantage achievable with the interpolation step is that information are inferred also for spatial frequencies corresponding to ``virtual'' subcollimators with angular resolution between the minimum and the maximum values available with the RHESSI hardware. Moreover, with the new (uniform) resampling on the visibilities in the Fourier plane, the FFT routine can be applied. As concerns the extrapolation step, the idea is the extraction of information also for higher frequencies, to get a super-resolution effect of the reconstructed image \cite{Bertero1998b,Youla1982}. This can be done with a projected Gerchberg-Papoulis method \cite{Gerchberg1974,Papoulis1975}, which consists in finding a positive distribution $f(x,y)$ such that
\begin{equation*}
V(u,v) = \chi_B(u,v)(\mathcal{F}f)(u,v),
\end{equation*}
where $(u,v)$ denotes a pair of spatial frequencies, $V(u,v)$ is the corresponding visibility, $B$ is the band in which RHESSI provides the visibilities, $\chi_B$ is the characteristic function of $B$ and $\mathcal{F}$ is the Fourier transform. The steps of the computational procedure adopted are summarized in Algorithm \ref{GPA}. The steplength parameter $\tau$ has to be properly chosen in order to assure the convergence of the algorithm. Regularization is achieved through an early stopping of the iterations based on the descent of $\|\chi_B \mathcal{F}f - V\|$.

\begin{algorithm}[ht]
\caption{The Gerchberg-Papoulis method}
\label{GPA}
Put the starting point $f^{(0)}$ equal to the null map, and set the parameter $\tau > 0$. \\[.2cm]
{\textsc{For}} $k=0,1,2,...$ \textsc{do the following steps:}
\begin{itemize}
\item[]
\begin{AlgorithmSteps}[4]
\item[1] Compute the Fourier transform $\mathcal{F}f^{(k)}$ of $f^{(k)}$;
\item[2] Compute
$$ \mathcal{F}f^{(k+1)} = \mathcal{F}f^{(k)} + \tau(V-\chi_B\mathcal{F}f^{(k)});$$
\item[3] Compute the inverse Fourier transform $f^{(k+1)}$ of $\mathcal{F}f^{(k+1)}$;
\item[4] Project $f^{(k+1)}$ on the set of the real positive numbers.
\end{AlgorithmSteps}
\end{itemize}
\textsc{End}
\label{uvsmooth}
\end{algorithm}

\subsection{Simulated datasets}

\begin{figure}
\begin{center}
\begin{tabular}{cccc}
\includegraphics[width=0.23\columnwidth]{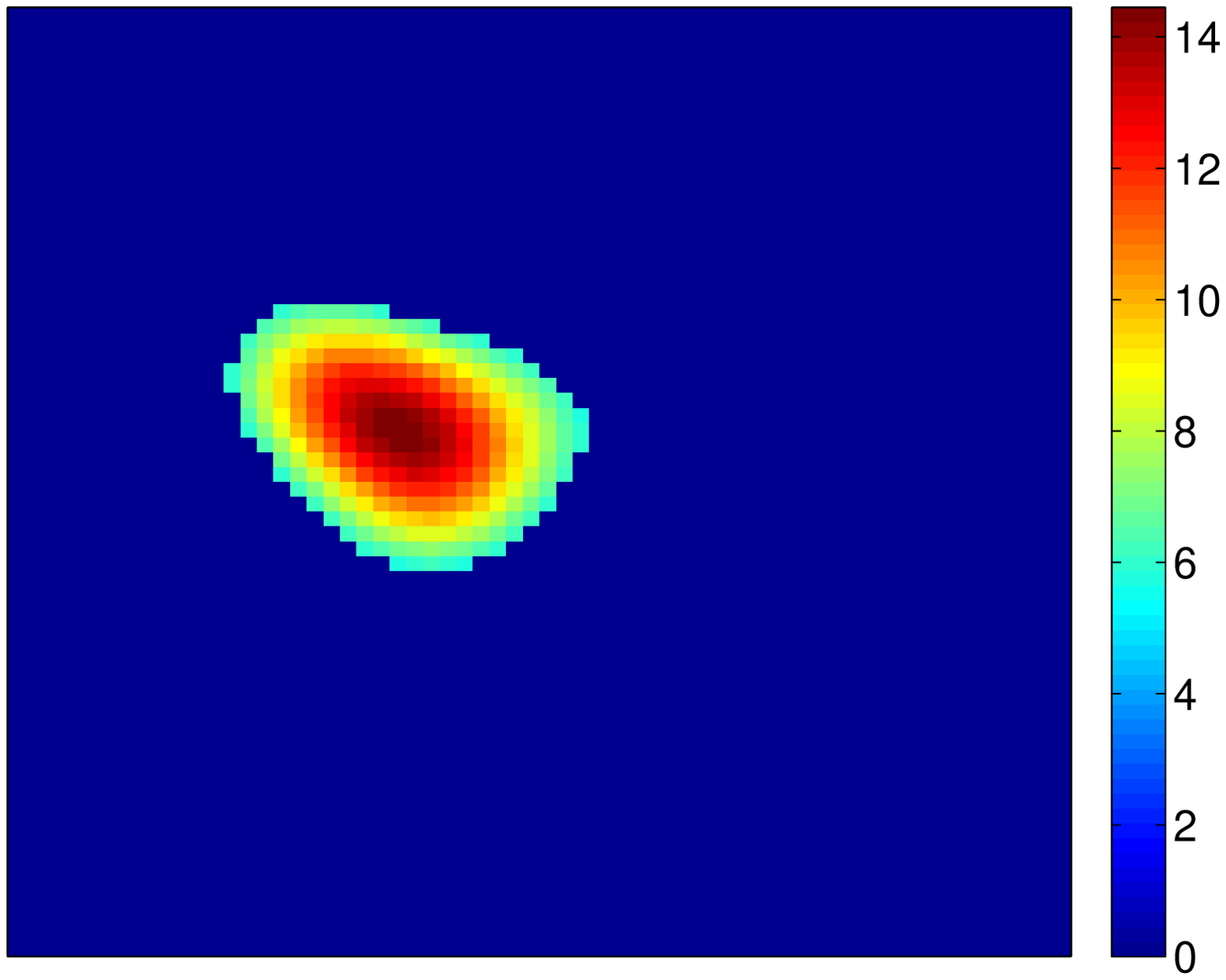} &
\includegraphics[width=0.23\columnwidth]{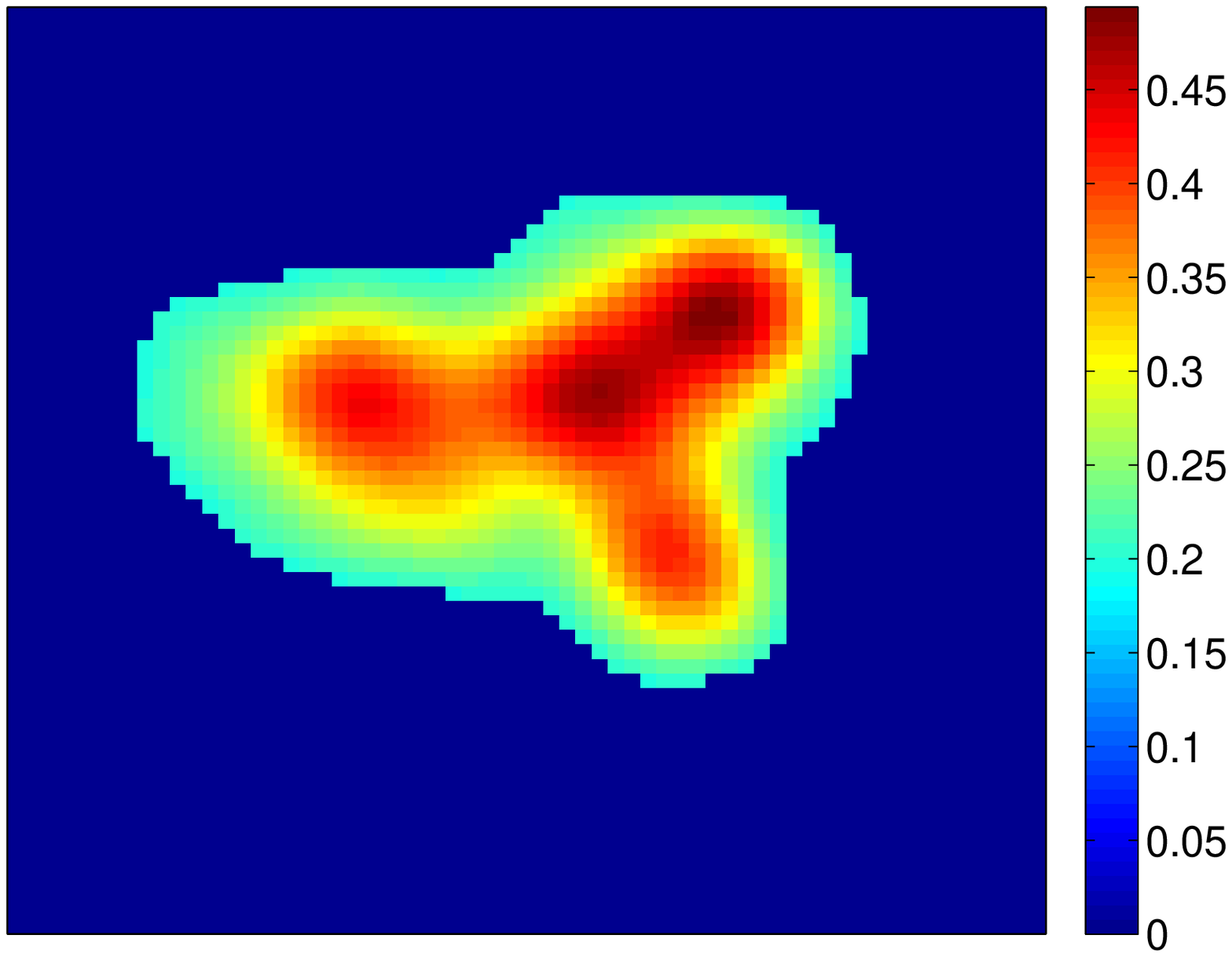} &
\includegraphics[width=0.23\columnwidth]{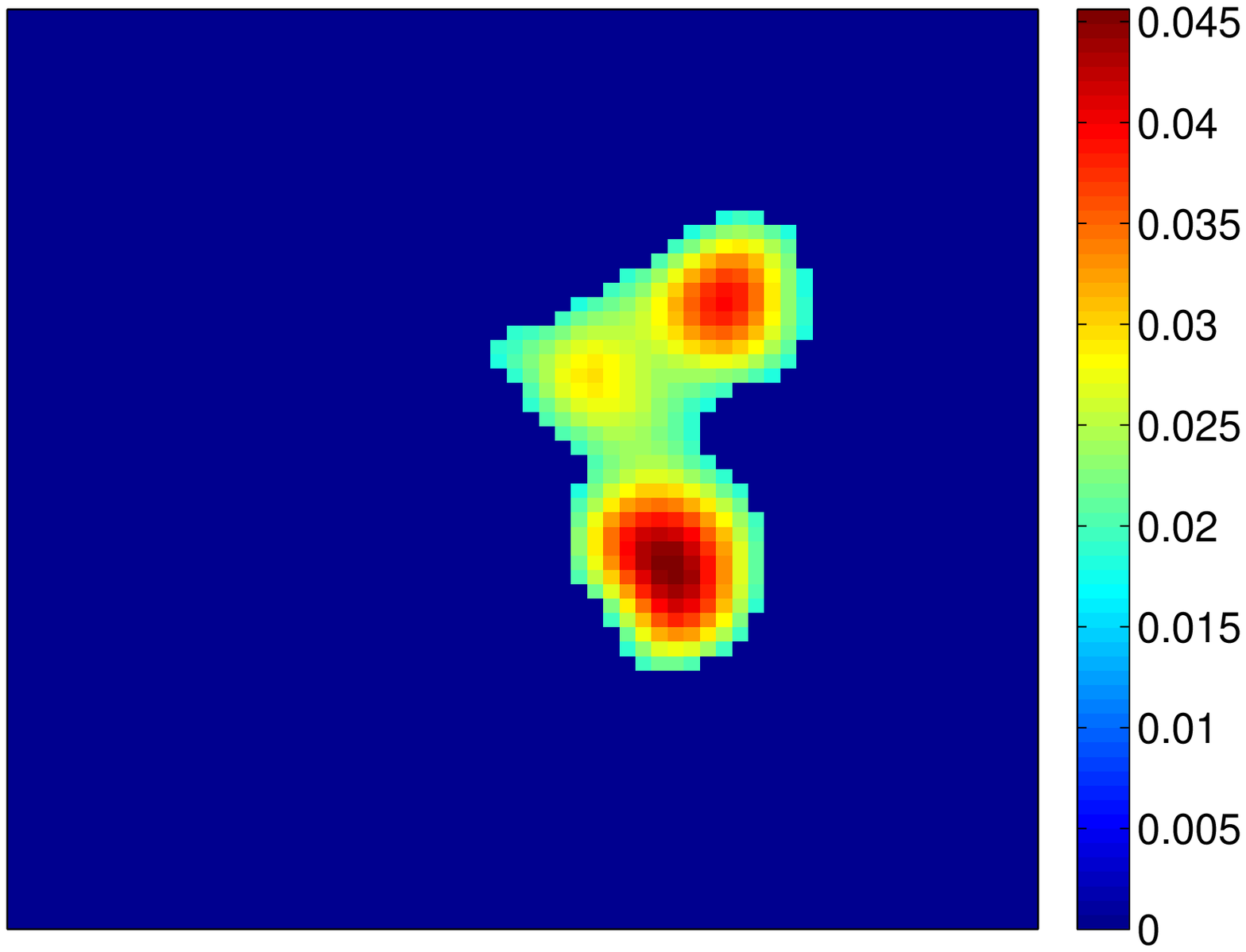} \\
\includegraphics[width=0.23\columnwidth]{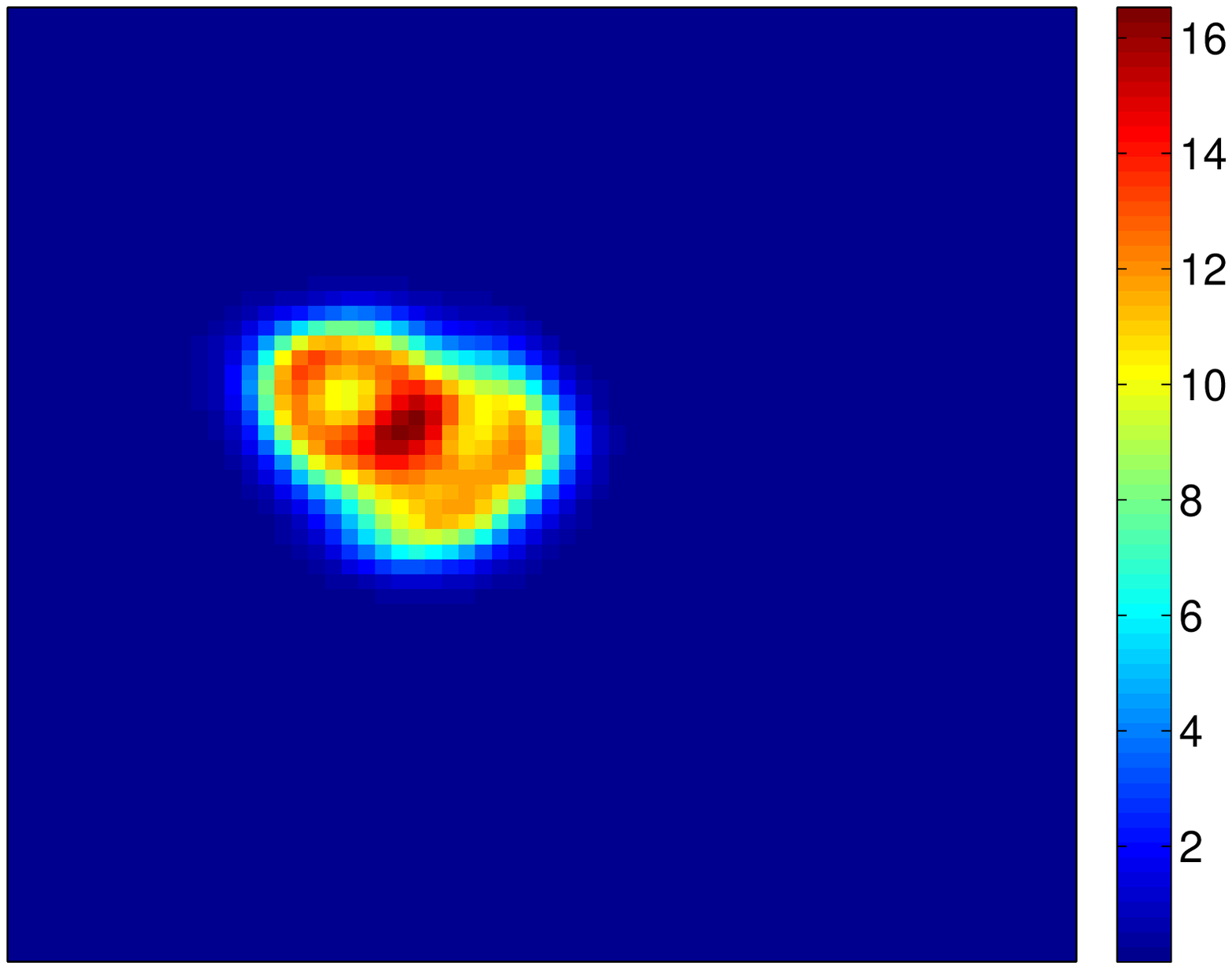} &
\includegraphics[width=0.23\columnwidth]{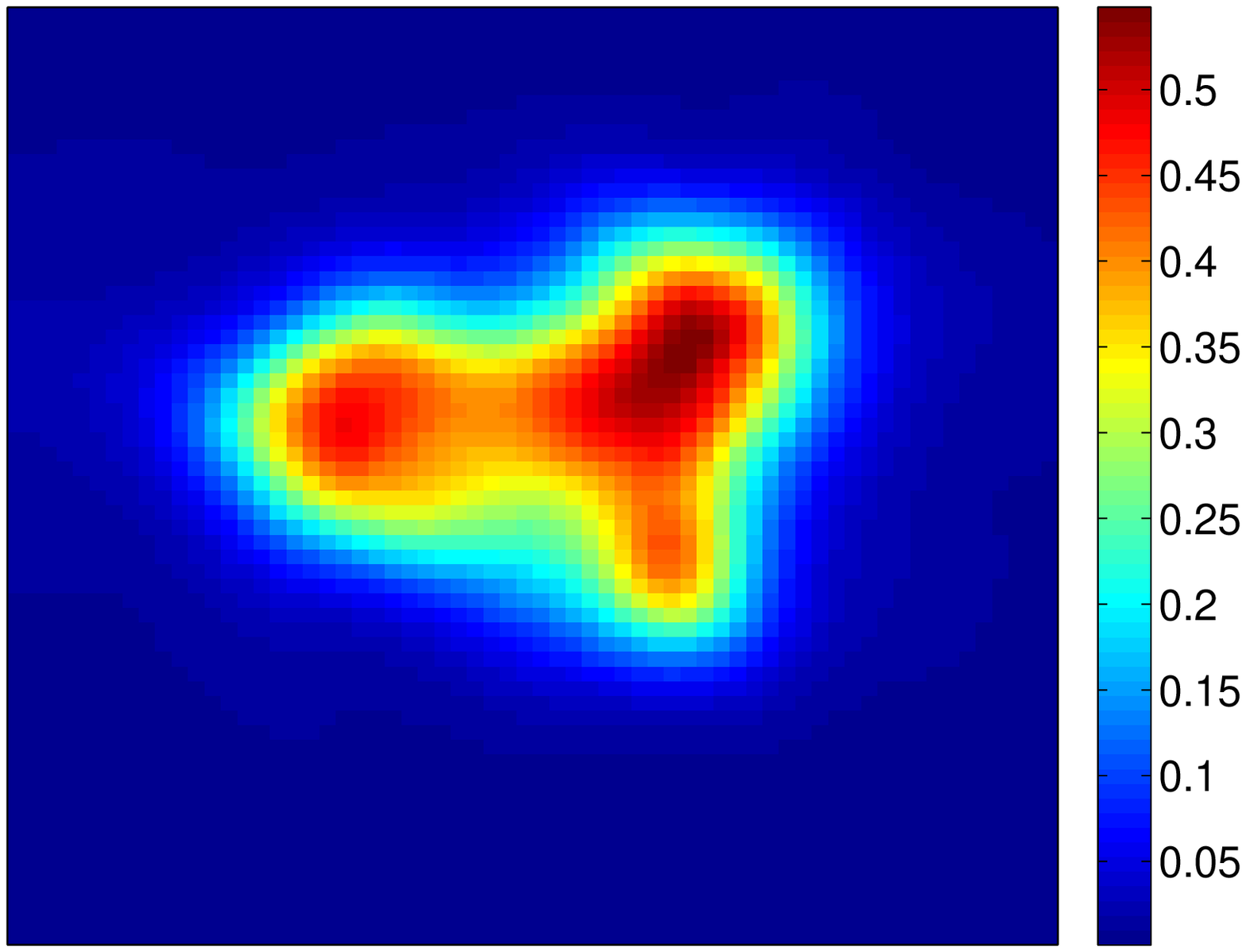} &
\includegraphics[width=0.23\columnwidth]{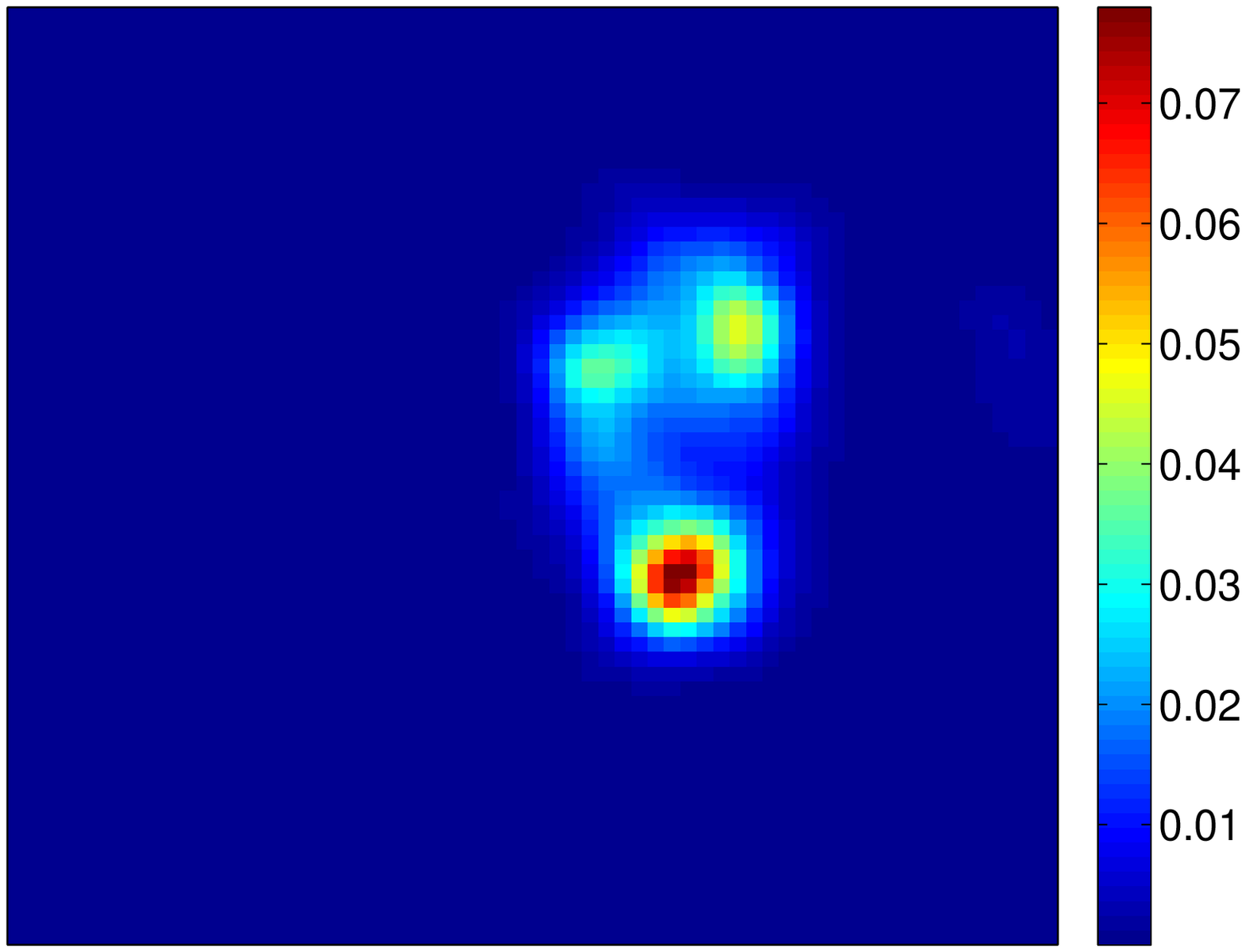} \\
\includegraphics[width=0.23\columnwidth]{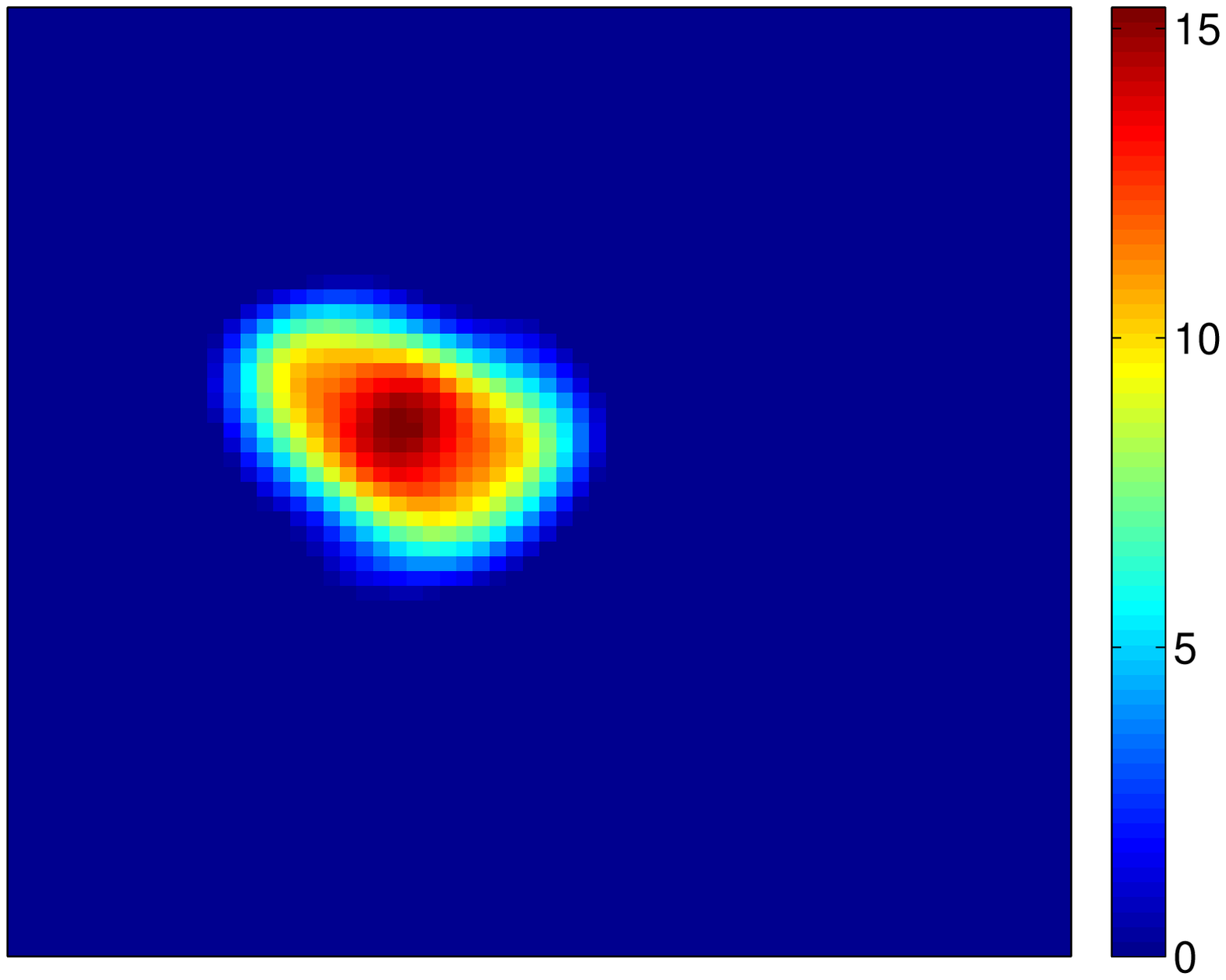} &
\includegraphics[width=0.23\columnwidth]{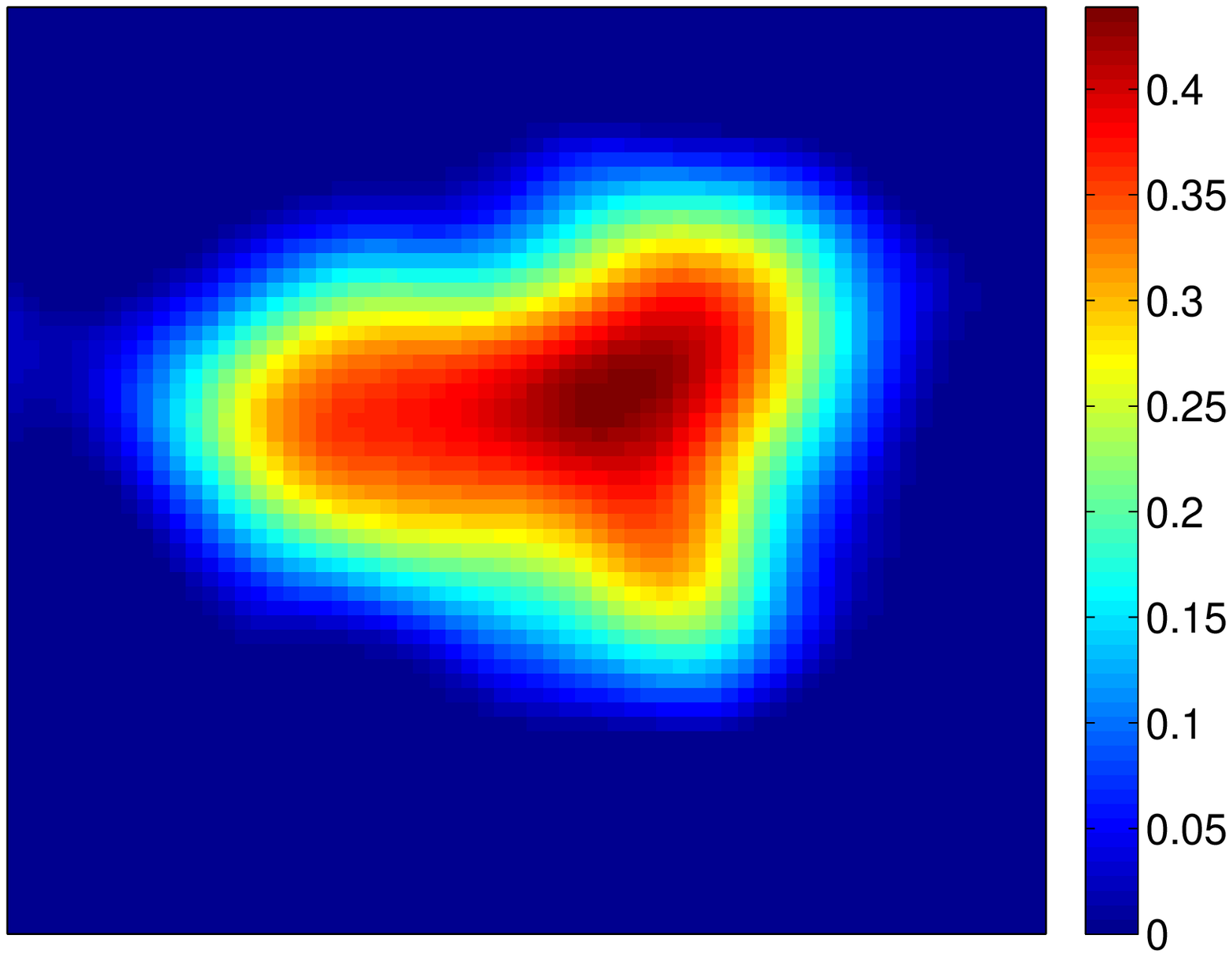} &
\includegraphics[width=0.23\columnwidth]{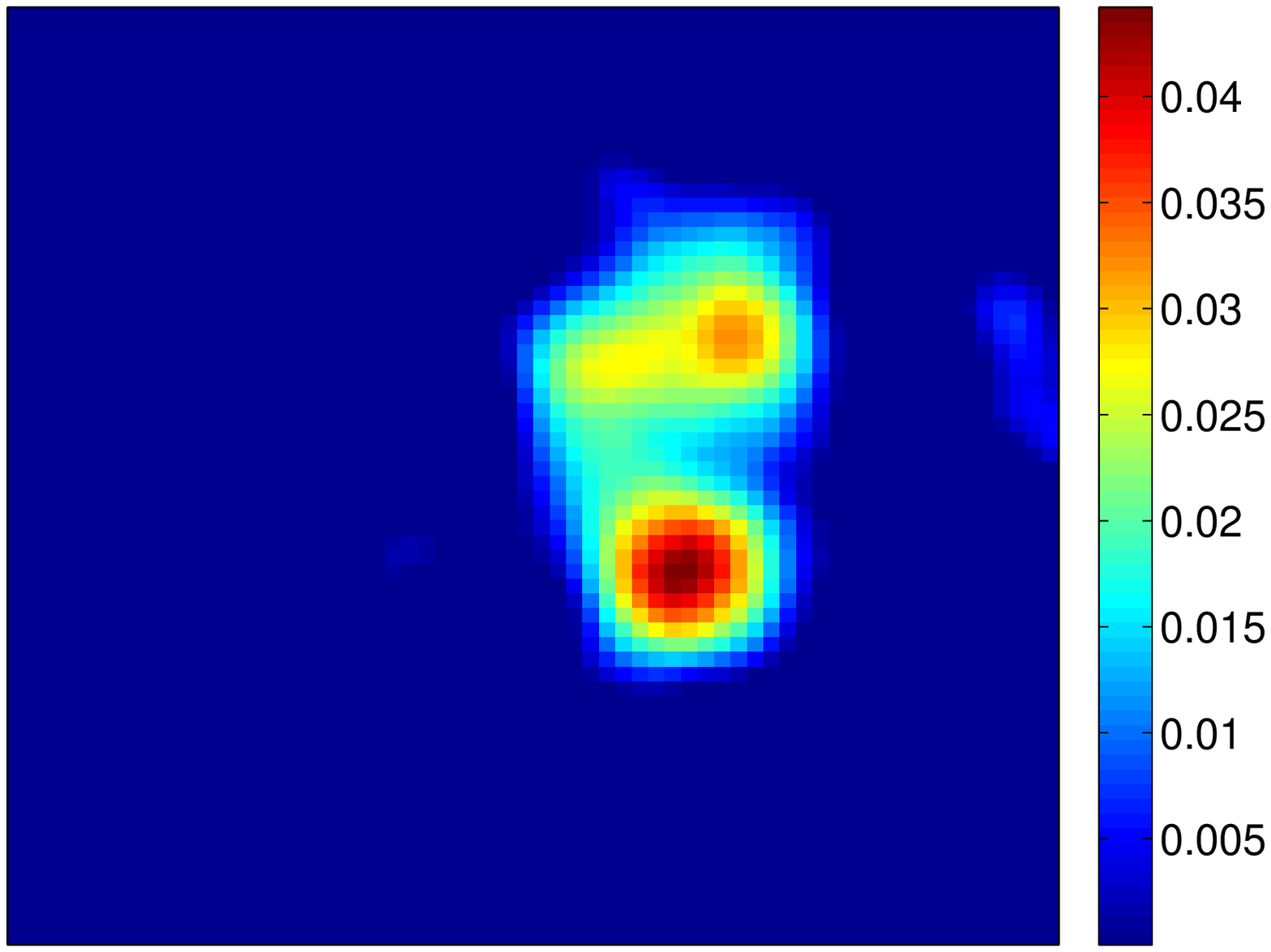} \\
\includegraphics[width=0.23\columnwidth]{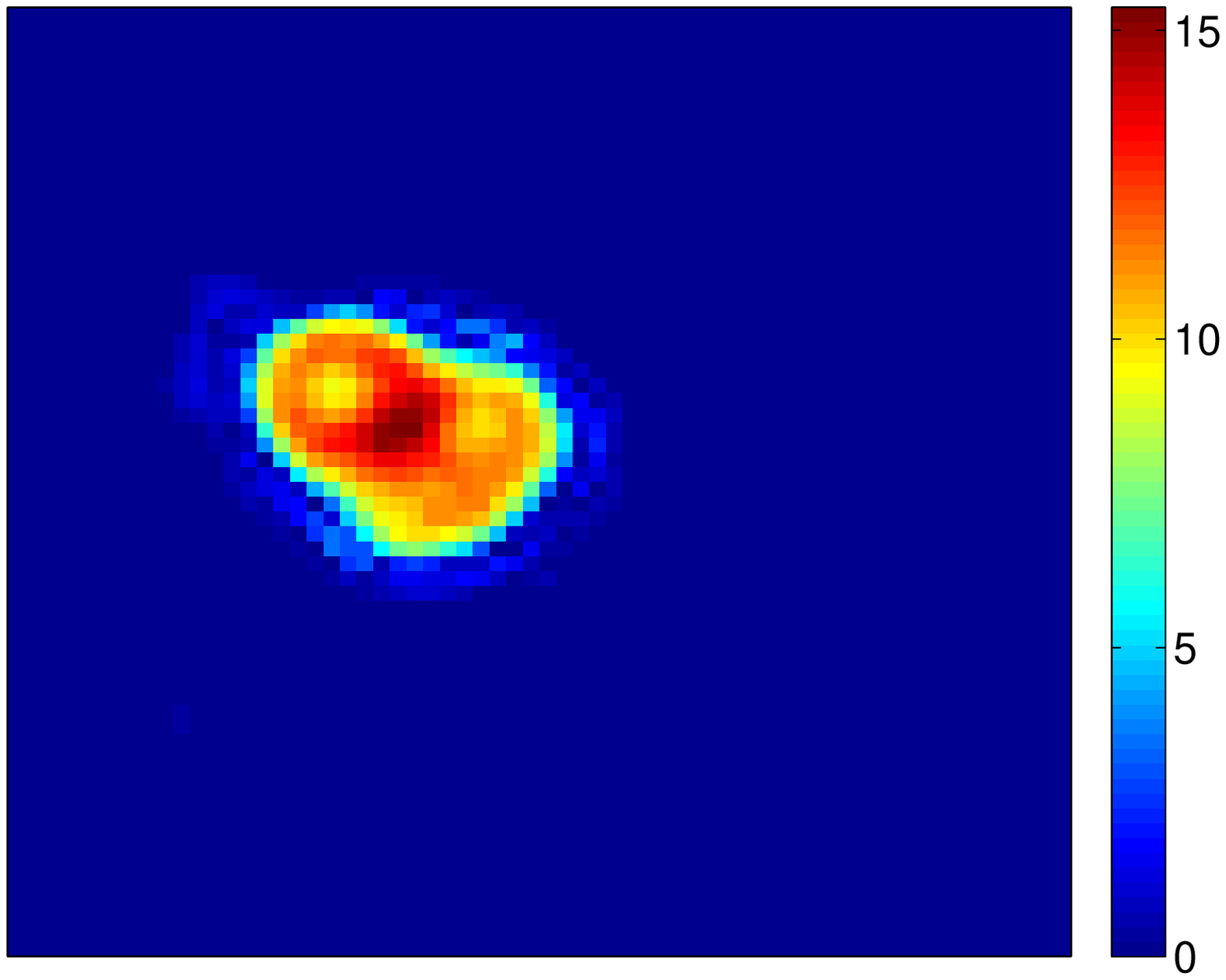} &
\includegraphics[width=0.23\columnwidth]{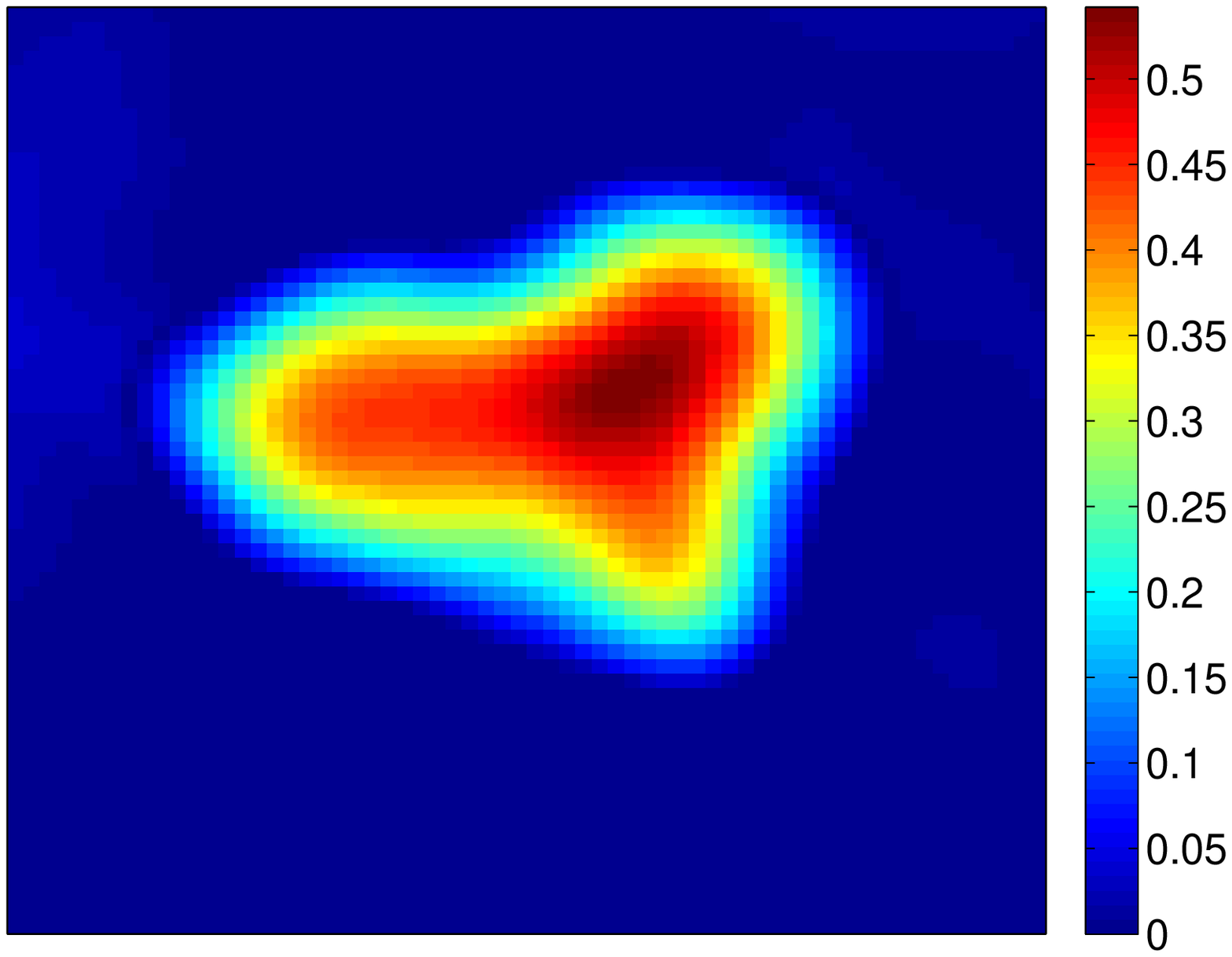} &
\includegraphics[width=0.23\columnwidth]{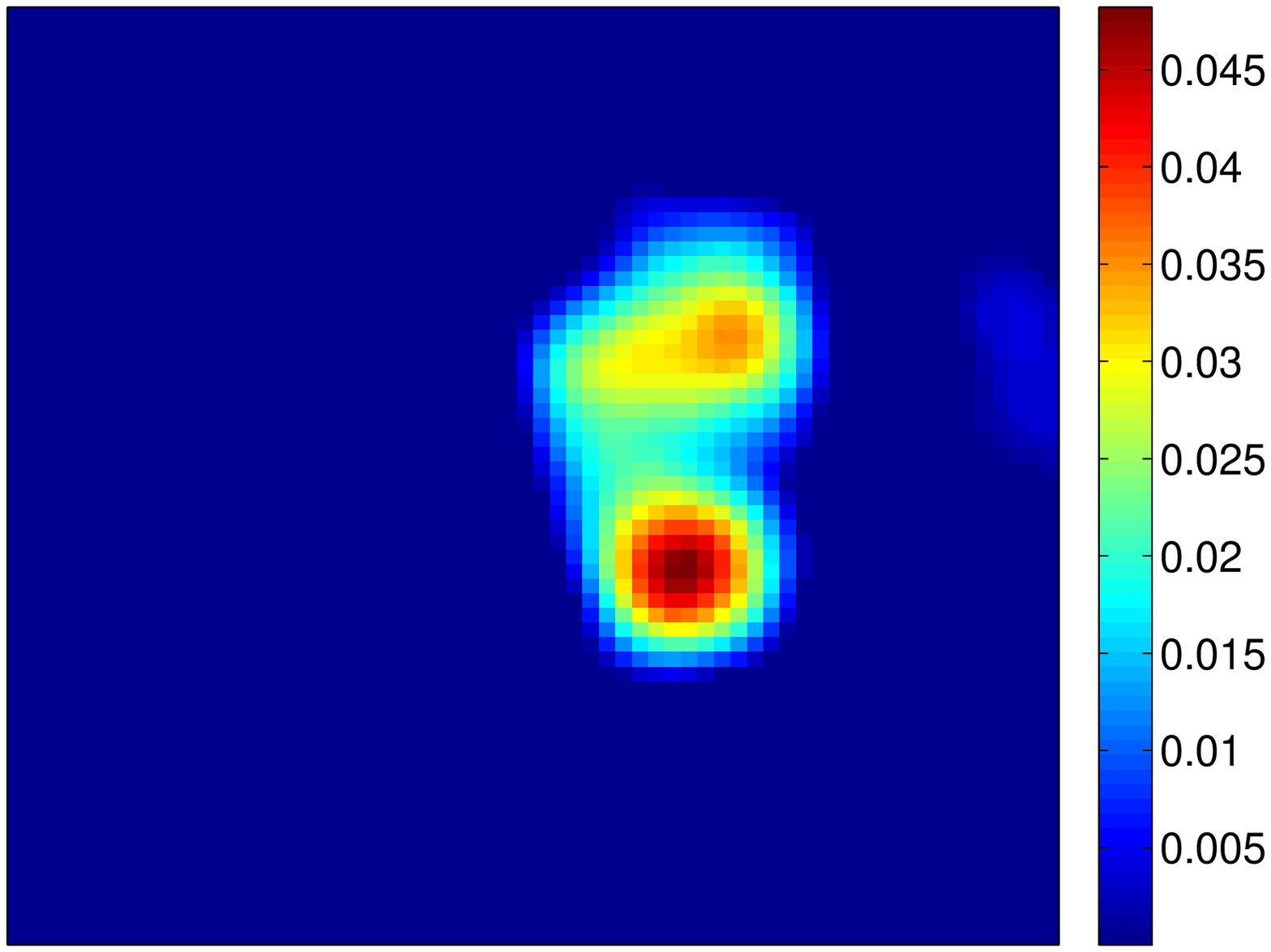} \\
\includegraphics[width=0.23\columnwidth]{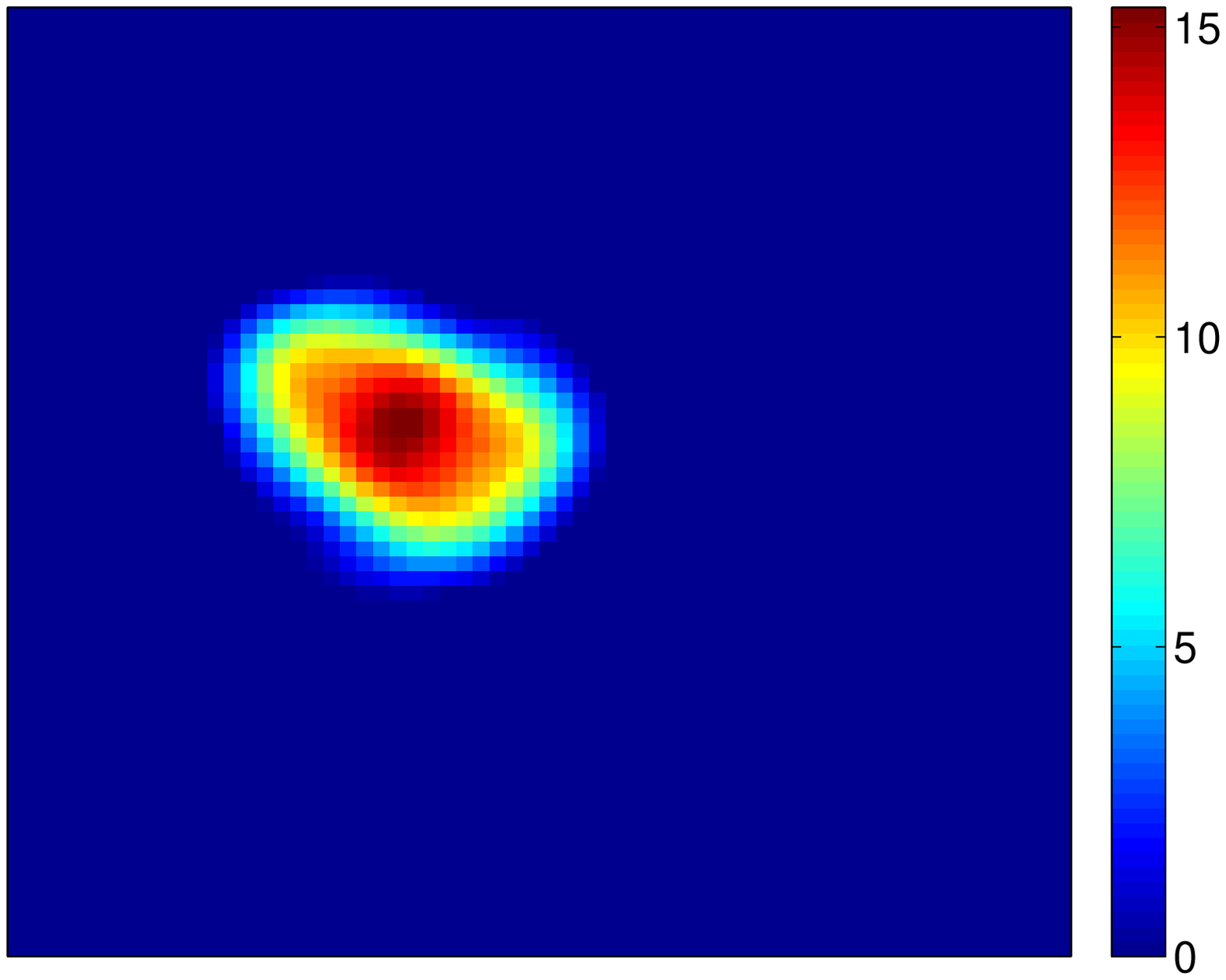} &
\includegraphics[width=0.23\columnwidth]{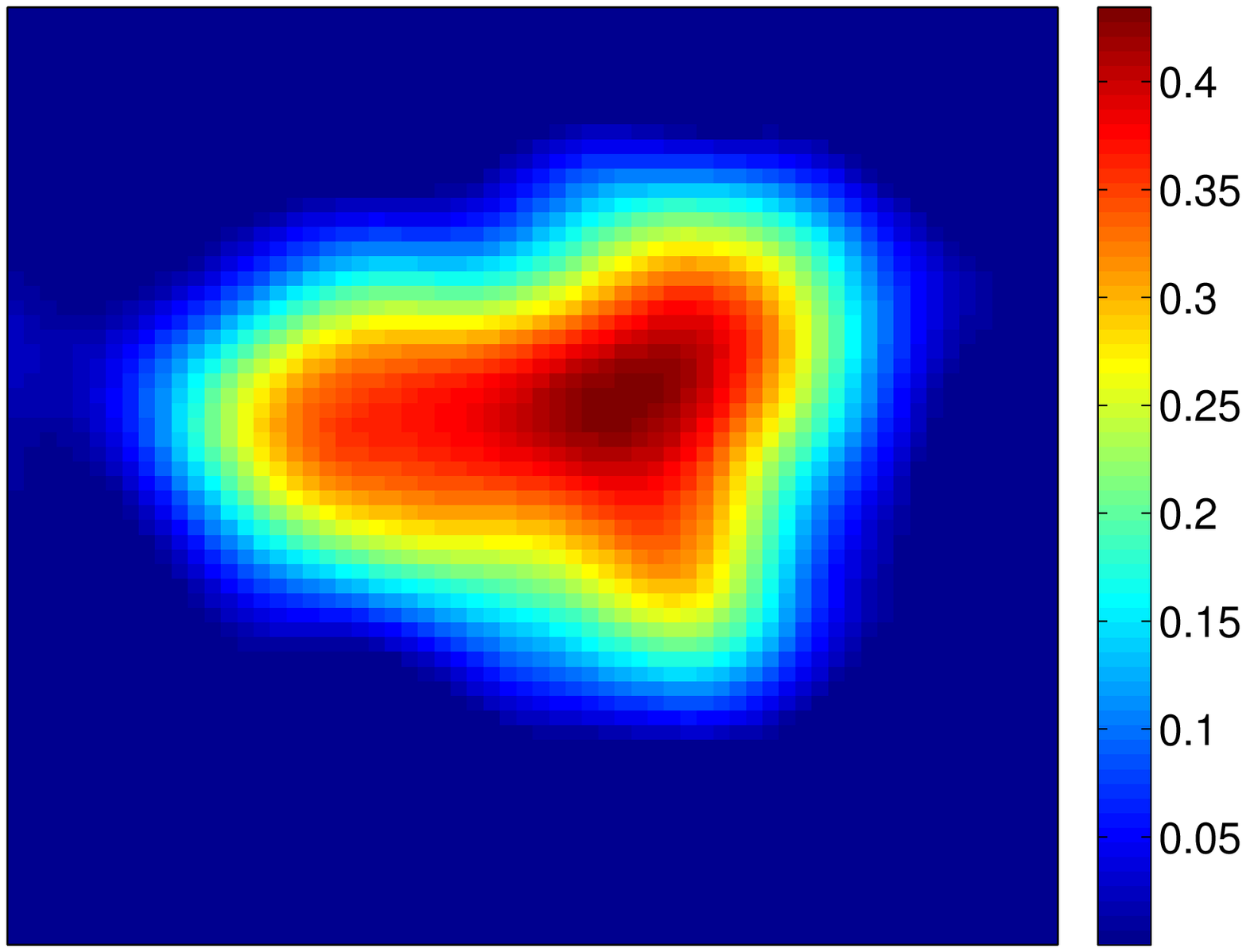} &
\includegraphics[width=0.23\columnwidth]{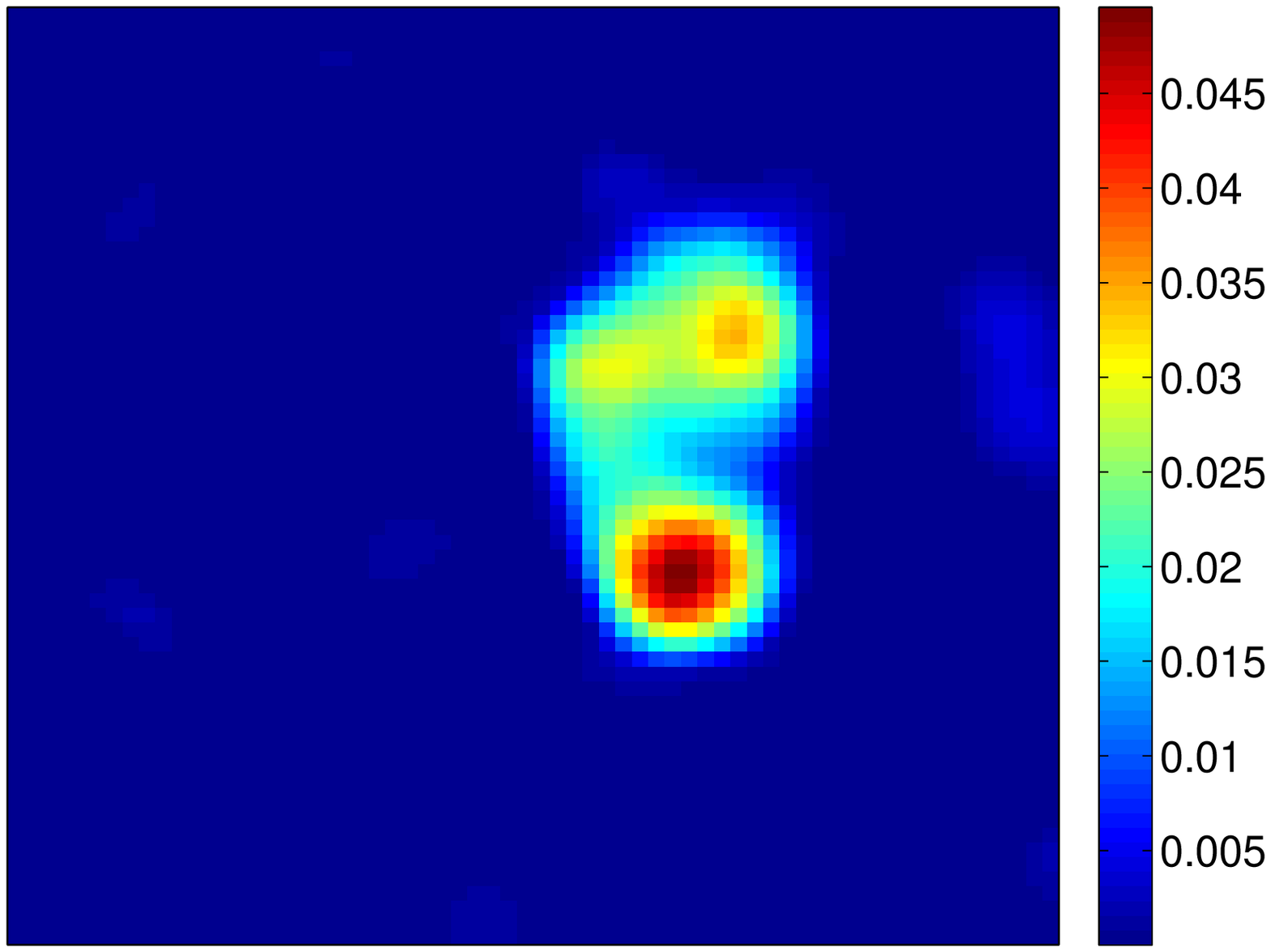} \\
\includegraphics[width=0.23\columnwidth]{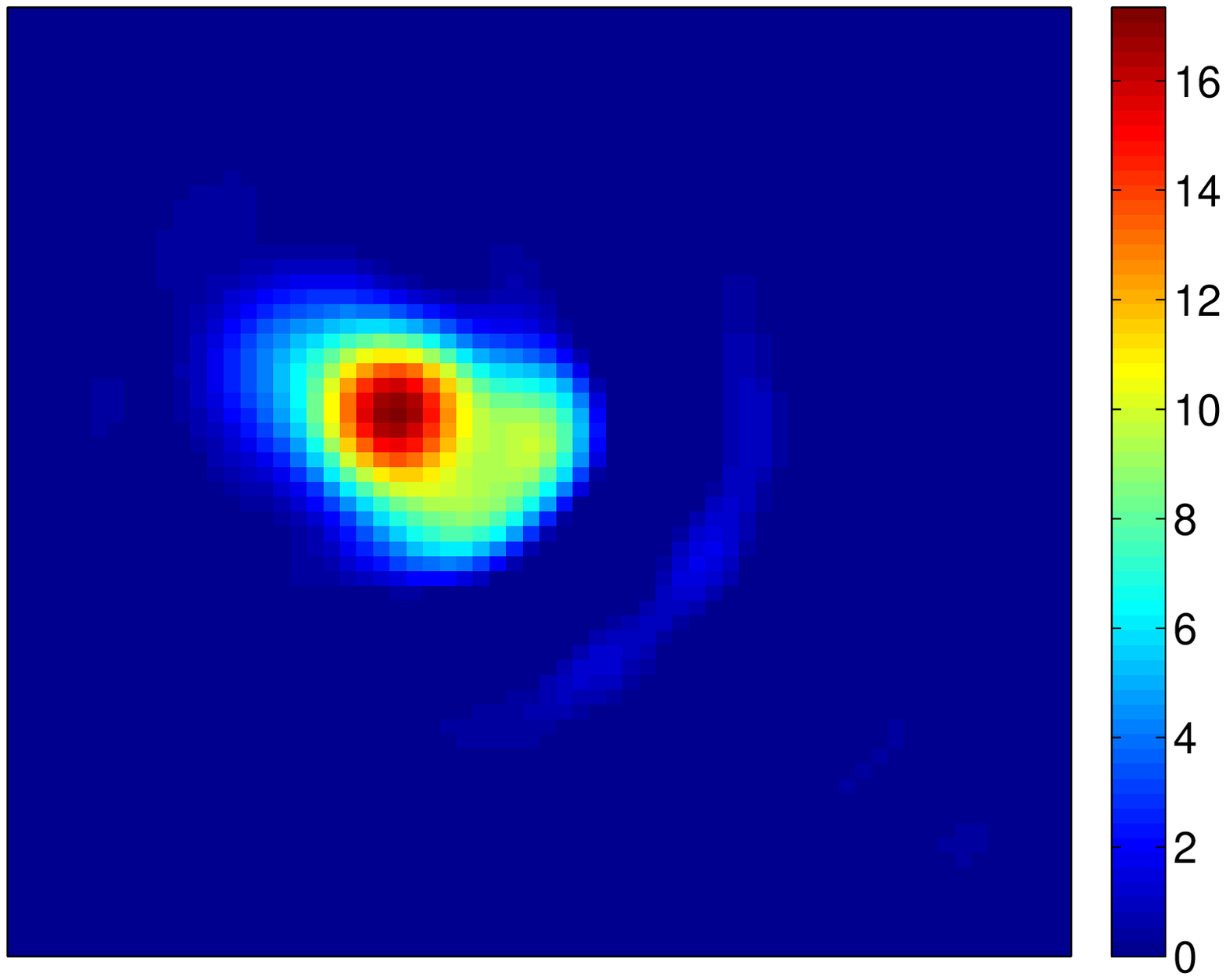} &
\includegraphics[width=0.23\columnwidth]{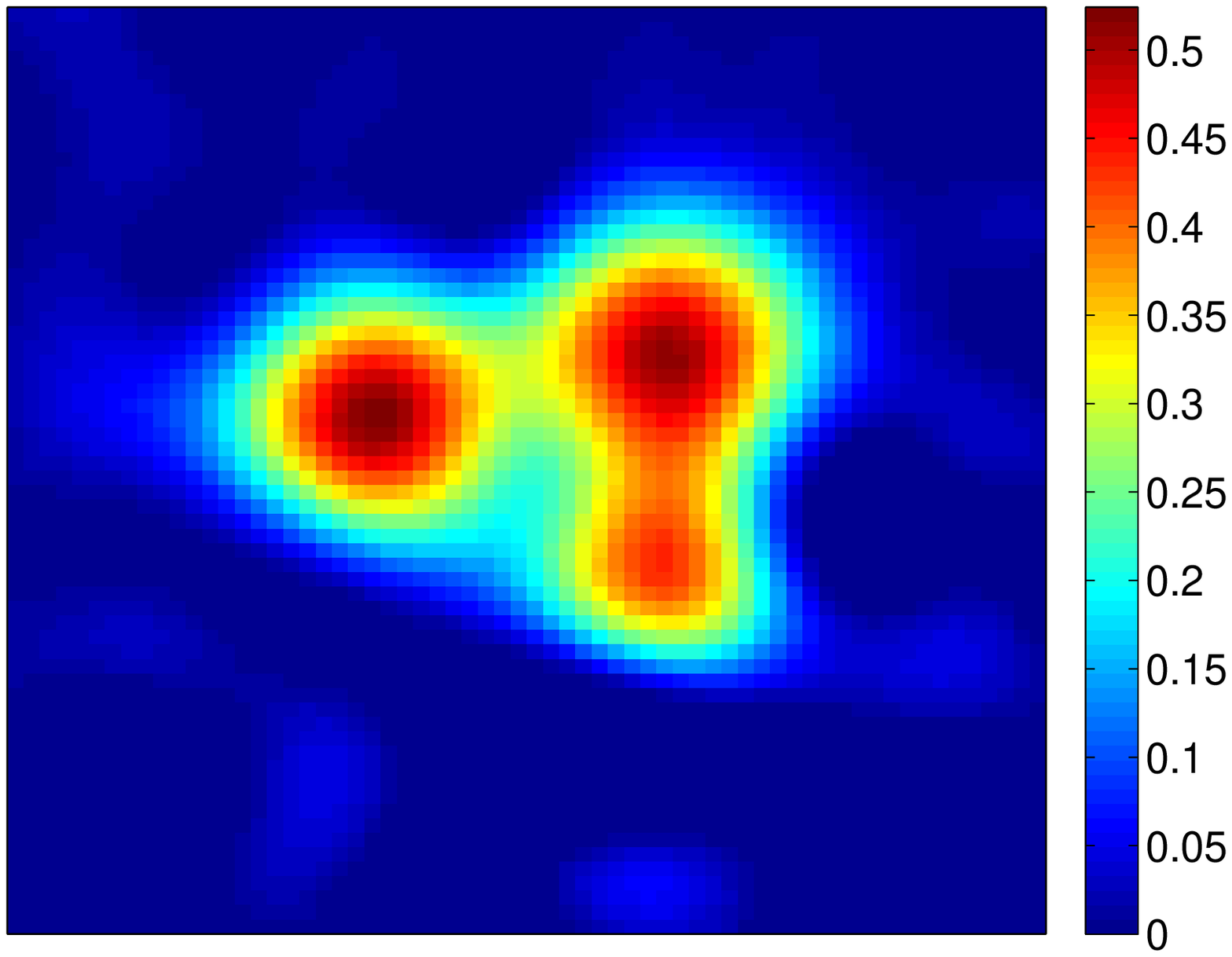} &
\includegraphics[width=0.23\columnwidth]{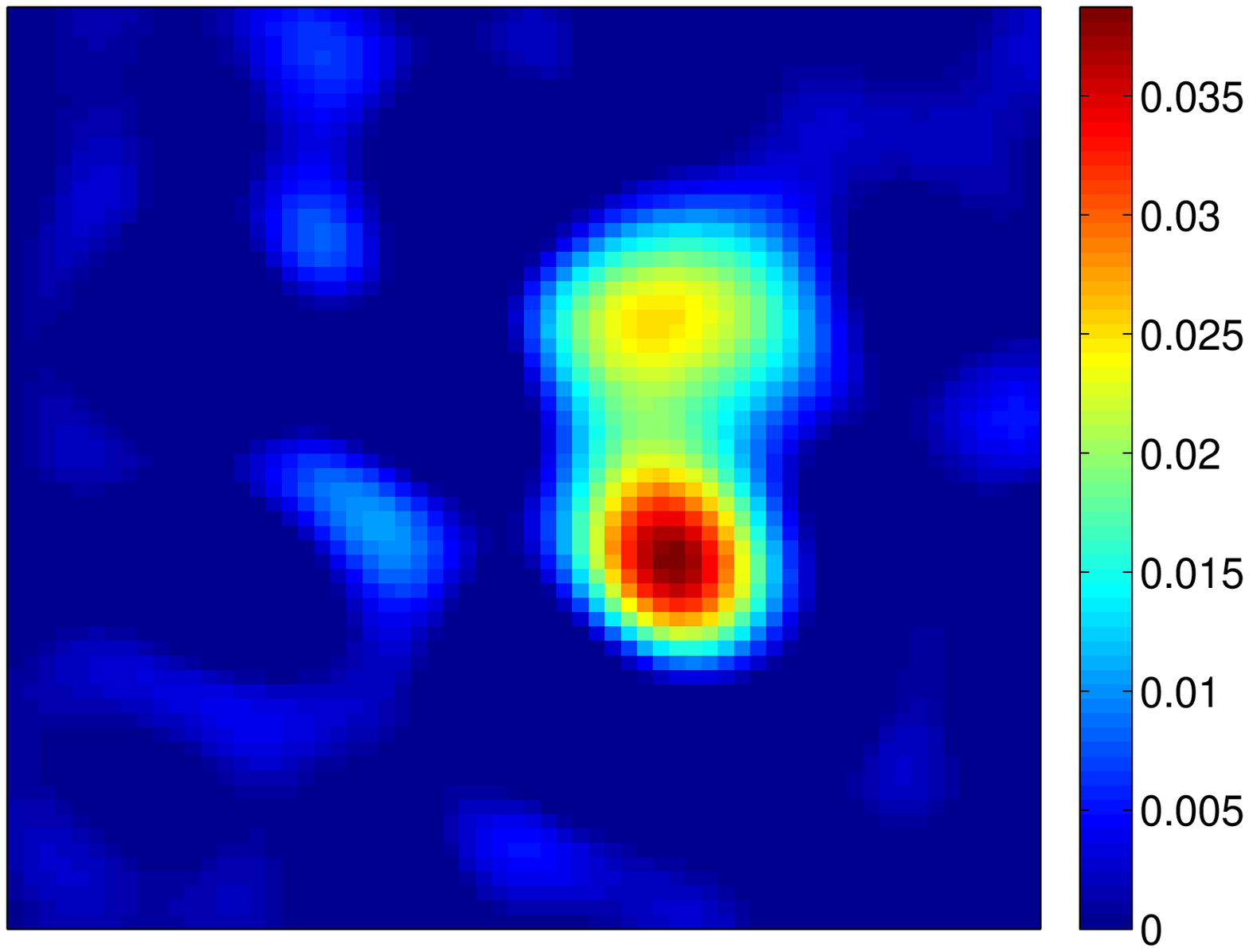} \\
\end{tabular}
\caption{The original images (first row) of the Sim1, Sim2 and Sim3 datasets with the reconstructions provided by EM (second row), SGP (third row), GPE (fourth row) and AS\_CBB (fifth row) as iterative regularization methods applied to \eqref{minpro} and combined with the discrepancy principle \eqref{discr1}. The images obtained with the visibility-based uv-smooth algorithm are also shown (last row).}
\label{figimm1}
\end{center}
\end{figure}

\begin{figure}
\begin{center}
\begin{tabular}{cccc}
\includegraphics[width=0.23\columnwidth]{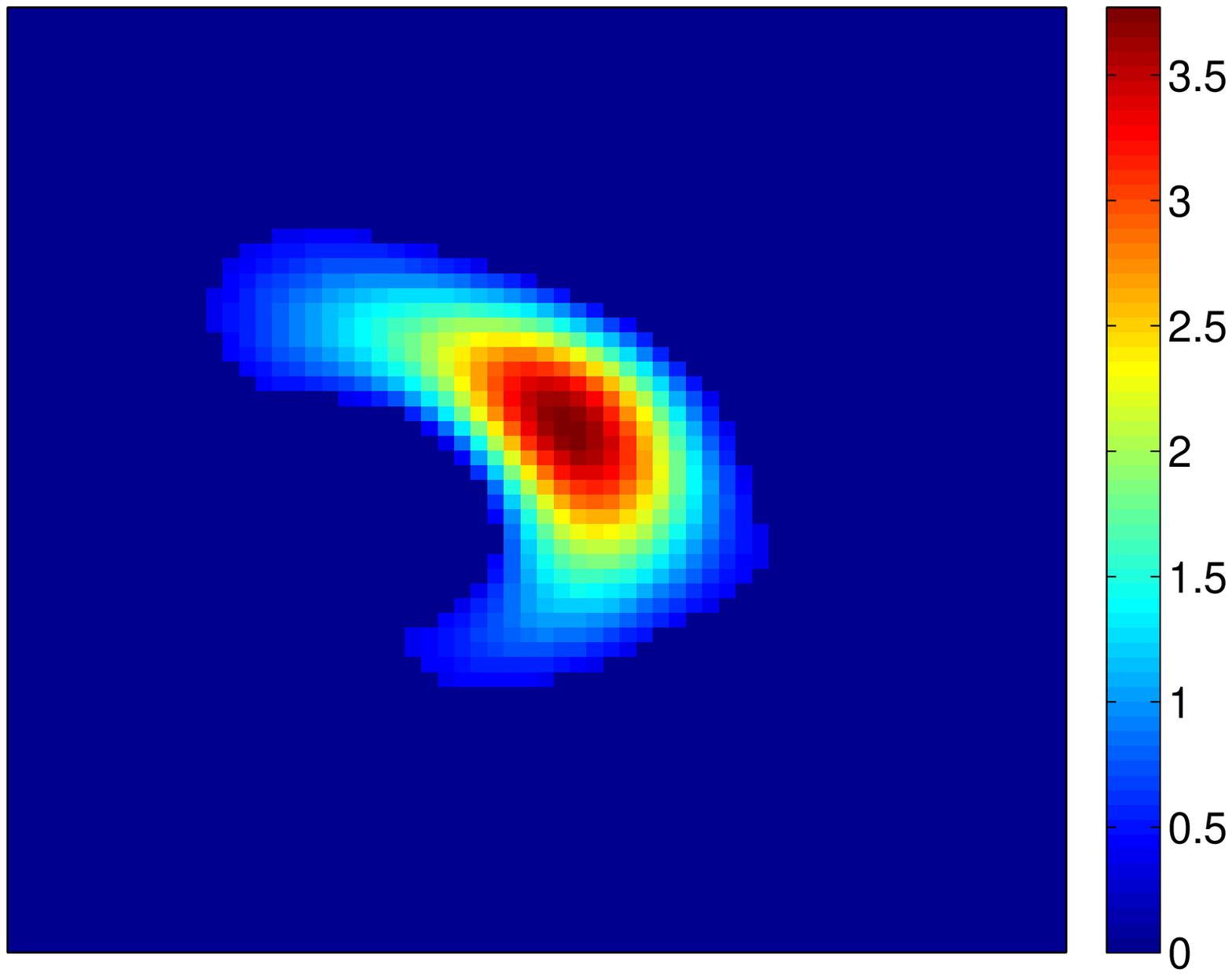} &
\includegraphics[width=0.23\columnwidth]{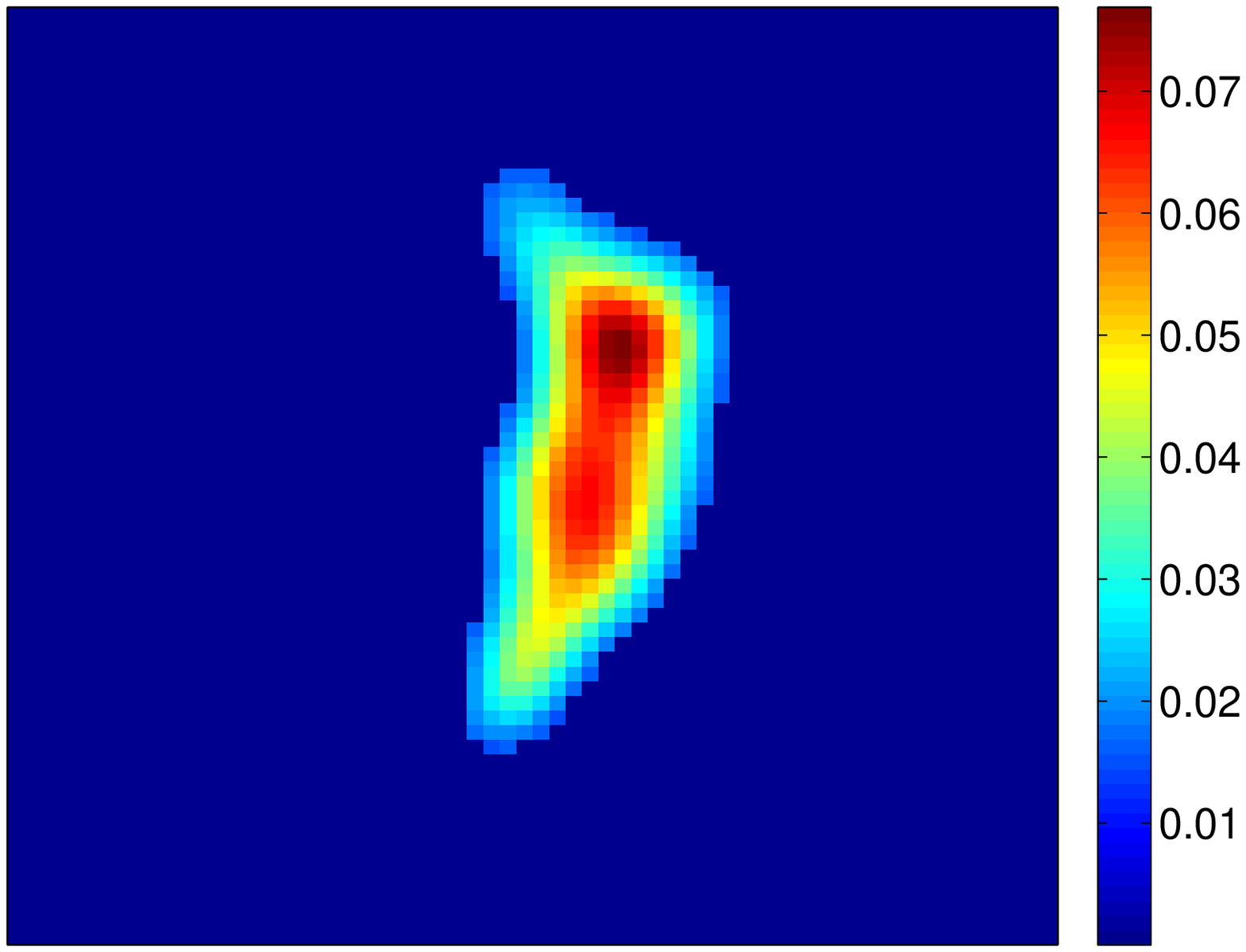} &
\includegraphics[width=0.23\columnwidth]{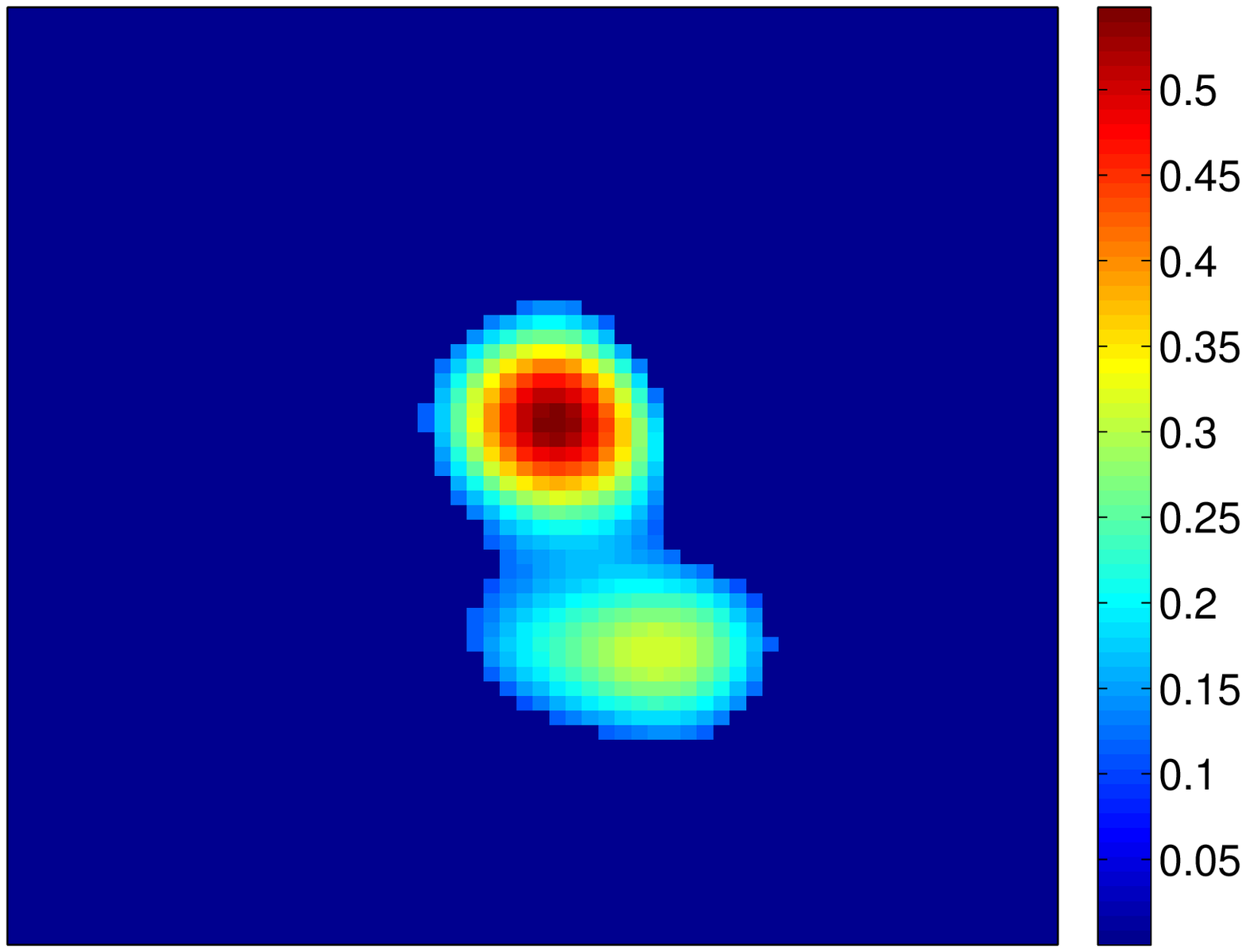} \\
\includegraphics[width=0.23\columnwidth]{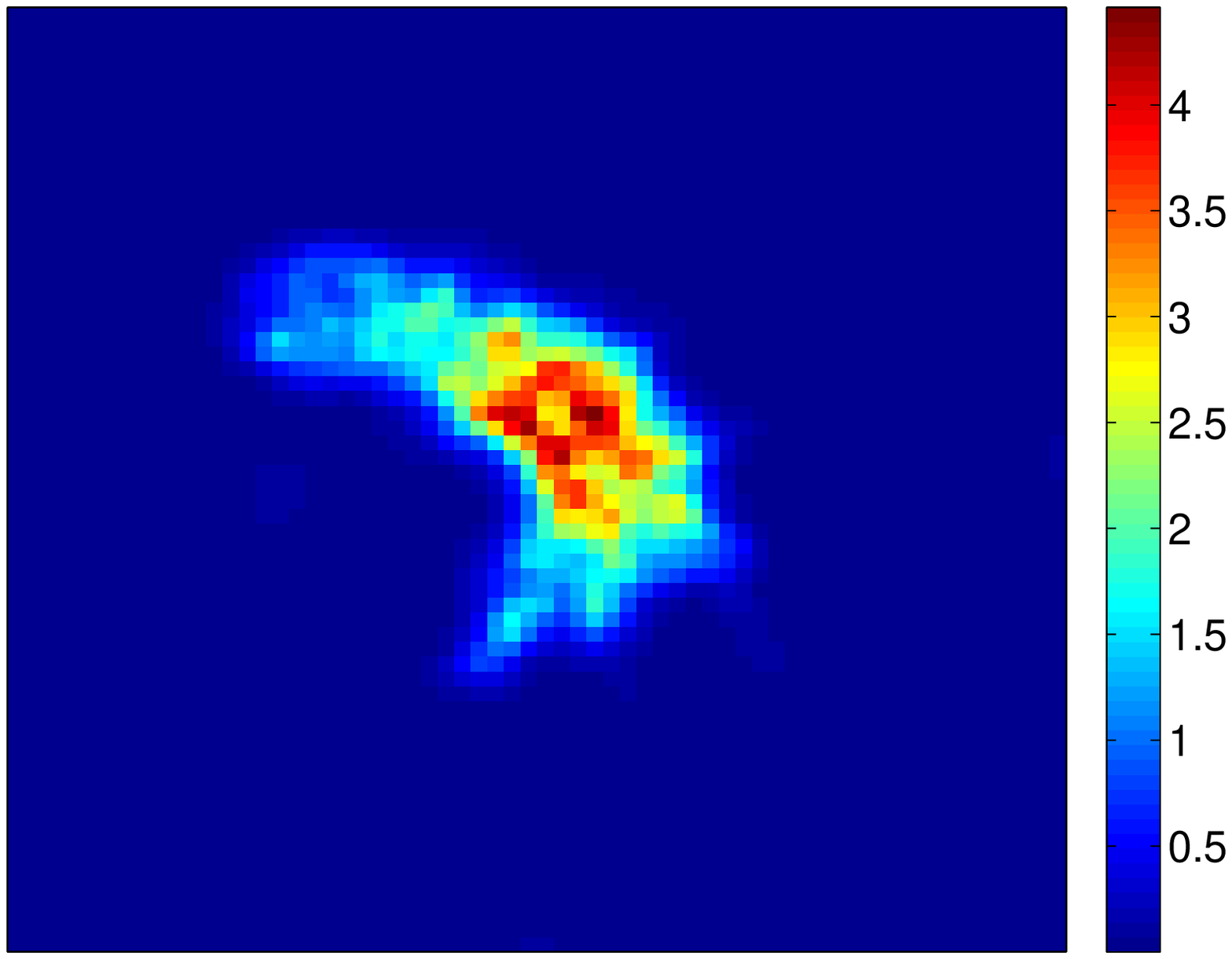} &
\includegraphics[width=0.23\columnwidth]{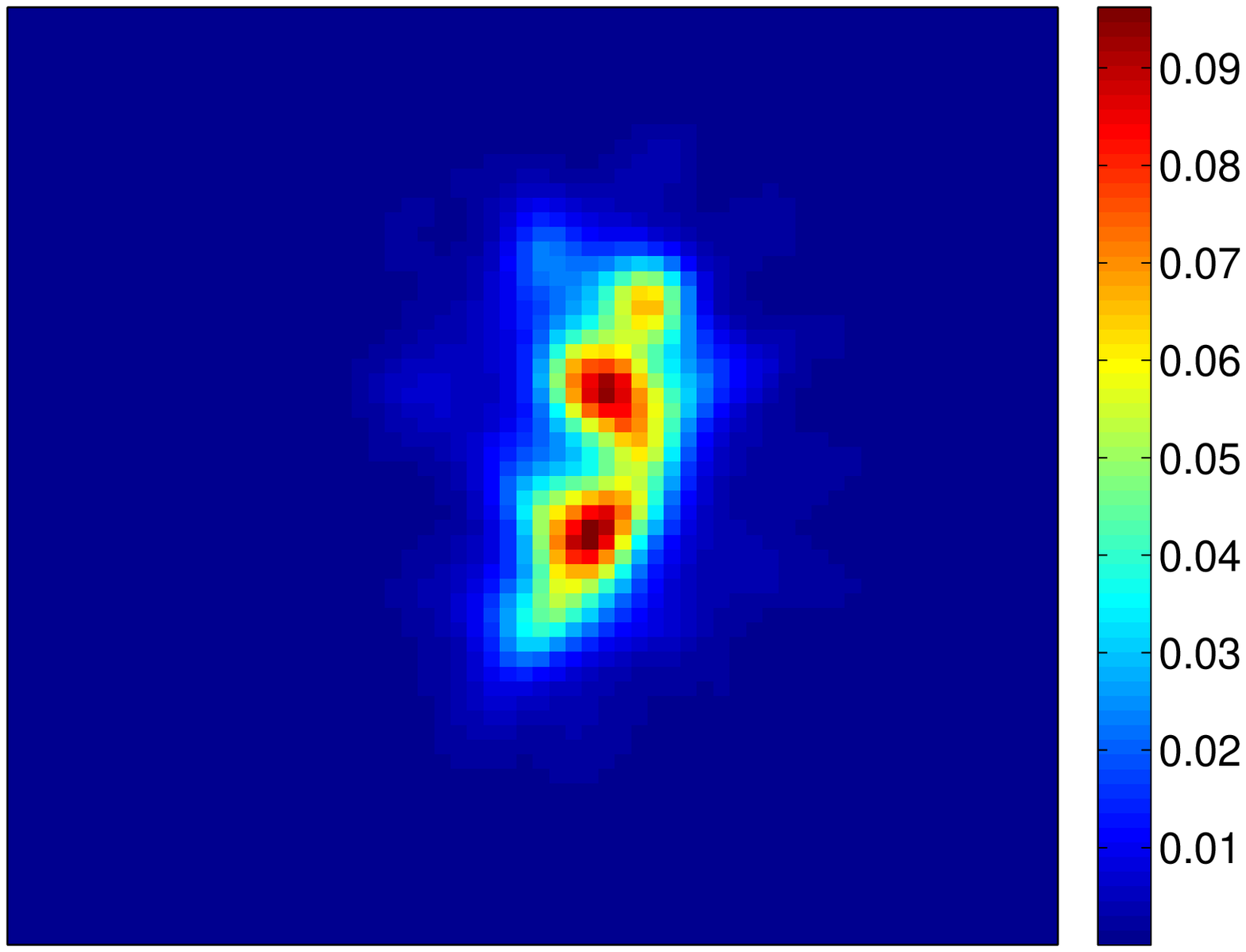} &
\includegraphics[width=0.23\columnwidth]{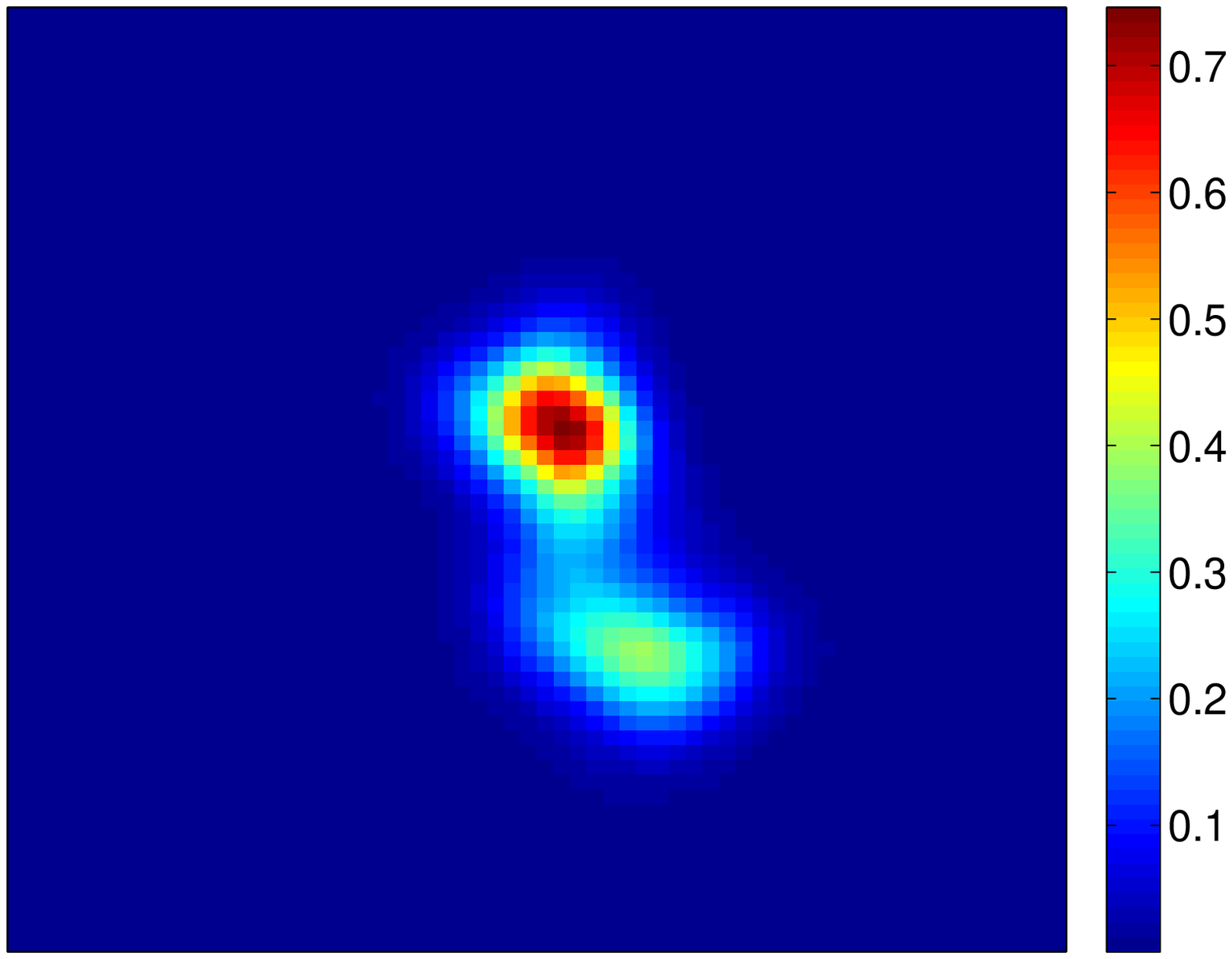} \\
\includegraphics[width=0.23\columnwidth]{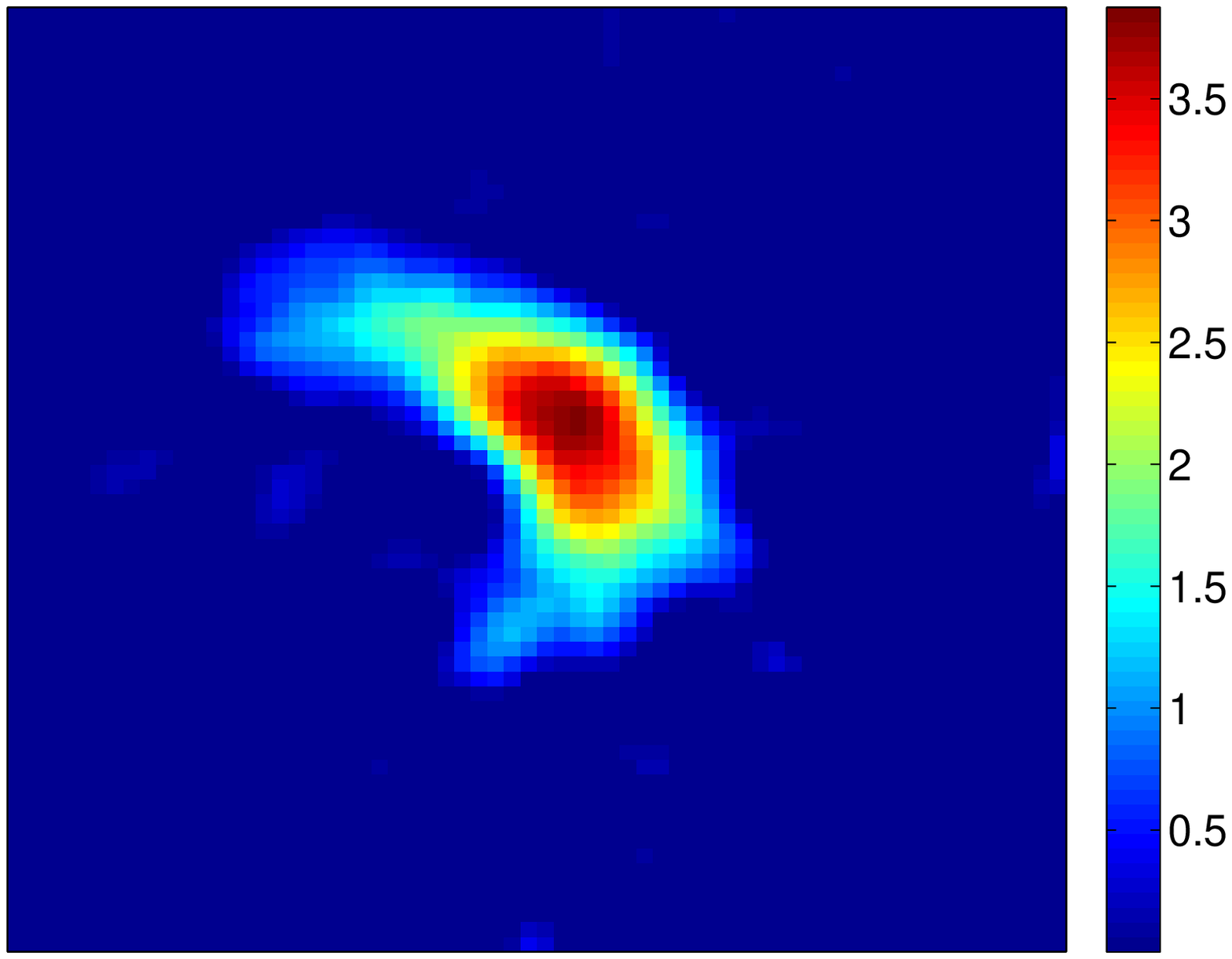} &
\includegraphics[width=0.23\columnwidth]{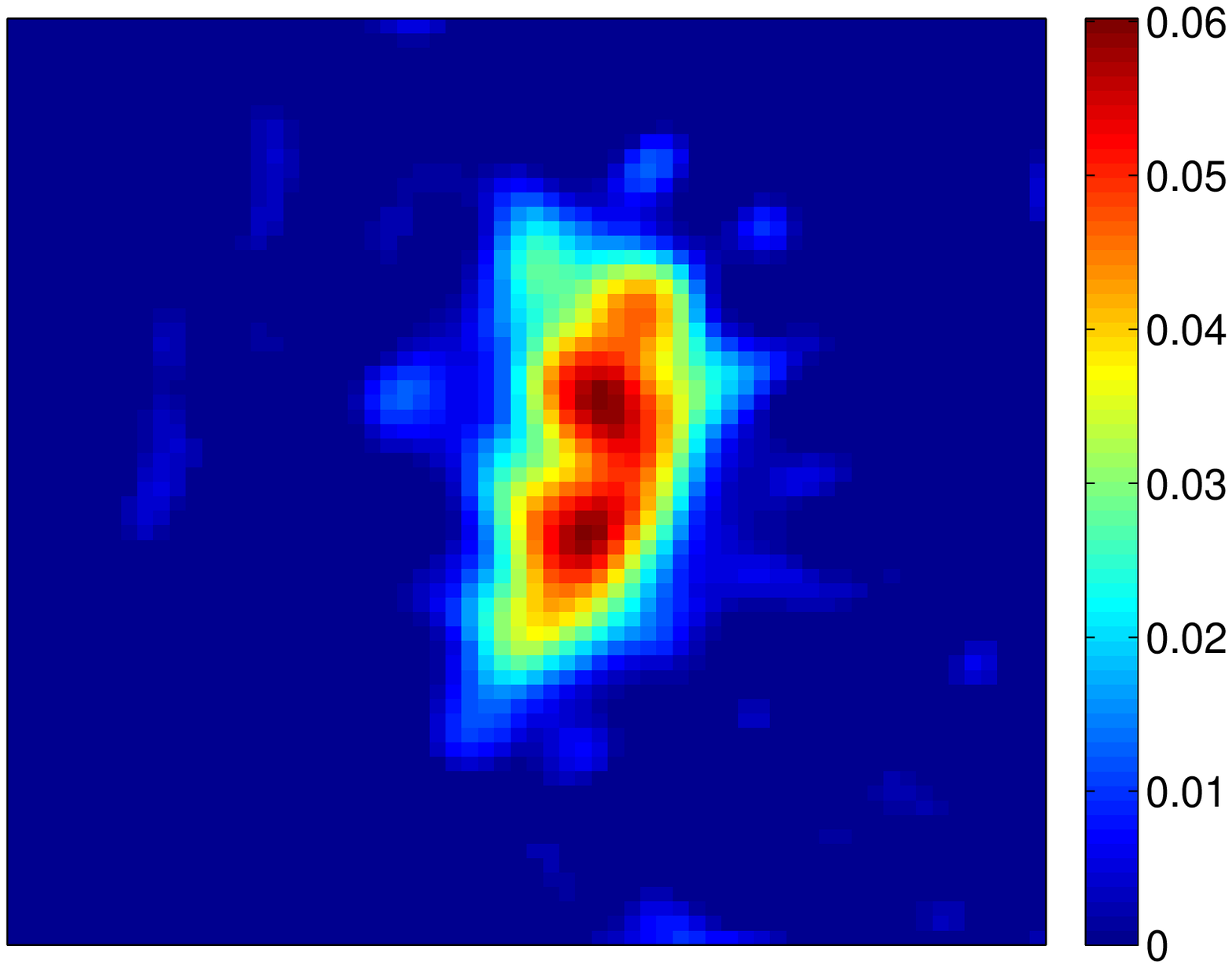} &
\includegraphics[width=0.23\columnwidth]{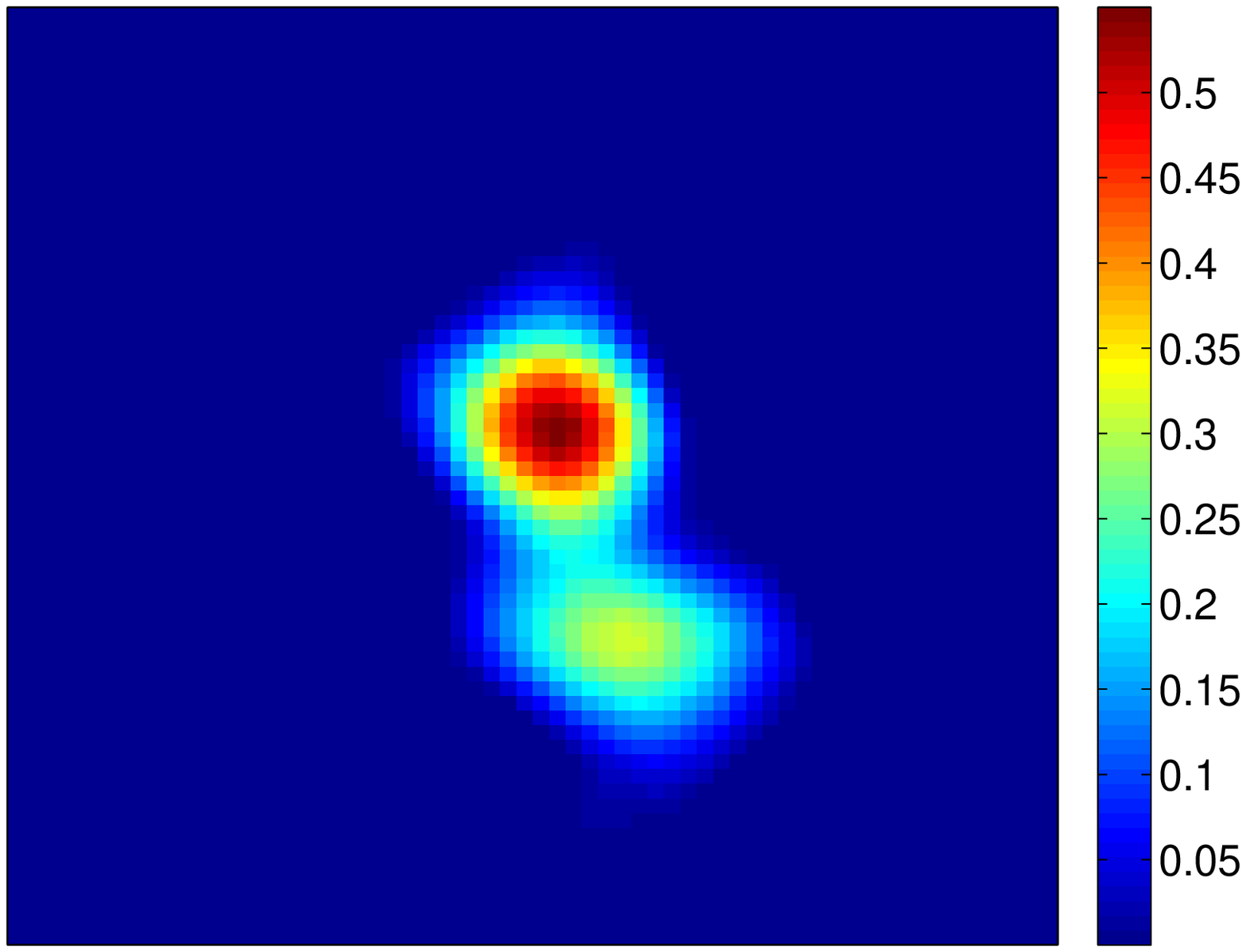} \\
\includegraphics[width=0.23\columnwidth]{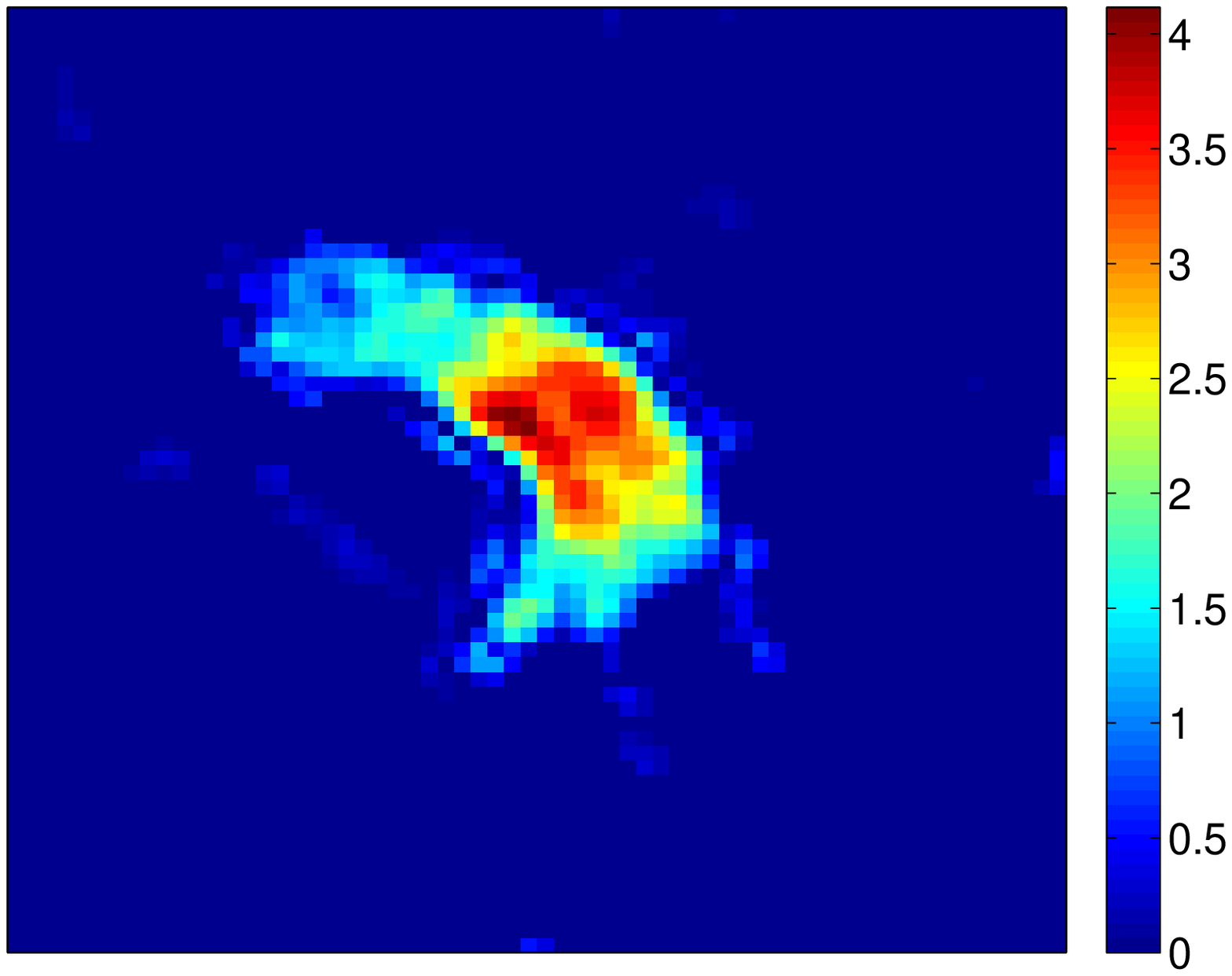} &
\includegraphics[width=0.23\columnwidth]{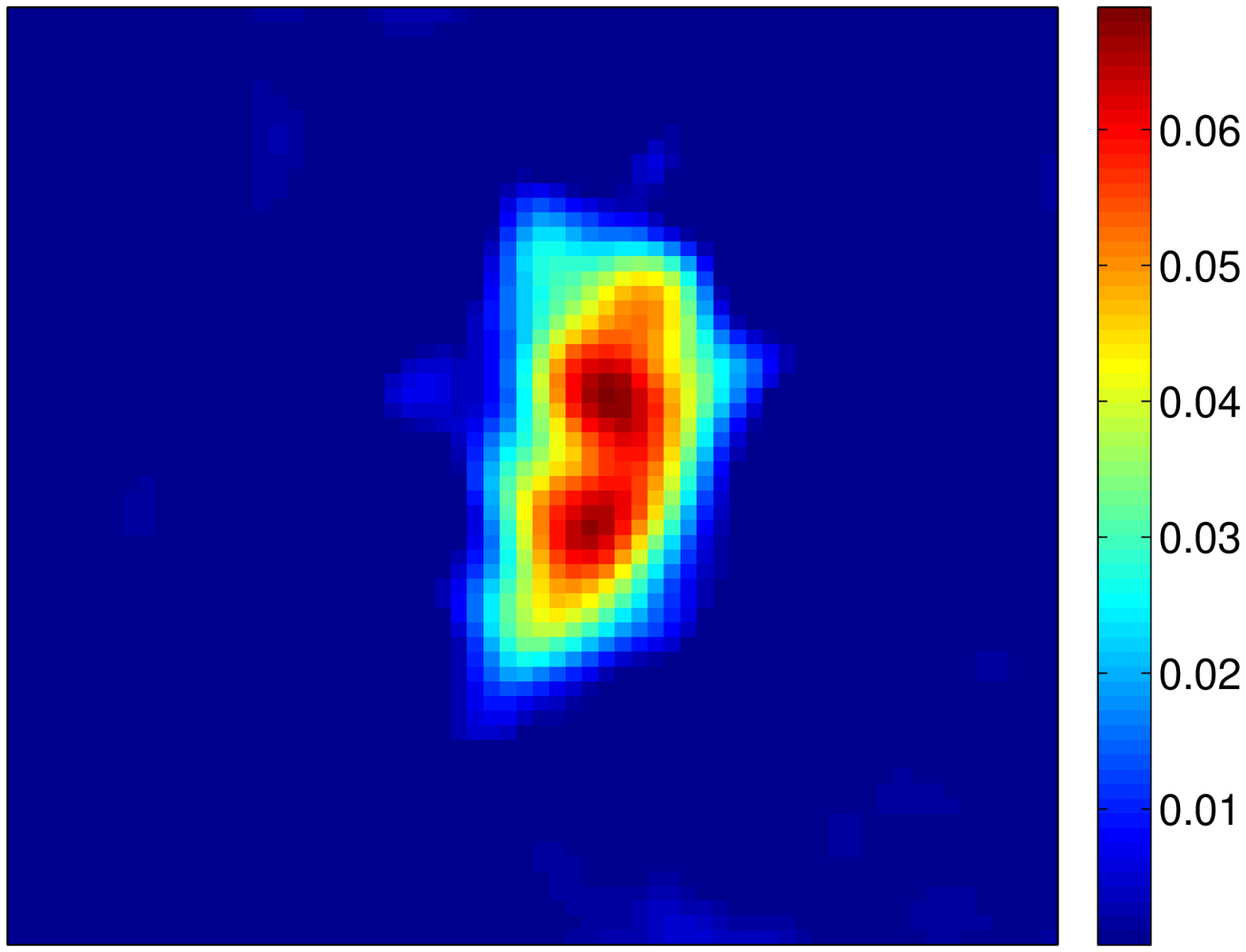} &
\includegraphics[width=0.23\columnwidth]{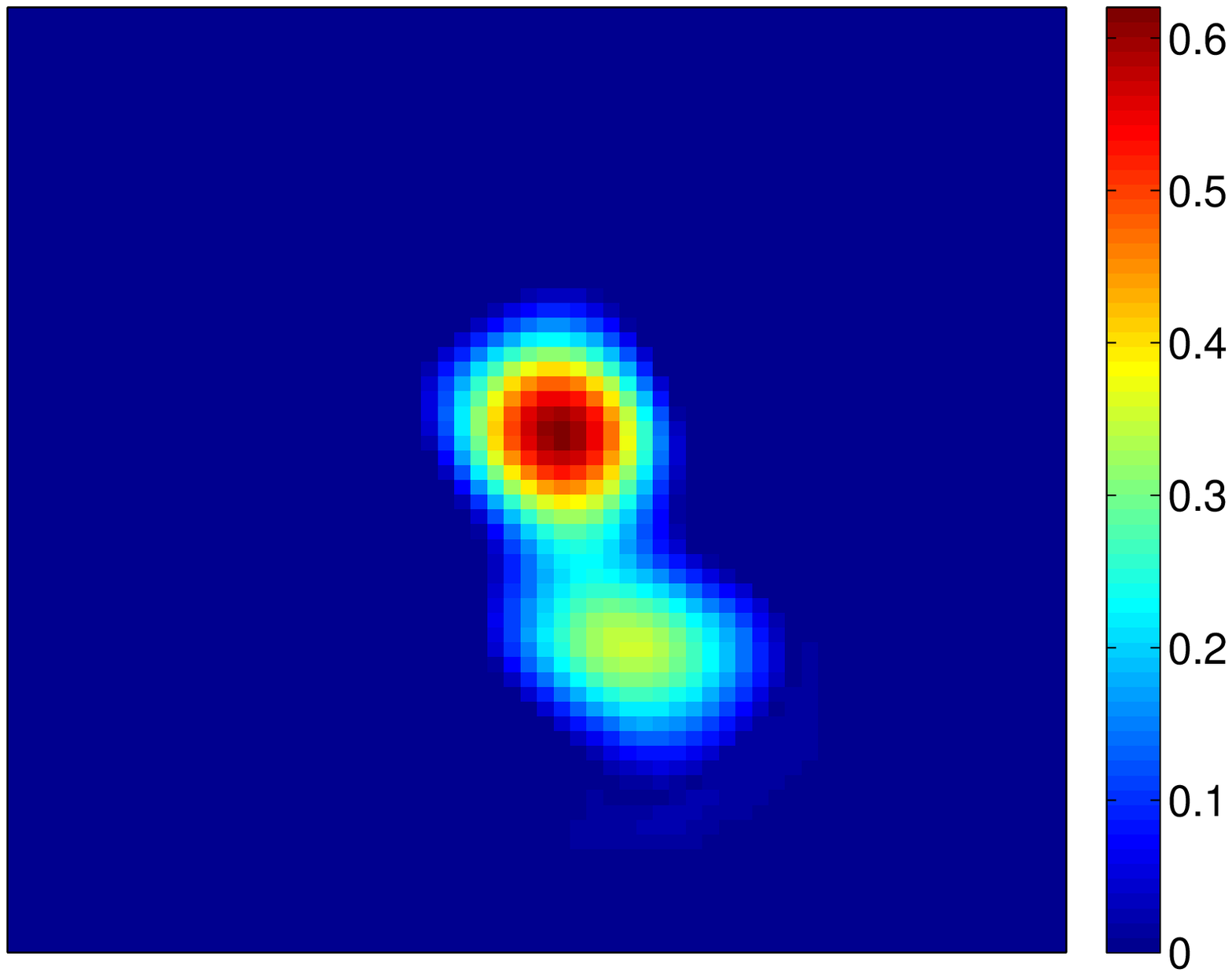} \\
\includegraphics[width=0.23\columnwidth]{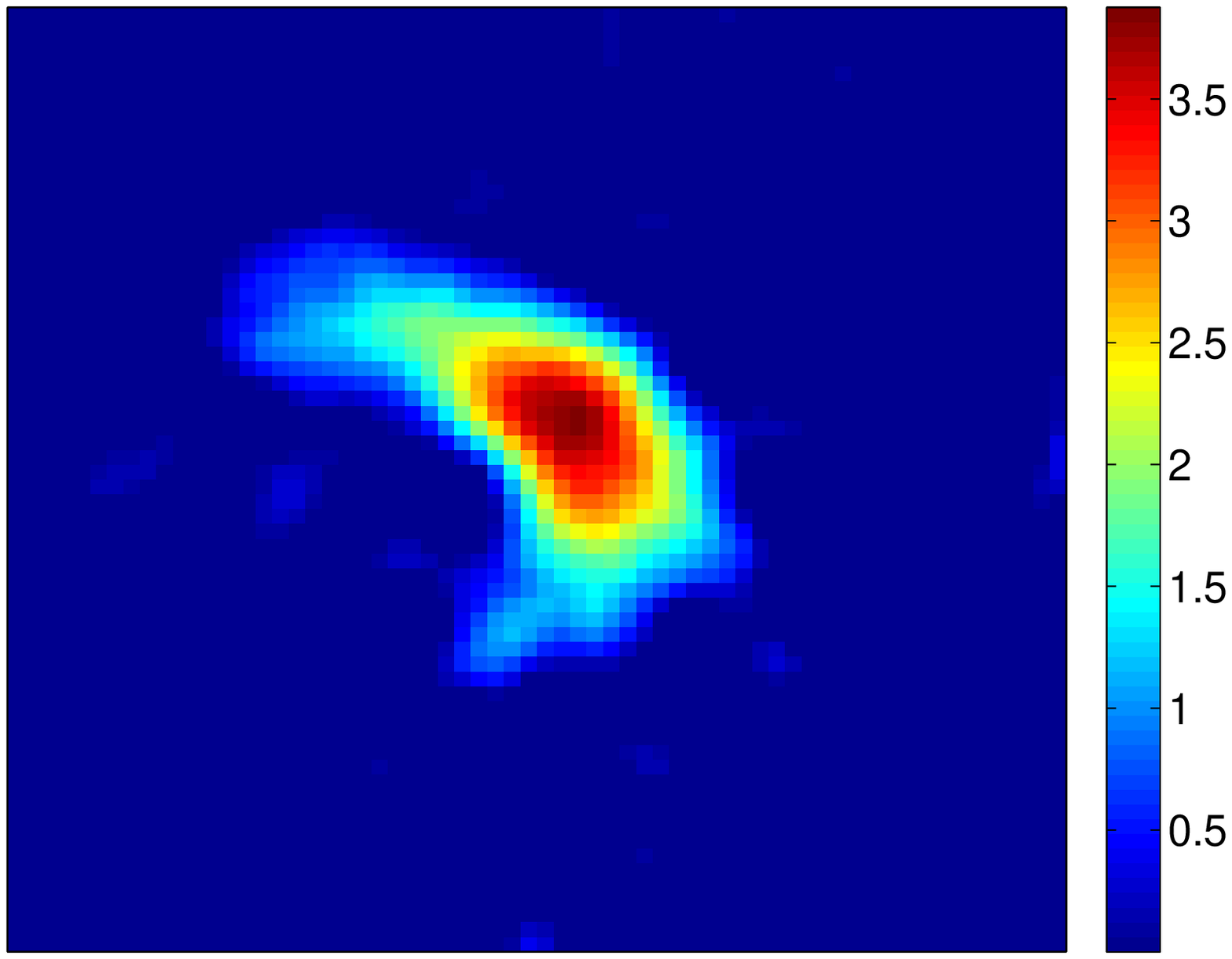} &
\includegraphics[width=0.23\columnwidth]{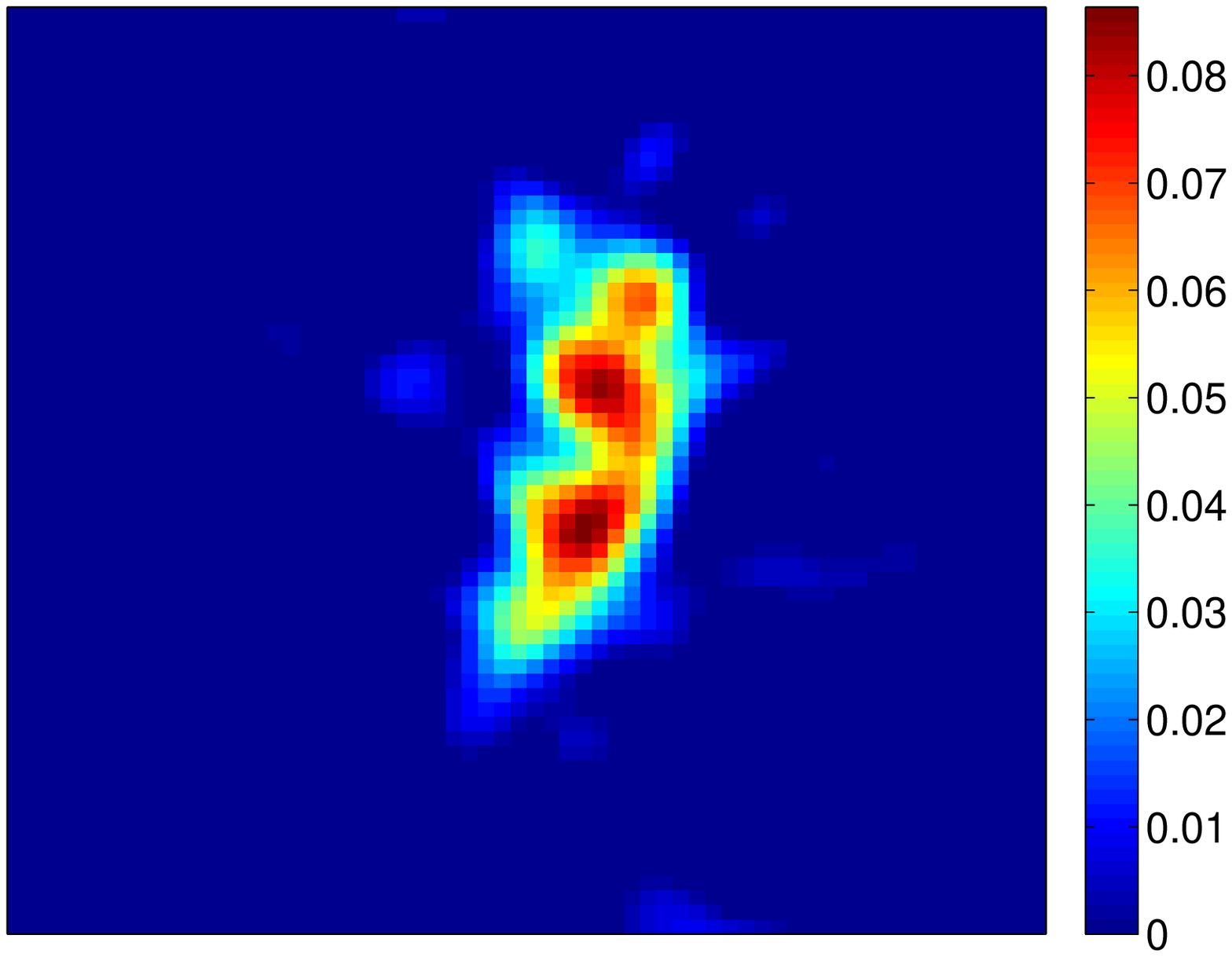} &
\includegraphics[width=0.23\columnwidth]{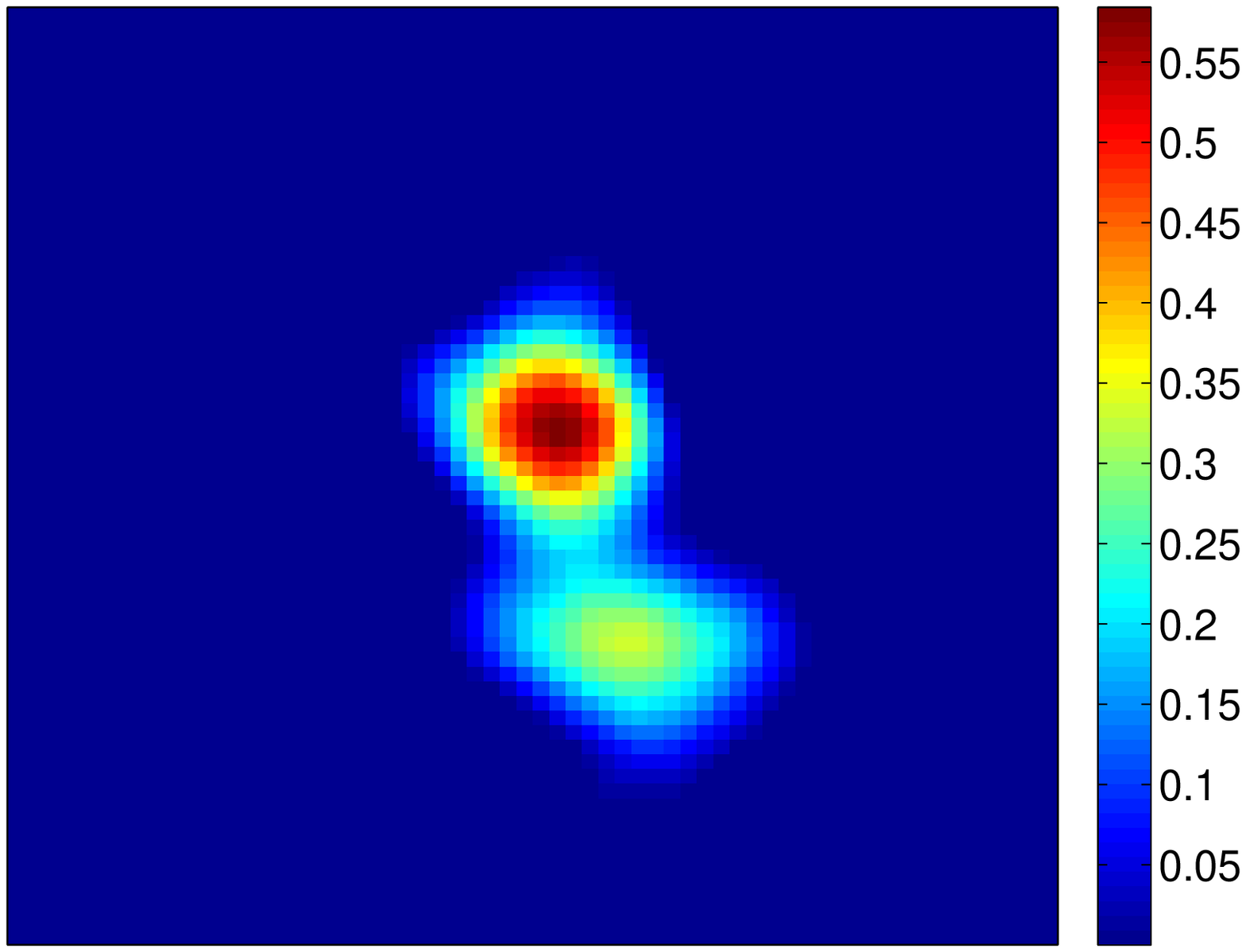} \\
\includegraphics[width=0.23\columnwidth]{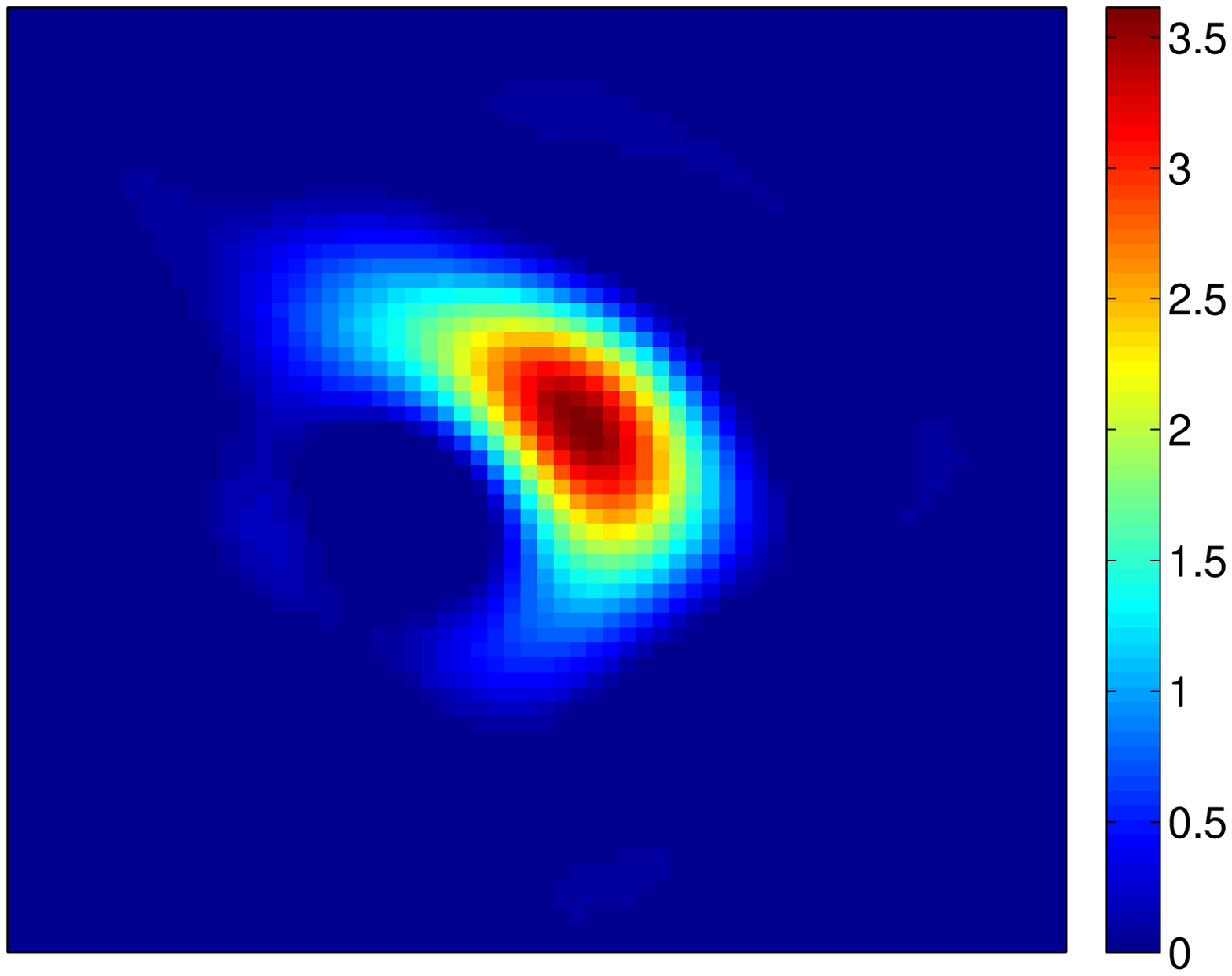} &
\includegraphics[width=0.23\columnwidth]{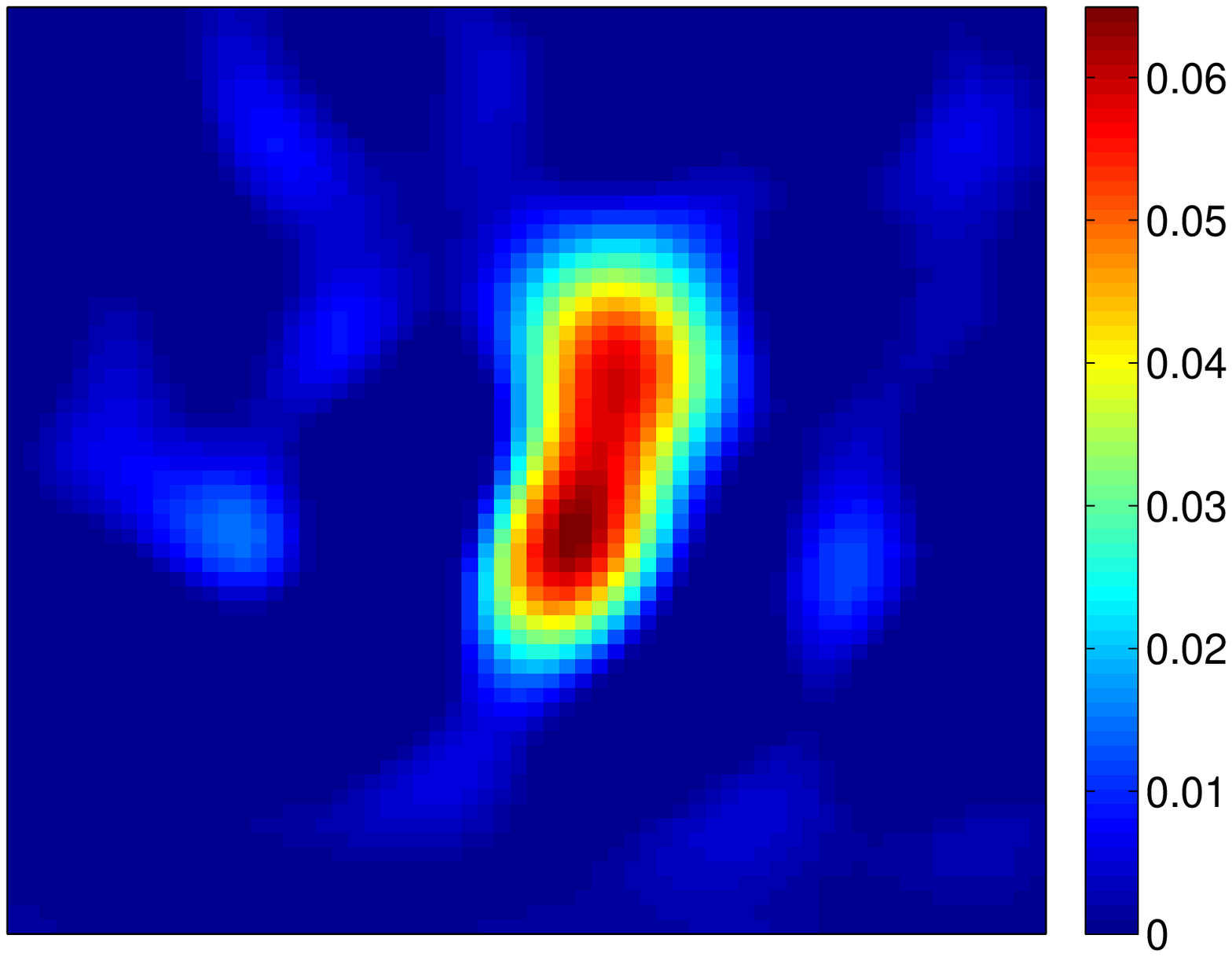} &
\includegraphics[width=0.23\columnwidth]{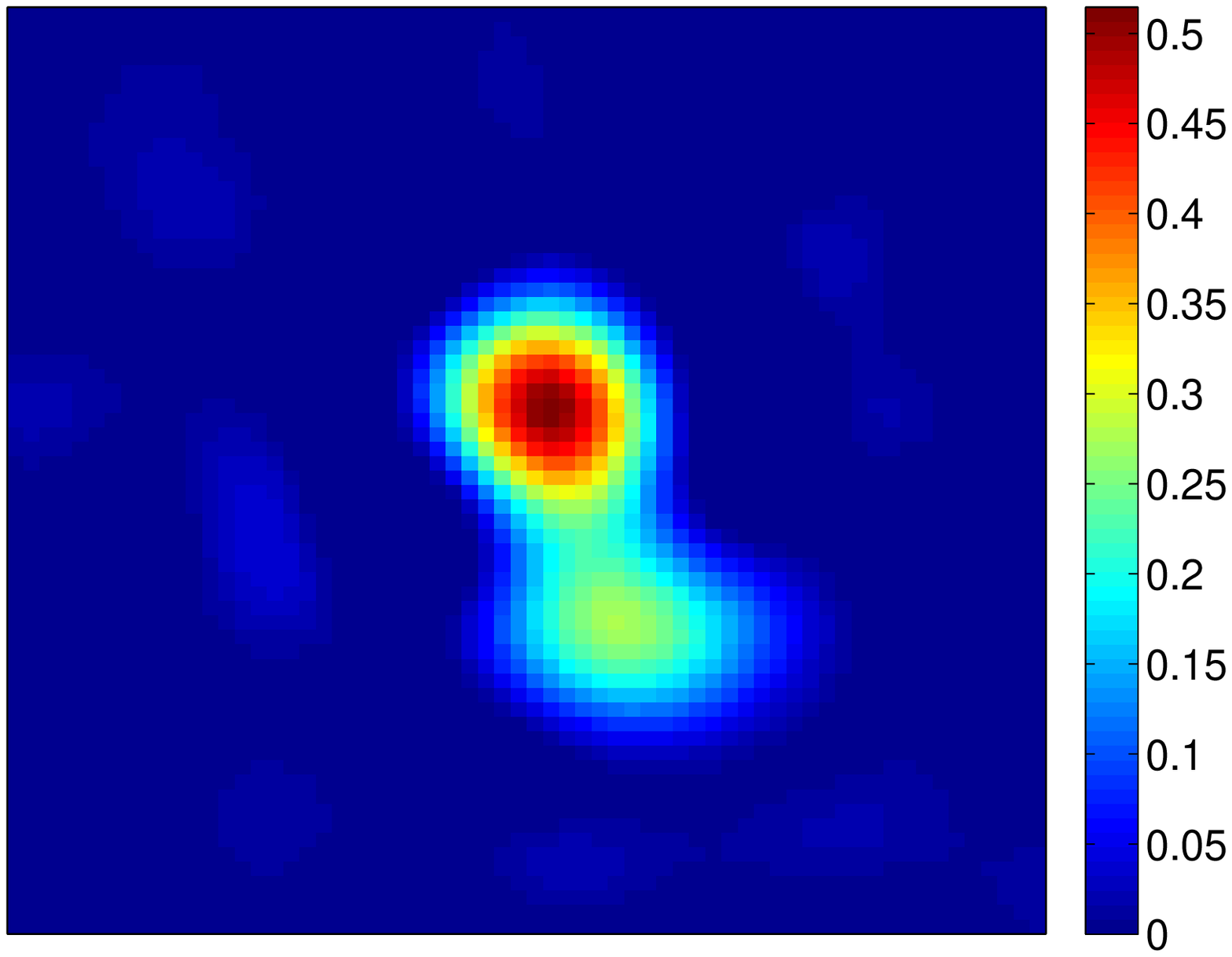} \\
\end{tabular}
\caption{The original images (first row) of the Sim4, Sim5 and Sim6 datasets with the reconstructions provided by EM (second row), SGP (third row), GPE (fourth row) and AS\_CBB (fifth row) as iterative regularization methods applied to \eqref{minpro} and combined with the discrepancy principle \eqref{discr1}. The images obtained with the visibility-based uv-smooth algorithm are also shown (last row).}
\label{figimm2}
\end{center}
\end{figure}

For the creation of the synthetic datasets, we employed the following procedure:
\begin{itemize}
\item We considered several real flare events and, starting from the radiation collected by RHESSI, we reconstructed the corresponding $64 \times 64$ images with the CLEAN algorithm \cite{Hogbom1974}, available in SSW. In particular, we considered three energy ranges of the July 23 2002 flare (00:29:10--00:30:19 UT, 20--22 keV, 41--46 keV and 156--177 keV) and one energy range of the April 15 2002 (00:05:00--00:10:00 UT, 12--14 keV), the August 31 2004 (05:33:00--05:38:00 UT, 10--12 keV) and the June 1 2005 flares (02:37:00--02:41:00 UT, 10--15 keV).
\item These images have been cleared of possible artifacts introduced by the reconstruction method, and suitable scaled in order to have different statistics (and, therefore, different noise levels). The resulting distributions are shown in the first row of figures \ref{figimm1} and \ref{figimm2}.
\item For the $j$-th detector ($j=1,\ldots,9$), we computed the corresponding modulated count profiles $c^{(j)}$ by means of a realistic transmission probability $P^{(j)}(x_s,y_s;t)$ (even if we simplified the procedure by neglecting the hardware-based parameter $K^{(j)}$ in \eqref{cphiell}). {{As concerns the discretization of the observational time, we used $N_j=640$ time bins for each detector.}}
\item We perturbed the count profiles with Poisson noise by using the \texttt{imnoise} function available in the Image Processing Toolbox of Matlab. We will denote the resulting six synthetic datasets as Sim1$,\ldots,$ Sim6, following the order of the original events described in the first item.
\end{itemize}

\noindent Once these six datasets have been created, we reconstructed the target distributions with the EM, SGP, GPE and AS\_CBB methods, by stopping the iterations according to the discrepancy principle described in Section \ref{sec4}. All the algorithms have been initialized with a constant image whose total flux has been deduced by the modulated count profiles (see the appendix of \cite{Bonettini2013b} for more details). The numerical experiments for these methods has been performed by means of routines implemented by ourselves in Matlab.\\
The choice of the subcollimators used in the inversion process requires a brief comment: due to the thin slits of the grids, detectors 1 and 2 are typically discarded since the corresponding count profiles are characterized by a very low SNR. Therefore, we performed three sets of simulations, by using all the subcollimators (set 1), subcollimators 2 to 9 (set 2) and 3 to 9 (set 3). We found the best results in the second case for the datasets Sim1, Sim4 and Sim5, while lower reconstruction errors have been registered in the third case for Sim2, Sim3 and Sim 6 (whose original images have lower flux).\\
As concerns the uv-smooth method, we found the best reconstructions by using always detectors 3 to 9, in agreement with the fact that a fitting step on the raw count profiles amplifies the effect of the noise on the processed data (and, therefore, the information provided by subcollimator 2 which can be exploited in the count-based strategies here have a corruptive effect on the reconstructions). We remark that in this case we used the original IDL routine available in SSW.

\subsection{Results}

\begin{figure}
\begin{center}
\begin{tabular}{ccc}
\includegraphics[width=0.3\columnwidth]{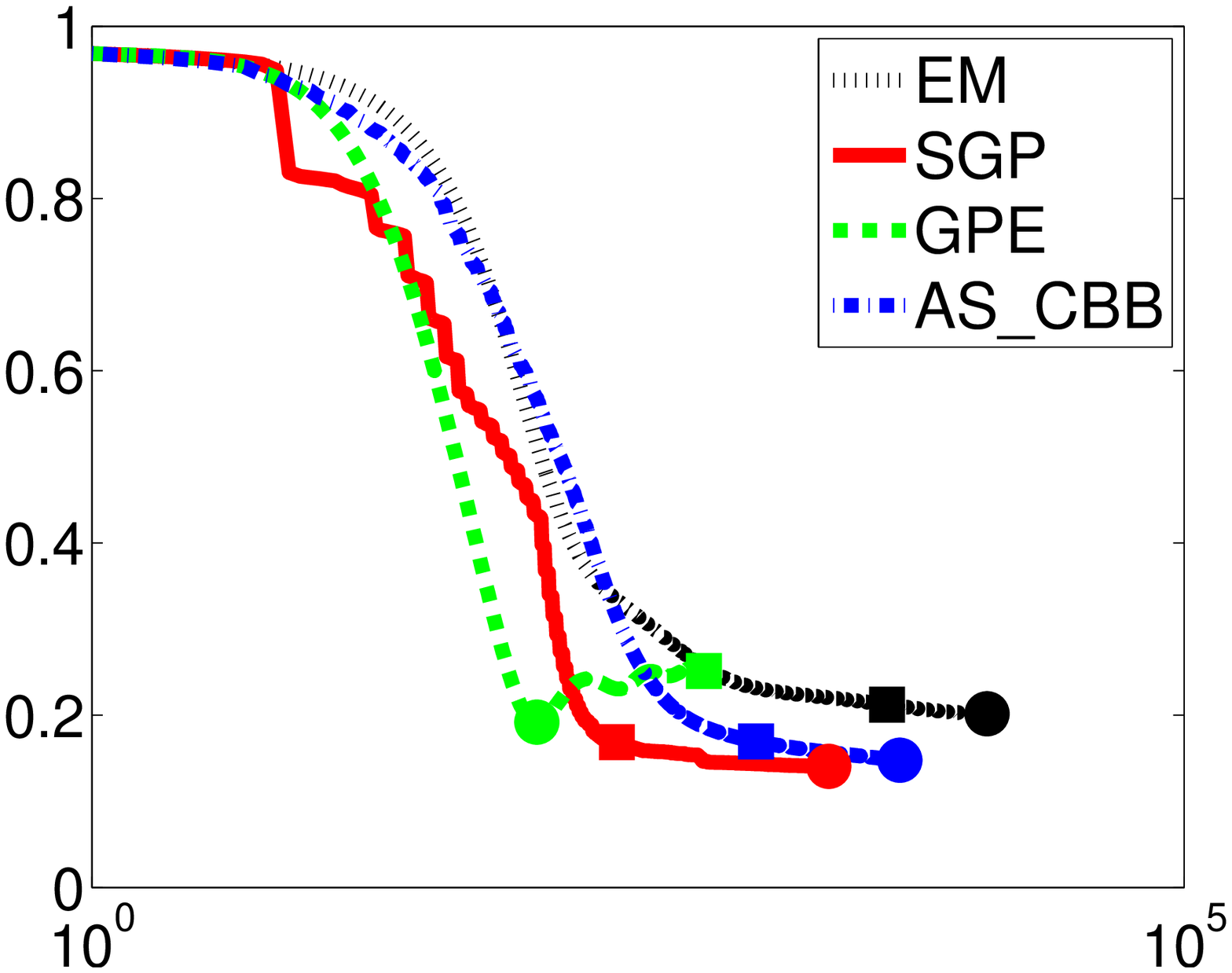} &
\includegraphics[width=0.3\columnwidth]{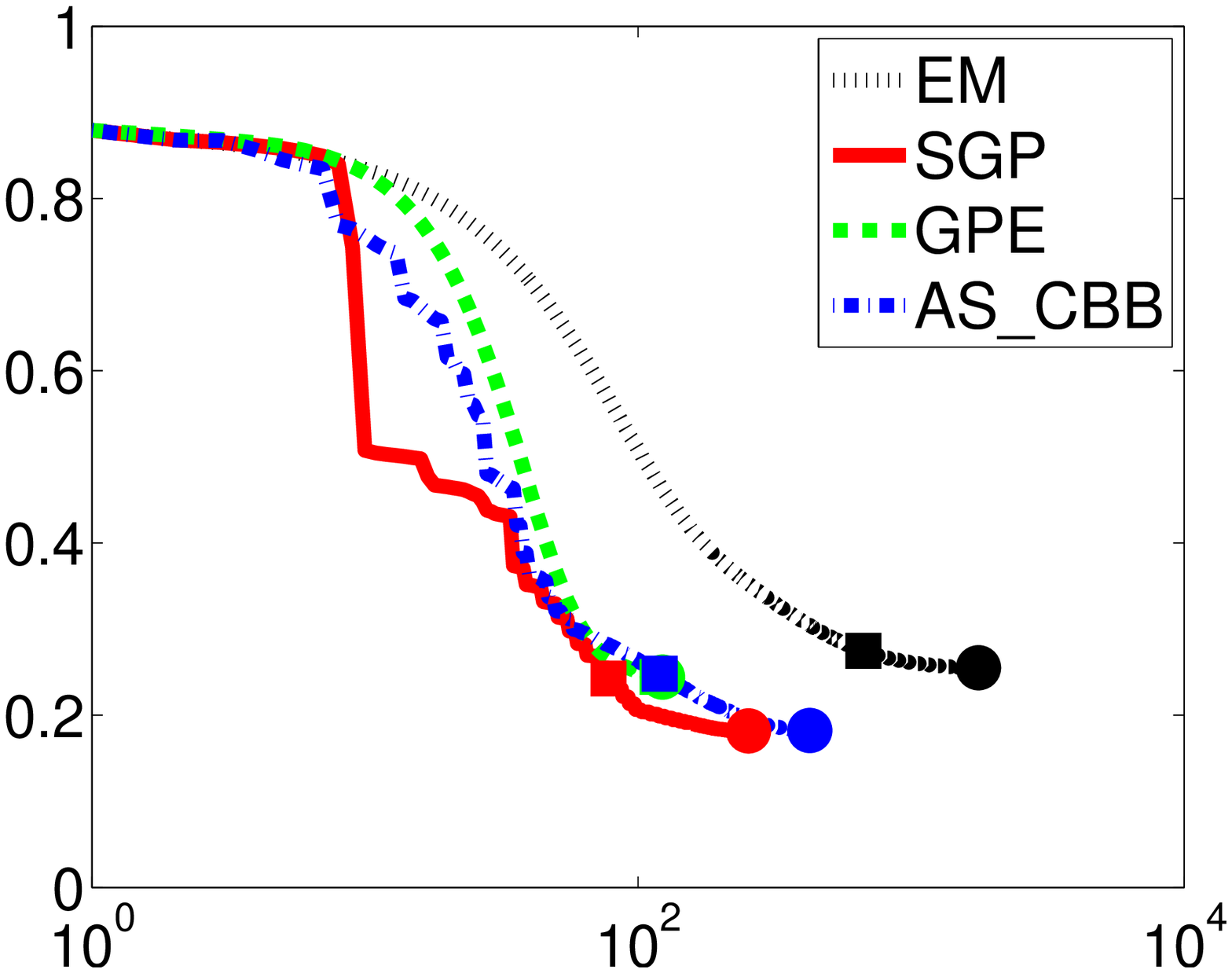} &
\includegraphics[width=0.3\columnwidth]{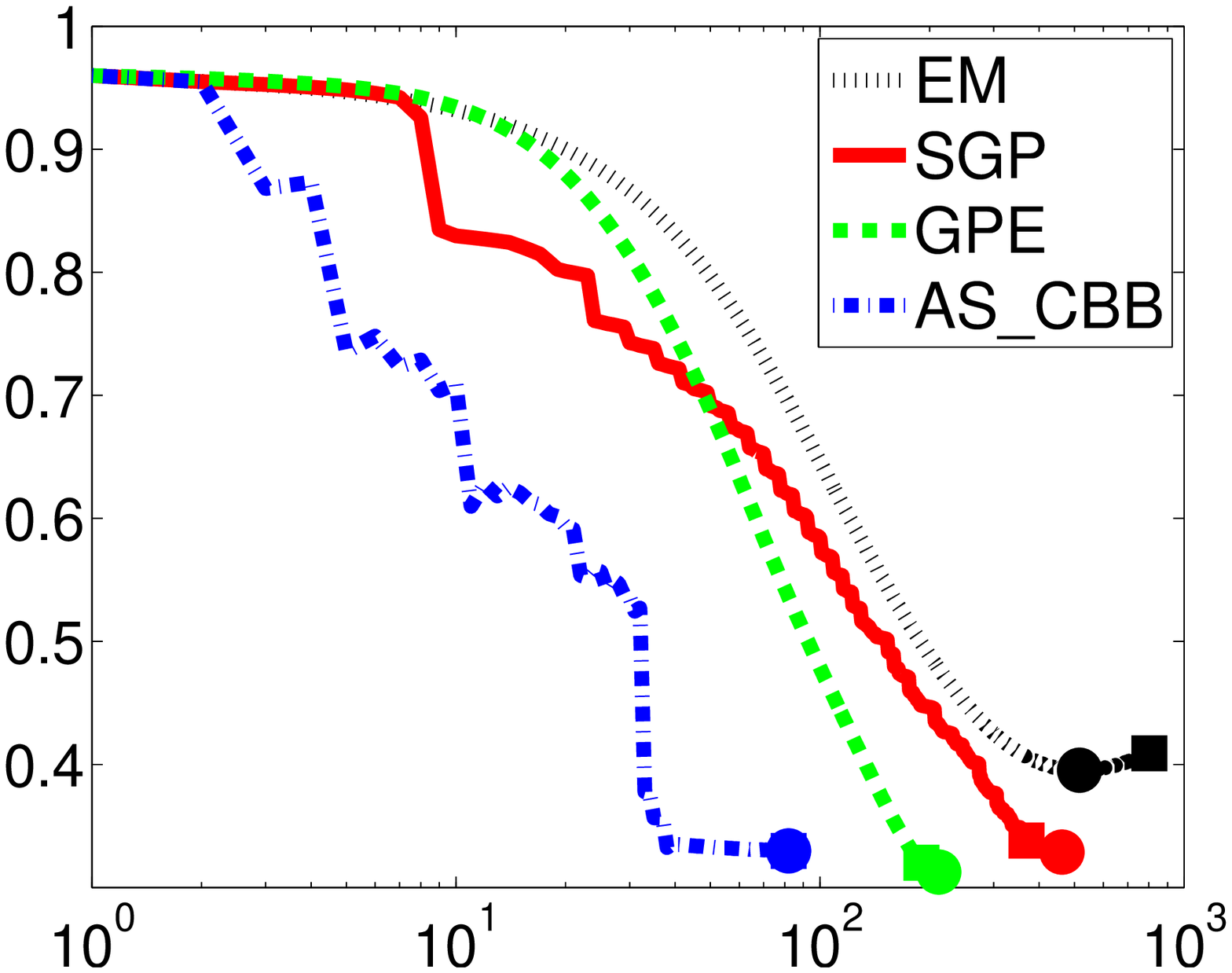} \\
\includegraphics[width=0.3\columnwidth]{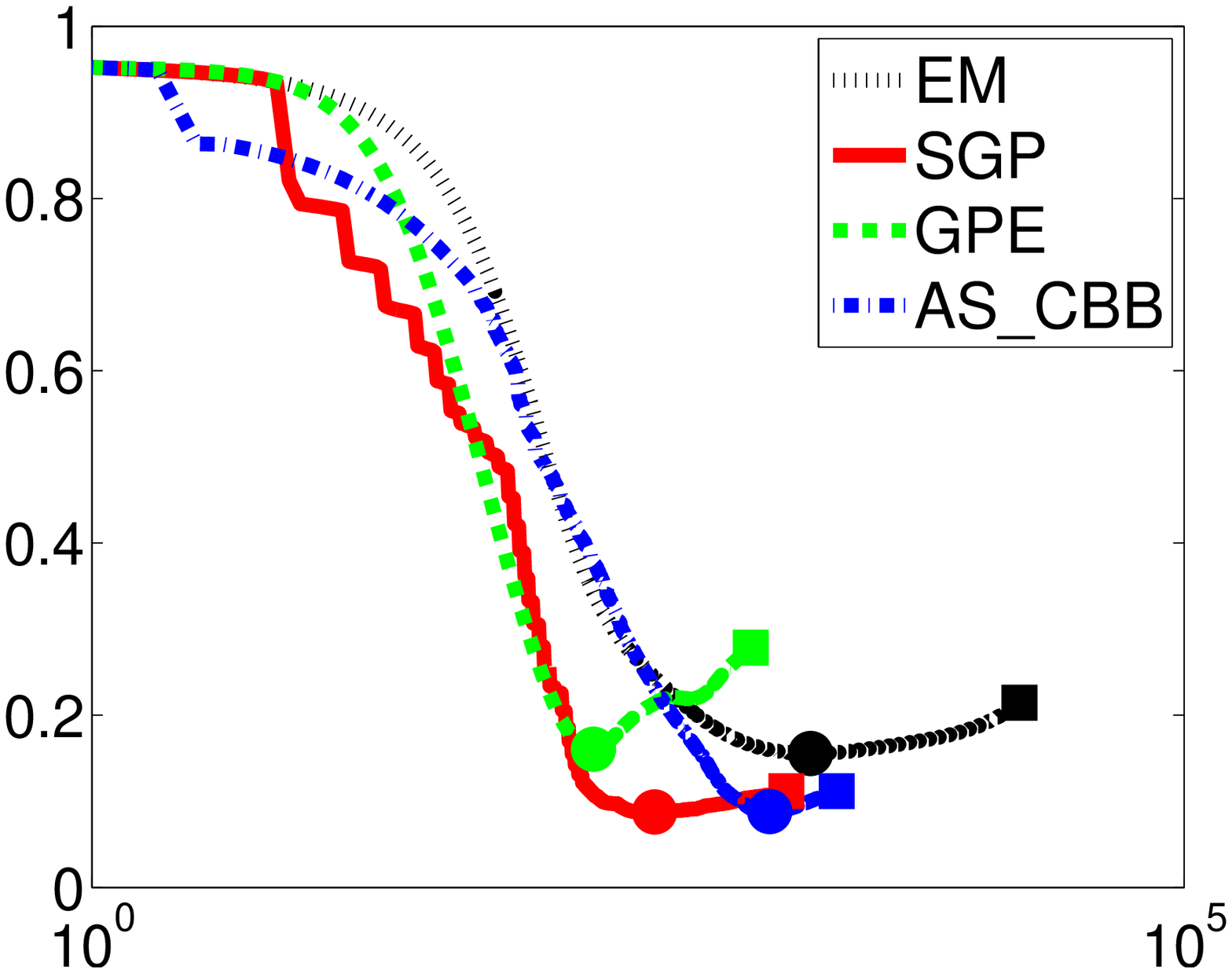} &
\includegraphics[width=0.3\columnwidth]{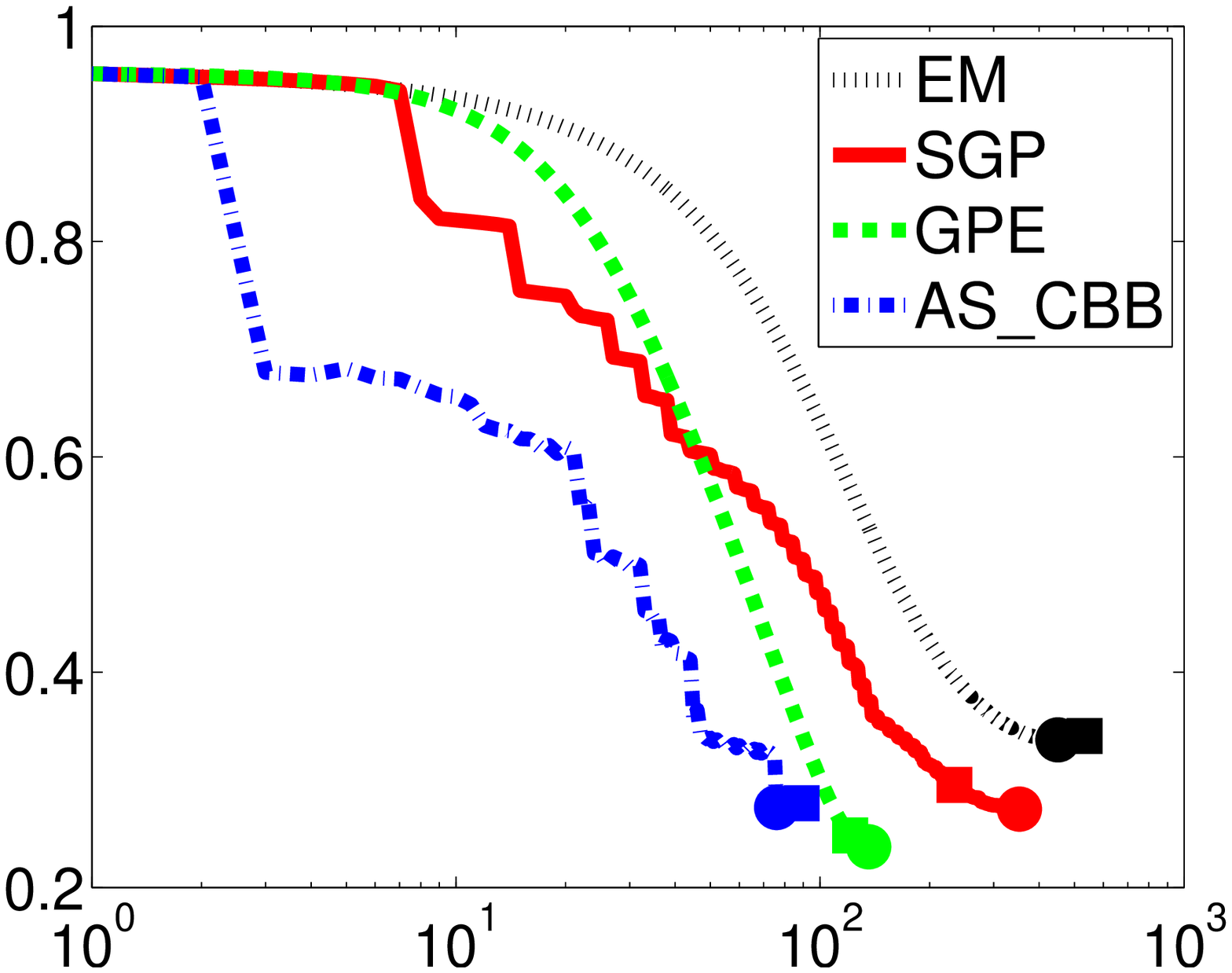} &
\includegraphics[width=0.3\columnwidth]{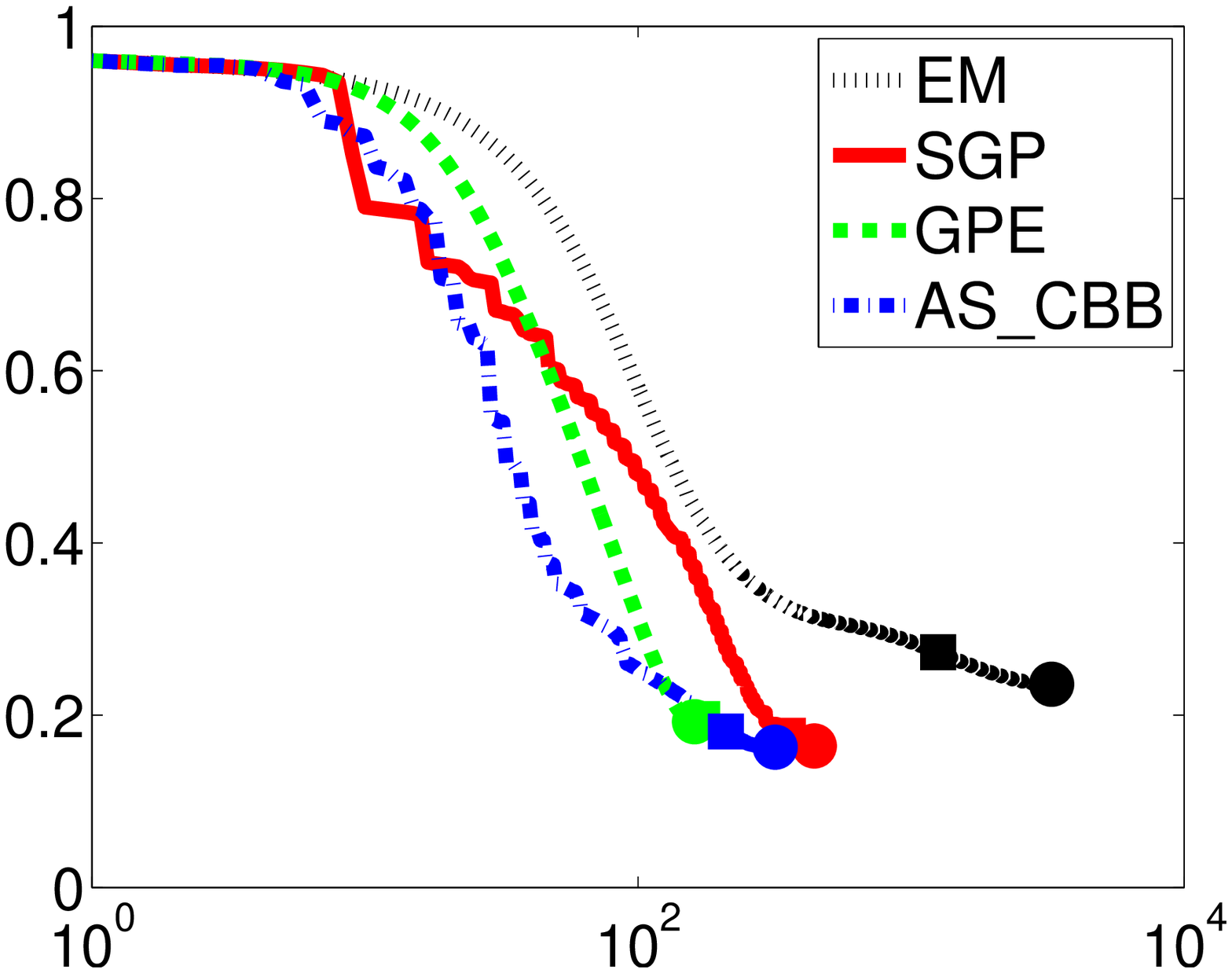} \\
\end{tabular}
\caption{Plots of the reconstruction errors provided by EM, SGP, GPE and AS\_CBB as functions of the iteration number for the six datasets described in the text. The squares indicate the values provided by the discrepancy principle \eqref{discr1}, while the circles those corresponding to the minimum errors.}
\label{figplots1}
\end{center}
\end{figure}

\begin{figure}
\begin{center}
\begin{tabular}{ccc}
\includegraphics[width=0.3\columnwidth]{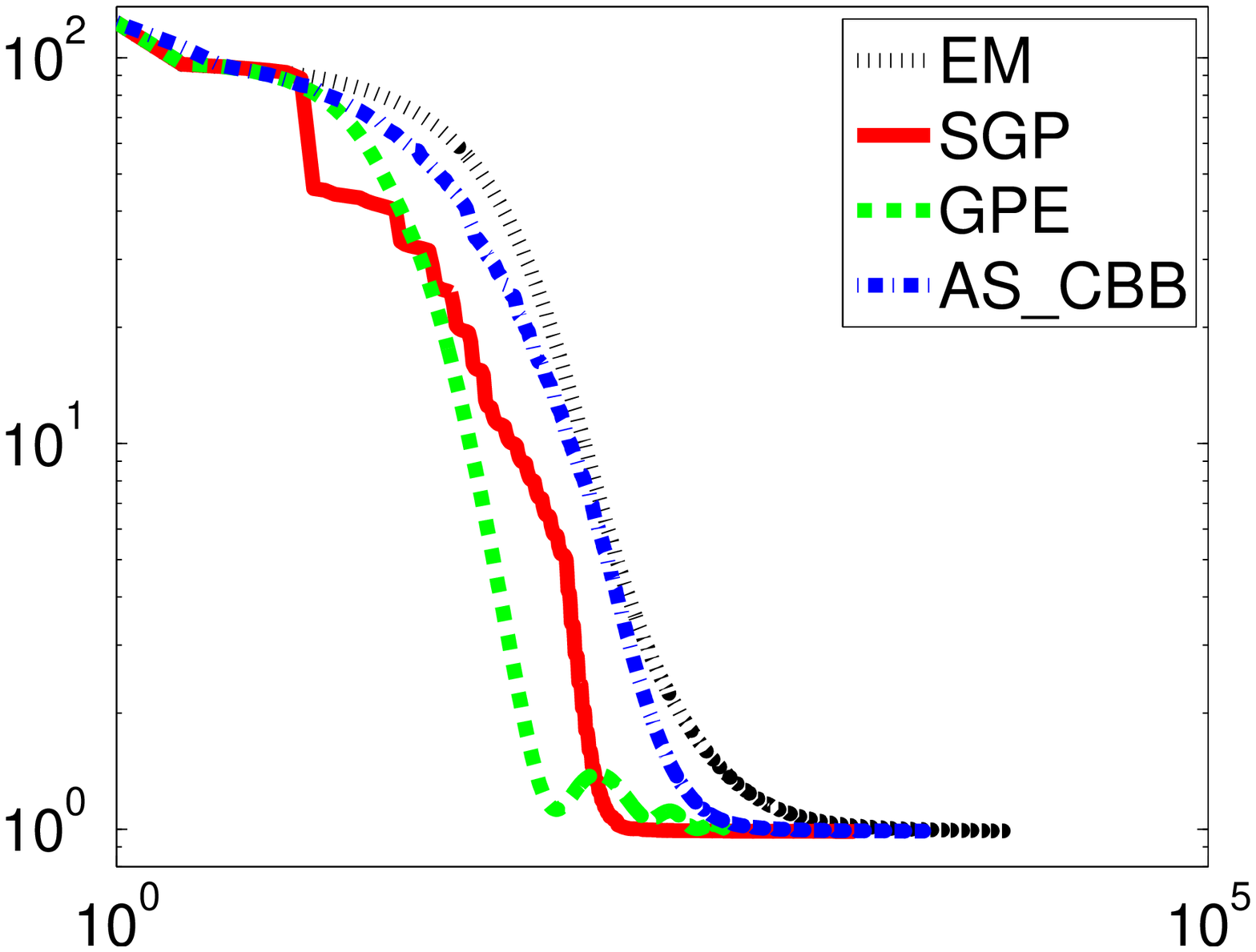} &
\includegraphics[width=0.3\columnwidth]{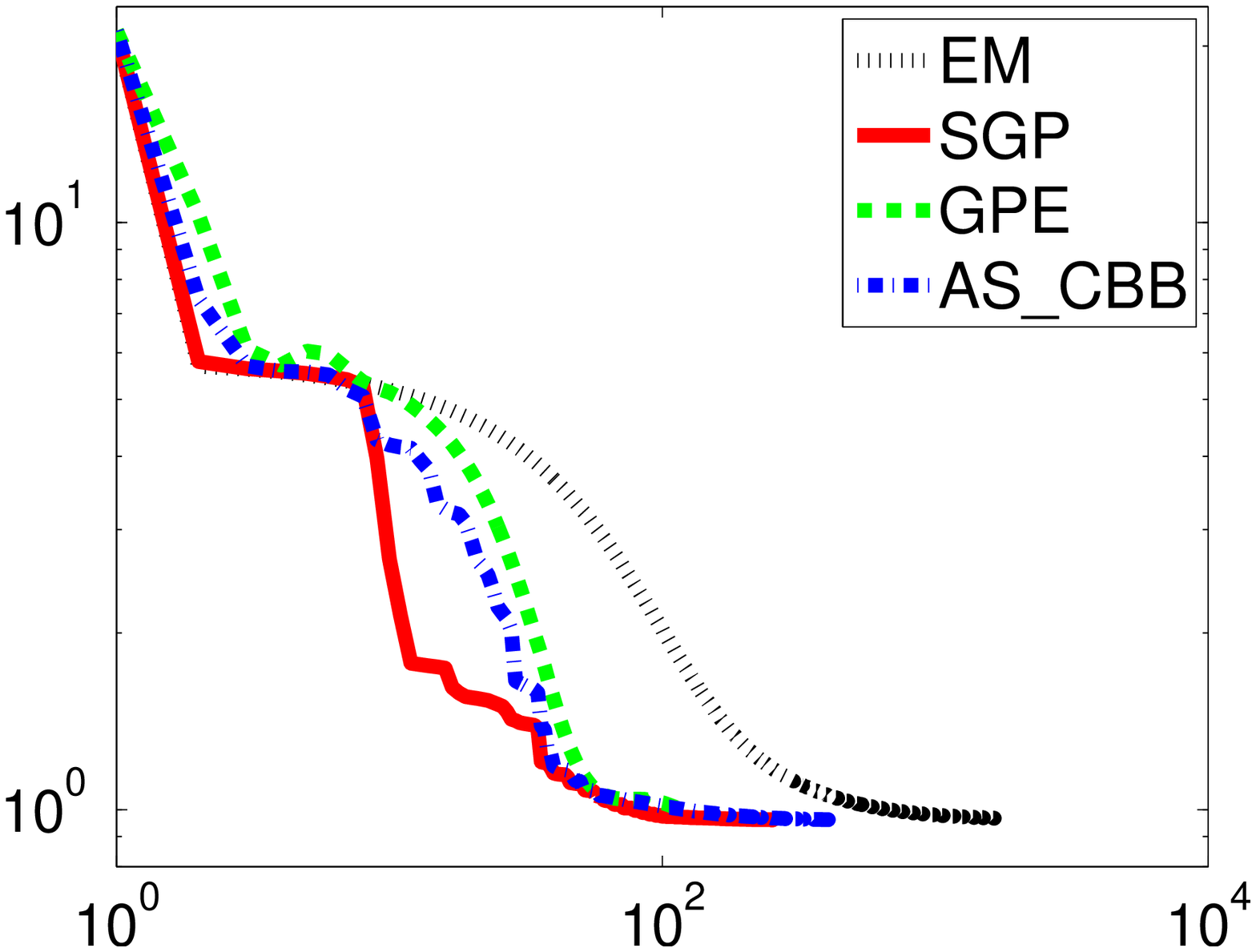} &
\includegraphics[width=0.3\columnwidth]{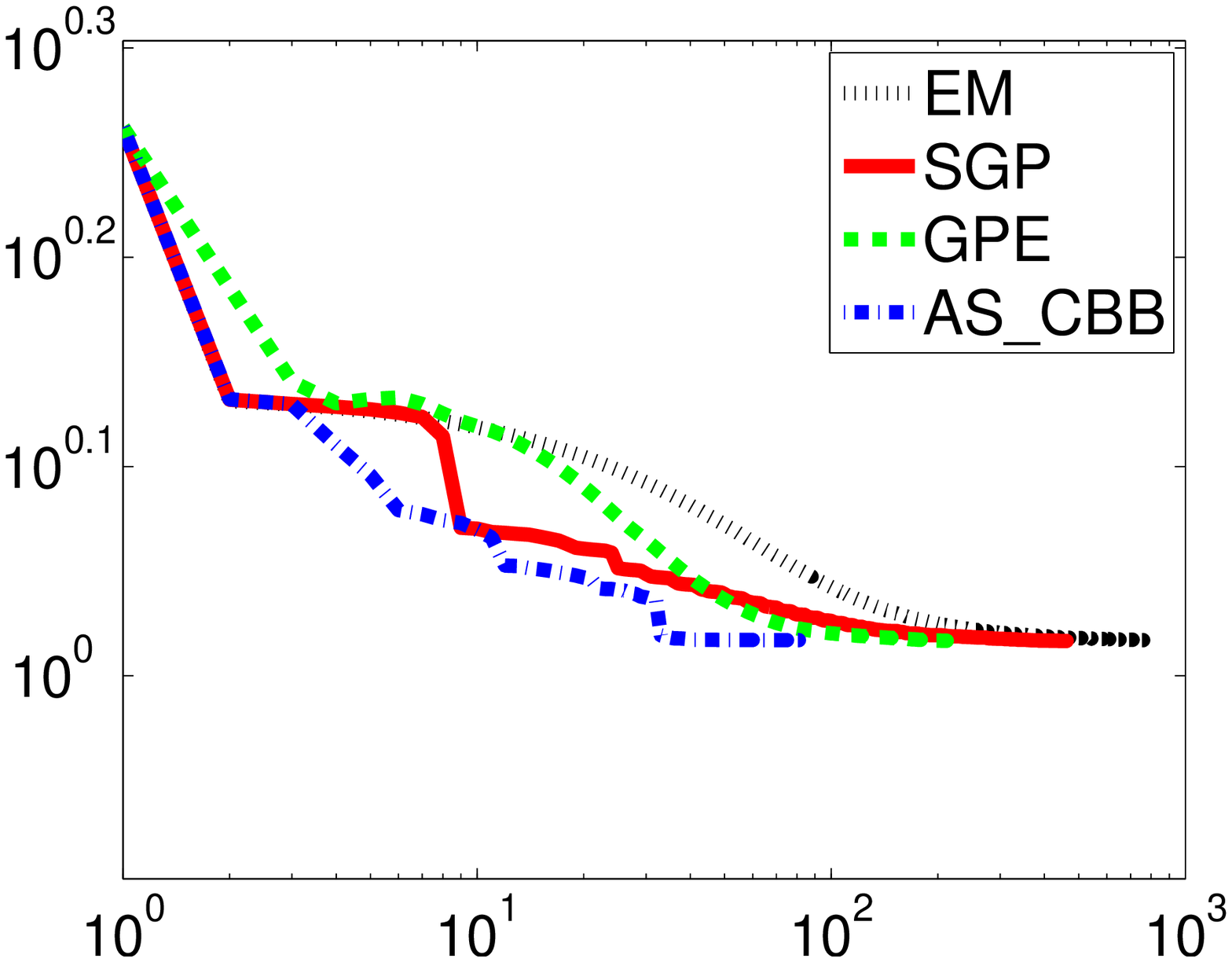} \\
\includegraphics[width=0.3\columnwidth]{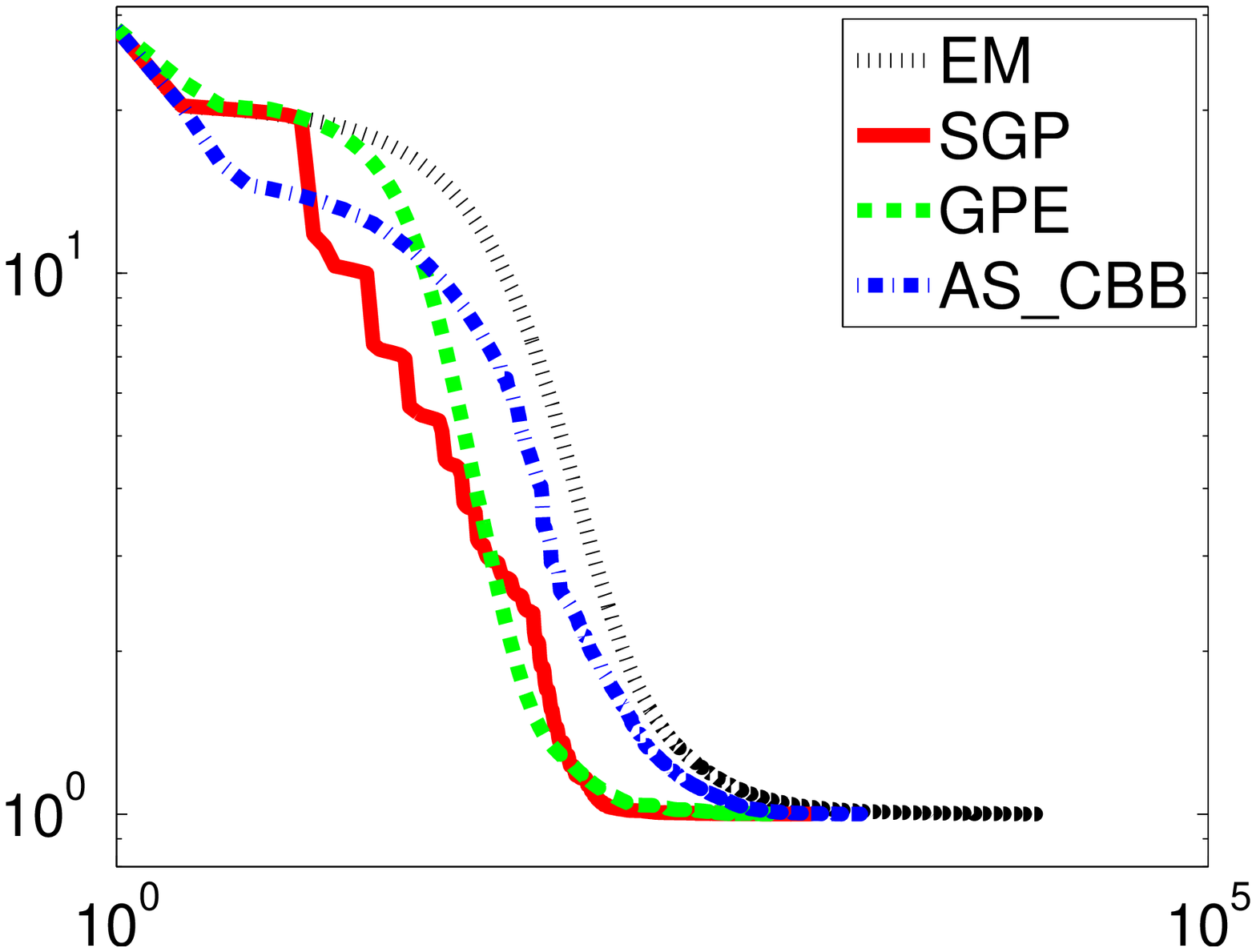} &
\includegraphics[width=0.3\columnwidth]{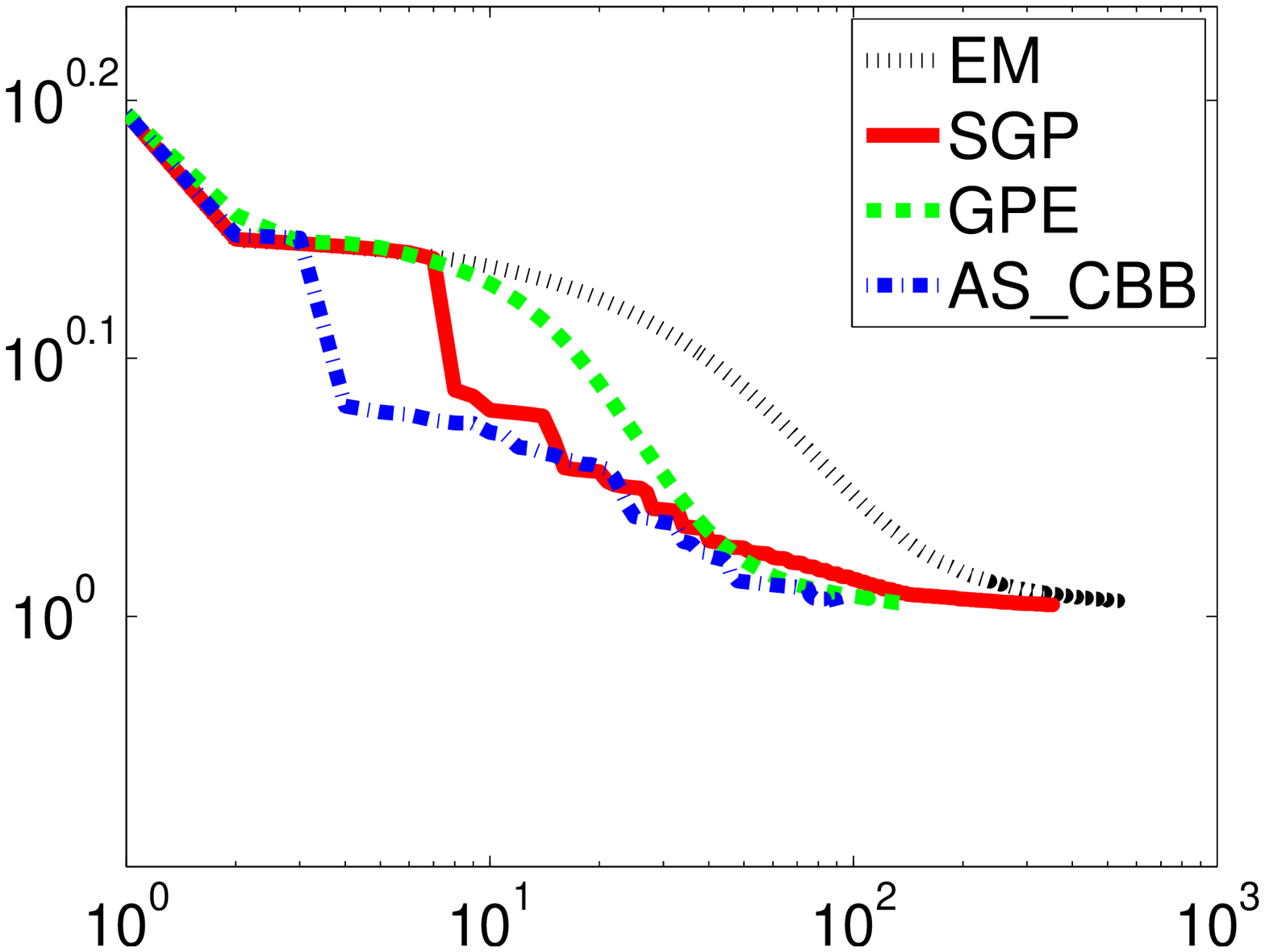} &
\includegraphics[width=0.3\columnwidth]{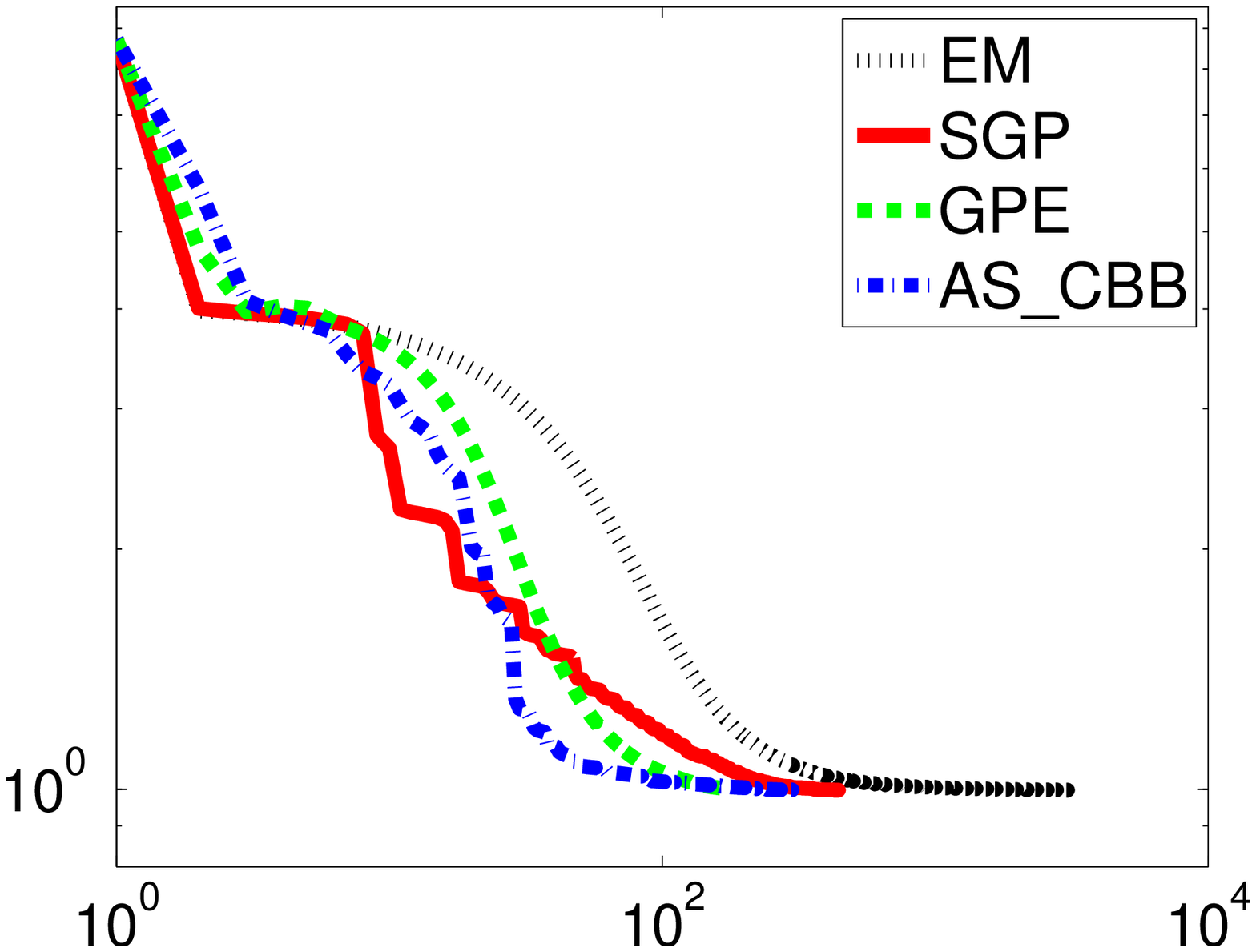} \\
\end{tabular}
\caption{Plots of the discrepancy functions provided by EM, SGP, GPE and AS\_CBB as functions of the iteration number for the six datasets described in the text.}
\label{figplots2}
\end{center}
\end{figure}

The reconstructed images are shown in the last five rows of figures \ref{figimm1} and \ref{figimm2}, while in table \ref{table} we reported the relative reconstruction errors (in Euclidean norm) and the iterations required by the gradient methods. In order to evaluate the effectiveness of the stopping criterion for the count-based algorithms, we also added in the table the analogous values corresponding to the best reconstructions (i.e., the ones with the lowest reconstruction error). {{Finally, in figures \ref{figplots1} and \ref{figplots2} we plotted the reconstruction errors and the discrepancy functions $\frac{2}{\sum_{j=d_{min}}^9N_j}D_{KL}(c,Pf^{(k)})$ against the iteration number, where $d_{min}$ is 2 or 3 according to the detectors used in our simulations (see discussion above).}}

\begin{table}
\caption{\label{table}Reconstruction errors provided by EM, SGP, GPE and AS\_CBB used as iterative regularization methods for solving \eqref{minpro} in the six simulations described in the text. Both the values obtained with the discrepancy principle \eqref{discr1} and the optimal values (i.e., the ones minimizing the reconstruction error) are shown. The reconstruction errors obtained with the visibility-based uv-smooth algorithm are also reported. The lower values for each simulation are highlighted in bold.}
\begin{indented}
\item[]\begin{tabular}{@{}cccccccc}
\br
                           &      & Sim1       & Sim2       & Sim3       & Sim4       & Sim5       & Sim6        \\
\mr
\multirow{2}{*}{EM Discr}  & Err  & 0.212      & 0.274      & 0.409      & 0.214      & 0.341      & 0.273       \\
											     & Iter & 4369       & 669        & 803        & 17632      & 533        & 1256        \\
\multirow{2}{*}{SGP Discr} & Err  & \bf{0.169} & \bf{0.243} & 0.338      & \bf{0.112} & 0.295      & \bf{0.176}  \\
											     & Iter & 253        & 78         & 369        & 1511       & 234        & 354         \\
\multirow{2}{*}{GPE Discr} & Err  & 0.251      & 0.245      & \bf{0.320} & 0.278      & \bf{0.248} & 0.196       \\
											     & Iter & 634        & 118        & 190        & 1038       & 121        & 172         \\
\multirow{2}{*}{AS\_CBB Discr} & Err  & 0.170      & 0.248      & 0.330      & \bf{0.112} & 0.278      & 0.181       \\
											     & Iter & 1098       & 120        & 82         & 2568       & 89         & 210         \\
\mr
\multirow{2}{*}{EM Best}   & Err  & 0.202      & 0.255      & 0.395      & 0.155      & 0.337      & 0.236       \\
											     & Iter & 12536      & 1766       & 516        & 1952       & 450        & 3271        \\
\multirow{2}{*}{SGP Best}  & Err  & \bf{0.141} & \bf{0.182} & 0.329      & \bf{0.088} & 0.273      & 0.164       \\
											     & Iter & 2361       & 254        & 462        & 377        & 353        & 442         \\
\multirow{2}{*}{GPE Best}  & Err  & 0.192      & 0.244      & \bf{0.312} & 0.161      & \bf{0.238} & 0.192       \\
											     & Iter & 109        & 123        & 212        & 198        & 136        & 161         \\
\multirow{2}{*}{AS\_CBB Best}  & Err  & 0.148      & \bf{0.182} & 0.330      & \bf{0.088} & 0.274      & \bf{0.163}  \\
											     & Iter & 5000       & 426        & 82         & 1267       & 76         & 318         \\
\mr
uv-smooth                  & Err  & 0.278      & 0.280      & 0.446      & 0.173      & 0.323      & 0.252       \\
\br											
\end{tabular}
\end{indented}
\end{table}

\noindent From the reconstructed images and the corresponding errors we can make the following observations:
\begin{itemize}
\item The accelerated methods are able to highly reduce the number of iterations required by EM to provide the reconstruction, with a gain ranging from a factor of about 2 to a factor of about 17. We remark that the cost per iteration of all the methods are essentially the same, since the main computations are the two matrix-vector products $Pf^{(k)}$ and $P^T\left(\frac{c}{Pf^{(k)}}\right)$ ({{few further scalar products are needed by SGP and AS\_CBB to compute the steplength, while few further matrix-vector products could be required by GPE due to its backtracking procedure}}). Therefore, since the IDL implementation of EM in SSW requires computational times of few tens of seconds \cite{Benvenuto2013}, we expect that, once the accelerated algorithms will be implemented in IDL and included in SSW by the RHESSI Software Team, the overall CPU effort required to produce the desired image will be very low, and comparable to that of the visibility-based methods (whose computational cost includes the time required to compute the visibilities by fitting the count profiles).
\item Quite surprisingly, besides the computational gain the accelerated methods are able to provide also some improvements in the quality of the final images, as attested numerically by the lower reconstruction errors in table \ref{table} and visually by the pictures in figures \ref{figimm1} and \ref{figimm2}. We remark that these improvements do not depend on the stopping criterion we adopted, since they still hold true also if one looks at the best errors achievable by the different methods.
\item The discrepancy principle we chose as stopping criterion seems to work in general quite well, since it is able to stop the iterates very close to the optimal ones. Of course some exceptions are present, and in some cases the algorithms are arrested when the reconstruction error is already growing back, while in other cases they are stopped too early (see figure \ref{figplots1}). However, we think that this behaviour is honestly unavoidable when a general automatic stopping procedure is adopted.
\item {{We did not find any clear prevalence of one accelerated scheme with respect to the others. The SGP and AS\_CBB algorithms provides similar reconstruction errors, while the performance of GPE seems to be more unstable (in Sim3 and Sim5 it provides the lowest errors, while in Sim1 and Sim4 the highest ones). Also in the number of iterations required we found different problem-dependent behaviours.}}
\item As concerns the comparison with uv-smooth, we can observe what, as it could be expected, the accuracy of the reconstructions is in general improved by the proposed algorithms, since no systematic error is present on the raw data.
\end{itemize}
As final test, we applied SGP to problem \eqref{minproreg} in order to investigate the effect of the addition of a zero order or first order Tikhonov term in the objective function to ensure a certain smoothness of the reconstructed image. In this case, the regularized solutions have been obtained according to the procedure described in Section \ref{sec4}. We reported the reconstruction errors in table \ref{table2}, together with the optimal values for both the regularization parameters $\eta^*$ and the tolerance for the stopping criterion $\varepsilon$. Our simulations do not show clear indications on the effectiveness of the presence of the Tikhonov term, since only in some cases improvements in the reconstruction can be noticed. We can conclude that, since the computational procedure to find a regularized solution of \eqref{minproreg} is much heavier than the use of SGP as iterative regularization method to solve \eqref{minpro}, the latter approach is definitely preferable.

\begin{table}
\caption{\label{table2}Reconstruction errors provided by SGP applied to \eqref{minproreg} with $\mathcal{L}$ equal to the identity matrix or the discrete gradient operator in the six simulations described in the text. The regularization parameters $\eta^*$ obtained with the discrepancy principle \eqref{discr2} are shown together with the values of $\varepsilon$ resulting from the procedure described in Section \ref{sec4}.}
\begin{indented}
\item[]\begin{tabular}{@{}cccccccc}
\br
                             &               & Sim1    & Sim2    & Sim3    & Sim4    & Sim5    & Sim6    \\
\mr
											       & $\varepsilon$ & 7.82e-5 & 7.31e-4 & 4.00e-2 & 2.62e-4 & 1.40e-2 & 2.43e-3 \\
\multirow{2}{*}{SGP + Tikh0} & Err           & 0.158   & 0.232   & 0.364   & 0.129   & 0.331   & 0.184   \\
											       & $\eta^*$      & 0.100   & 21.54   & 316.2   & 0.316   & 316.2   & 14.68   \\
\multirow{2}{*}{SGP + Tikh1} & Err           & 0.164   & 0.237   & 0.331   & 0.120   & 0.265   & 0.172   \\
											       & $\eta^*$      & 0.681   & 464.2   & 1468    & 0.4642  & 2154    & 100     \\
\br											
\end{tabular}
\end{indented}
\end{table}

\section{Conclusions}\label{sec6}

This paper deals with the real-world problem of reconstructing an X-ray image of a solar flare by means of the data collected by RHESSI. This problem can be addressed in two ways, both of which present advantages and drawbacks. The visibility-based methods are very fast but acts on preprocessed data, characterized by systematic errors. The count-based strategies can be applied straightly on the raw data, but require the computation of the transmission probabilities, which overloads the cost per iteration of the reconstruction algorithms. Our work belongs to the latter category and, starting from the statistical model proposed in \cite{Benvenuto2013}, aims to decrease significantly the number of iterations needed by means of very powerful optimization schemes, thus leading to a computational effort comparable with that of the methods acting on the visibilities.\\
In order to achieve a meaningful solution we introduced regularization both by early stopping the proposed methods applied to the minimization of the KL divergence and by using SGP as optimization method to exactly minimize a weighted sum of the KL and the Tikhonov functional. As automatic criterion for the choice of the number of iterations or the regularization parameter, we adopted a recently proposed discrepancy principle for Poisson data, suitable adapted to take into account different levels of noise.\\
Our numerical experiments on synthetic datasets show that the accelerated gradient algorithms used as iterative regularization methods instead of EM are able to reduce substantially not only the number of iterations, but also the reconstruction error, thus leading to a faster recovery of higher quality images. An analysis on the same data showed that the approach of adding a Tikhonov term to the KL does not provide sufficient improvements to justify the much higher computational effort required. Finally, the comparison with a state-of-the-art visibility-based method showed that the reconstruction errors provided by the proposed methods are systematically lower. Starting from the analysis of the execution times carried out in \cite{Benvenuto2013} and taking into account the gain in the iterations shown in this paper with respect to EM, we are confident that these methods implemented within SSW will run in a computational time comparable with that of the visibility-based methods. Moreover, the integration of these algorithms in SSW will allow an intensive comparison with all the other reconstruction algorithms on the real datasets provided by RHESSI.\\
Future work will involve the application of an accelerated gradient method to the RHESSI spectroscopy problem, in which the measured count spectra are related to the electron spectra which emitted those counts by a Volterra integral equation (see e.g. \cite{Prato2009a}).

\section*{Acknowledgments}

This work has been partially supported by MIUR (Italian Ministry for University and Research), under the projects FIRB - Futuro in Ricerca 2012, contract RBFR12M3AC, and PRIN 2012, contract 2012MTE38N. The Italian GNCS - INdAM (Gruppo Nazionale per il Calcolo Scientifico - Istituto Nazionale di Alta Matematica) is also acknowledged.

\section*{References}

\bibliographystyle{unsrt}
\bibliography{biblio_Marco}

\end{document}